\documentclass[%
pre,
twocolumn,
 amsmath,amssymb,
 aps,
]{revtex4-2}
\UseRawInputEncoding
\usepackage[normalem]{ulem}
\usepackage{soul}
\usepackage{amsmath}
\usepackage{amssymb}
\usepackage{color}
\usepackage{graphicx}
\usepackage{dcolumn}
\usepackage{bm}
\usepackage{verbatim}

\begin{document}

\title{Particles on Demand for flows with strong discontinuities }

\author{N. G. Kallikounis}
\affiliation
{Department of Mechanical and Process Engineering, ETH Zurich, 8092 Zurich, Switzerland}
\author{B. Dorschner}
\affiliation
{Department of Mechanical and Process Engineering, ETH Zurich, 8092 Zurich, Switzerland}
\author{I. V. Karlin}\thanks{Corresponding author}
 \email{ikarlin@ethz.ch}
\affiliation
{Department of Mechanical and Process Engineering, ETH Zurich, 8092 Zurich, Switzerland}

\date{\today}

\begin{abstract}

Particles on Demand formulation of kinetic theory [B. Dorschner, F. B\"{o}sch and I. V. Karlin, {\it Phys. Rev. Lett.} {\bf 121}, 130602 (2018)] is used to simulate a variety of compressible flows with strong discontinuities in density, pressure and velocity. Two modifications are applied to the original formulation of the Particles on Demand method. First, a regularization by Grad's projection of particles populations is combined with the reference frame transformations in order to enhance stability and accuracy. Second, a finite-volume scheme is implemented which allows tight control of mass, momentum and energy conservation. The proposed model is validated with an array of challenging one- and two-dimensional benchmarks of compressible flows, including hypersonic and near-vacuum situations, Richtmyer-Meshkov instability, double Mach reflection and astrophysical jet. Excellent performance of the modified Particles on Demand method is demonstrated beyond the limitations of other lattice Boltzmann-like approaches to compressible flows.

\end{abstract}

\maketitle

\section{Introduction}

The lattice Boltzmann method (LBM) is a recast of fluid dynamics into a fully discrete kinetic system of designed particles with the discrete velocities $\bm{c}_i$, $i=0,\dots,Q-1$, fitting into a regular space-filling lattice, with the kinetic equation for the populations $f_i(\bm{x},t)$ following a simple algorithm of ``stream along links $\bm{c}_i$ and collide at the nodes $\bm{x}$ in discrete time $t$". Since its inception  \citep{Higuera_1989_SucciBenzi,Higuera_1989}, LBM has evolved into a versatile tool for the simulation of complex flows including  transitional flows \citep{Dorschner2017JFM}, flows in complex moving geometries \citep{dorschner2016entropic}, thermal and convective flows \cite{he1998,guo2007twopop,karlinConsistent}, multiphase and multicomponent flows \citep{Mazloomi2015prl,Mazloomi2017JFM,MultiPhase_Wohrwag,NileshJFM1}, reactive flows \cite{NileshReactive} and rarefied gas {\citep{shan2006kinetic}}, to mention a few recent instances; see \cite{SHARMAReview,KrugerBook,SucciBook} for a discussion of LBM and its application areas.

Arguably, LBM is most advantageous for nearly-incompressible fluid flow due to  exact (lattice) propagation combined with relatively simple discrete velocities, referred to as standard lattices. However, the same features become an obstacle for compressible flows. Several avenues of extending LBM to high-speed flows have been explored. First, a number of proposals proceed with extended sets of discrete velocities while retaining exact lattice propagation \cite{Chikatamarla2006,Chikatamarla2009,Alexander1993}. LBM with extended sets of discrete velocities demonstrates excellent results \cite{Frappoli_2015,Frappoli_2016}. However, the gain in both the temperature and Mach number is perceived as moderate when compared to the increased complexity of the higher-order lattices. A second approach uses standard lattices, while LBM is augmented with corrections terms tailored to eliminate error terms in momentum and energy equations \cite{Prasianakis2007,Saadat2019,HosseinExtended1,HosseinExtended2}. While trans- and supersonic flows can be captured efficiently with this approach, it too remained limited to moderate Mach number and discontinuities.

Another line of research abandons the restriction of lattice-fitting discrete velocities and proposes a less rigid off-lattice propagation. While originally conceived for standard velocity sets in order to gain geometrical flexibility \cite{Bardow_2006,InterBasedLB_Shu,InterpoBasedLB_cHENG_HUNG}, off-lattice propagation schemes received some traction more recently in the context of compressible flows. Interesting realizations for compressible flows at moderate Mach numbers using a semi-Lagrangian advection  have recently been reported in \cite{SemiLag2017,Kramer_SemiLagrangian2020,Wilde2020_Turbulence}. Another class of numerical schemes are finite volume methods, such as the discrete unified gas kinetic scheme (DUGKS) \cite{Guo_DUGKS_IsoT,Guo_DUGKS_Compres}. Finite-difference propagation was used in the discrete Boltzmann model (DBM) targeting compressible flows with applications to combustion and detonation  \cite{DBM_2018_XU,DBM_2021_LUO,DBM_SUCCI_XU2015,DBM_SUCCI_XU2016,DBM_SUCCI2015}.

Regardless of the propagation scheme (on- or off-lattice), a common feature of the above models is the use of fixed discrete velocities which amounts to choosing the reference frame "at rest". While the latter is viable and even advantageous (due to lattice propagation) for nearly-incompressible, slow flows, it impedes  the use of kinetic-theory based solvers for high Mach number situations. Examples of a better reference frame are readily available. For instance, in \cite{Frapolli_ShiftedLattices}, the formulation of LBM in a co-moving Galilean reference frame demonstrated excellent performance for predominantly unidirectional compressible flows. Subsequently, the Particles on Demand (PonD) method \cite{Pond} addressed the problem of finding the optimal reference frame. In PonD, the discrete particle velocities are constructed relative to the local reference frame, which is defined by the local flow velocity and temperature, and which varies in space and time. This necessitates an off-lattice propagation scheme, together with a reference frame transformation to resolve the implicitness of the PonD scheme.

In this work, we aim at a further development of PonD in order to enable simulations at extreme cases of compressible flows. First, the reference frame transformation is supplemented with a regularization procedure based on Grad's projection. In the framework of PonD, regularization was recently suggested in \cite{RegPond}, with the purpose of reducing the computational cost. Here we show that a carefully tailored moment system, serving as a basis for the transformation, enables PonD to simulate flows with  strong discontinuities. Second, in addition to the semi-Lagrangian realization, we propose a finite-volume version of the PonD, based on the appropriate extension of the DUGKS method. This enables a tight control of the conservation laws which becomes especially important for flows with near-vacuum components.	It is noted that high Mach number flows, with near-vacuum regions and strong discontinuities, remain active area of research in classical CFD, with recent development of higher-order schemes such as targeted essentially non-oscillatory (TENO), besides more established shock-capturing methods  \cite{LinFuAllSpeed,ZhangShu2010, ZhangShu2012}. The proposed model is able to capture complicated flows robustly and accurately, without the need of positivity preserving schemes and sophisticated limiters \cite{ZhangShu2010}.

The paper is organized as follows. In Sec.\ \ref{Model_Description}, the two-population kinetic model is introduced and the reference frame transformation is explained. Subsequently, in Sec. \ref{Numerical_discretization} the semi-Lagrangian and the finite-volume discretization schemes are presented. In Sec.\ \ref{Results}, the model is validated against extreme one-dimensional Riemann problems and various two-dimensional benchmarks. Finally, concluding remarks are provided in Sec.\ \ref{Conclusion}.

\section{Model Description}
\label{Model_Description}

\subsection{Discrete velocities}

Without a loss of generality, we consider discrete speeds in two dimensions formed by tensor products of roots of Hermite polynomials $c_{i\alpha}$,
\begin{equation}
	\label{eq:ci}
	\bm{c}_i=(c_{ix}, c_{iy}).
\end{equation}
The model is characterized by the lattice temperature $T_L$ and the weights $W_i$ associated with the vectors (\ref{eq:ci}),
\begin{equation}
	\label{eq:wi}
	W_i=w_{ix}w_{iy},
\end{equation}
where $w_{i\alpha}$ are weights of the Gauss--Hermite quadrature. In this work we use the $D2Q16$ velocity set, where $D=2$ stands for two dimensions and $Q=16$ is the number of the discrete velocities. The discrete velocities and the associated weights are shown in Tab. \ref{tab:GaussHermiteVelSets}. With the discrete speeds (\ref{eq:ci}), the particles' velocities $	\bm{v}_i$ are defined relative to a reference frame,
specified by the frame velocity $\bm{u}_{{\rm ref}}$ and the reference temperature $T_{{\rm ref}}$,
\begin{equation}\label{eq:veli}
	\bm{v}_i=\sqrt{\frac{T_{\rm ref}}{T_L}}\bm{c}_i+\bm{u}_{\rm ref}.
\end{equation}
The optimal reference frame is the co-moving reference frame, which is specified by the \emph{local} temperature $T_{\rm ref}=T(\bm{x},t)$ and the \emph{local} flow velocity $\bm{u}_{\rm ref}=\bm{u}(\bm{x},t)$.

\begin{table}[h] \centering
	\caption{Lattice temperature $T_L$, roots of Hermite polynomials $c_{i\alpha}$ and weights $w_{i\alpha}$ of the $D=1$ Gauss--Hermite quadrature, and nomenclature.}
	\label{tab:GaussHermiteVelSets}
	\begin{tabular}{l|l|l|l|l}
		Model & $T_L$   &$c_{i\alpha}$      & $w_{i\alpha}$    & $D=2$     \\ 
		      &         &                   &                   &    \\
		$D1Q3 $ & $1$ &  $0,$     & $2/3$   & $D2Q9$          \\ 
		&    &$\pm\sqrt{3}$   & $1/6$       &       \\ 
		&        &                &          &           \\
		$D1Q4$ & $1$  &$\pm \sqrt{3-\sqrt{6}} $     & $(3+\sqrt{6})/12$   &   $D2Q16$         \\
		&   &$\pm \sqrt{3+\sqrt{6}}$     & $(3-\sqrt{6})/12$           & 
	\end{tabular}
\end{table}

\subsection{Kinetic equations}

In this paper, we restrict our consideration to a single relaxation time, two-population kinetic model for ideal gas with variable adiabatic exponent \cite{Frappoli_2015},
\begin{align}
\label{eq:f_equation}
{\partial_t f_i}+\bm{v}_i \cdot \nabla f_i=\Omega_{f,i}=\frac{1}{\tau}(f_i^{\rm eq}-f_i),\\
{\partial_t g_i}+\bm{v}_i \cdot \nabla g_i=\Omega_{g,i}=\frac{1}{\tau}(g_i^{\rm eq}-g_i), 
\label{eq:g_equation}
\end{align}
where $f_i^{\rm eq}$ and $g_i^{\rm eq}$ are local equilibrium populations and $\tau$ is the relaxation time. Local conservation laws for the density $\rho$, momentum $\rho\bm{u}$ and the total energy $\rho E$ are,
\begin{align}
    \rho &= \sum_{i=0}^{Q-1}f_i = \sum_{i=0}^{Q-1}f_i^{\rm eq}, \\
    \rho\bm{u} &=  \sum_{i=0}^{Q-1}\bm{v}_if_i = \sum_{i=0}^{Q-1}\bm{v}_if_i^{\rm eq}, \\
     \rho E &= \sum_{i=0}^{Q-1} \frac{{v}_i^2}{2}f_i
     	+  \sum_{i=0}^{Q-1} g_i  =  \sum_{i=0}^{Q-1} \frac{{v}_i^2}{2}f_i^{\rm eq}+  \sum_{i=0}^{Q-1} g_i^{\rm eq},
\end{align}
where the total energy of ideal gas is,
\begin{equation}
	\label{eq:E}
	\rho E=C_v\rho T + \frac{\rho{u}^2}{2},
\end{equation}
with $C_v$ the specific heat at constant volume. In the co-moving reference frame, the equilibrium populations depend only on the density and temperature,
\begin{align}
	\label{feqPond}
	&  f_i^{\rm eq} = \rho W_i, \\
	\label{geqPond}
	&  g_i^{\rm eq} = \left(C_v-\frac{D}{2}\right)T\rho W_i.
\end{align}
The single relaxation time Bhatnagar-Gross-Krook (BGK) model \eqref{eq:f_equation} and \eqref{eq:g_equation} results in a Prandtl number equal to one. This restriction is adopted in the present study for the sake of presentation since the benchmark cases considered below refer to non-dissipative compressible flow. Extension of the present model to a variable Prandtl number can be found in \cite{Frappoli_2015,Multiscale2021}.

\subsection{Regularized reference frame transformation}
\label{sec:transformation}

Let us consider a reference frame $\lambda$ defined by a reference temperature $T$ and frame velocity $\bm{u}$,
\begin{equation}\label{eq:reflambda}
    \lambda=\{\bm{u},T\}.
\end{equation}
Discrete velocities relative to the reference frame $\lambda$ (\ref{eq:reflambda}) are defined as,
\begin{equation}\label{eq:vlambda}
\bm{v}_i^\lambda = \sqrt{\frac{T}{T_L}}\bm{c}_i+\bm{u}.
\end{equation}
In order to keep the notation simple, we shall consider $f$-populations ($g$-populations are considered in the same fashion).
A key element of PonD is the transformation of populations $f_i^{\lambda}$, defined with respect to a $\lambda$-reference (\ref{eq:reflambda}), to a different reference frame $\lambda'$,
\begin{equation}\label{eq:refprime}
	\lambda'=\{\bm{u}',T'\},
\end{equation}
with the discrete velocities $\bm{v}_i^{\lambda'}$,
\begin{equation}\label{eq:vprime}
	\bm{v}_i^{\lambda'}= \sqrt{\frac{T'}{T_L}}\bm{c}_i+\bm{u}'.
\end{equation}
Below, the transformation proposed in \cite{Pond} is supplemented by a regularization procedure \cite{RegPond,PondReg2}.
Specifically, the transformed populations $f_i^{\lambda'}$ are sought as a third-order Grad's projection,
\begin{equation}
\label{eq:f_transformed}
\begin{split}
    f_i^{\lambda'}= W_i\left(a_0 +\frac{\bm{a}_1 \cdot \bm{c}_{i}}{T_L}+\frac{\bm{a}_2 \cdot  (\bm c_{i} \otimes \bm c_{i}-T_L \bm I)}{2T_L^2} \right. \\ 
    \left. +\frac{\bm{a}_3 \cdot \left(\bm{c}_{i} \otimes \bm{c}_{i} \otimes \bm{c}_{i}-T_L \overline{\bm{c}_{i} \otimes \bm{I}}\right)}{6T_L^3}\right),
    	\end{split}
\end{equation}
where overline denotes symmetrization, coefficients $\bm{a}_k$ are tensors of rank $k=0$ to $k=3$, while the dot stands for the full contraction.
Let us denote $\bm{M}_k^{\lambda}$ a moment tensor of order $k$,
\begin{equation}
	\bm{M}_k^{\lambda} =\sum_{i=0}^{Q-1}f_i^{\lambda} \underbrace{ \bm{v}_{i}^\lambda \otimes \bm{v}_{i}^\lambda \cdots \otimes \bm{v}_{i}^\lambda}_{k}.
\end{equation}
Then the regularized transformed populations are defined by the condition of invariance of the moments of orders $k=0,1,2,3$ with respect to the reference frame:
\begin{equation}\label{eq:Mcondition}
    \bm{M}_k^{\lambda'}=\bm{M}_k^{\lambda},\ k=0,1,2,3.
\end{equation}
Upon substitution, 
	\[\bm{M}_k^{\lambda'}=\sum_{i=0}^{Q-1}f_i^{\lambda'} \underbrace{ \bm{v}_{i}^{\lambda'} \otimes \bm{v}_{i}^{\lambda'} \cdots \otimes \bm{v}_{i}^{\lambda'}}_{k},\ k=0,1,2,3,\]
and using \eqref{eq:f_transformed} and \eqref{eq:vprime}, the linear system \eqref{eq:Mcondition} is solved to find Grad's coefficients $\bm{a}_k$, $k=0,1,2,3$ in terms of the new frame velocity $\bm{u}'$, reference temperature $T'$ and the moments in the old reference frame $\bm{M}_k^{\lambda}$,

\begin{widetext}
\begin{align}
    a_0 &= M_0^{\lambda}, \label{eq:a0}\\
\bm{a}_1 &= \left(\frac{T'}{T_L}\right)^{-{1}/{2}}\left(\bm{M}_1^{\lambda}-M_0^{\lambda}\bm{u}'\right), \label{eq:a1}\\
\bm{a}_2 &= \left(\frac{T'}{T_L}\right)^{-1}\left(\bm{M}_2^{\lambda}-M_0^{\lambda}T'\bm{I}-\left(\frac{T'}{T_L}\right)^{{1}/{2}} \overline{\bm{u}' \otimes \bm{a}_1} -M_0^{\lambda}\bm{u}' \otimes \bm{u}'\right), \label{eq:a2} \\
\begin{split}
 \bm{a}_3 &= \left(\frac{T'}{T_L}\right)^{-3/2}
\left(\bm{M}_3^{\lambda} -\left(\frac{T'}{T_L}\right)  \overline{\bm{u}' \otimes (M_0T_L\bm{I}+\bm{a}_2)} -T'\left(\frac{T'}{T_L}\right)^{{1}/{2}} \overline{\bm{a}_1 \otimes \bm{I}} -\left(\frac{T'}{T_L}\right)^{{1}/{2}} \overline{\bm{a}_1 \otimes \bm{u}' \otimes \bm{u}' } -M_0^{\lambda}\bm{u}' \otimes \bm{u}' \otimes \bm{u}'\right).\label{eq:a3}
\end{split}
\end{align}
\end{widetext}

Thus, the regularized transformation of the $f$-populations from a reference frame $\lambda$ to a reference frame $\lambda'$  is uniquely defined by the third-order Grad's projection with the coefficients \eqref{eq:a0}, \eqref{eq:a1}, \eqref{eq:a2} and \eqref{eq:a3}. For the regularized transformation of the $g$-populations, it is sufficient to use a second-order Grad's projection of the form \eqref{eq:f_transformed} where the third-order term is dropped while the coefficients $a_0$, $\bm{a}_1$ and $\bm{a}_2$ are defined by \eqref{eq:a0}, \eqref{eq:a1} and \eqref{eq:a2}, respectively, with the corresponding moments $M_k^{\lambda}$, $k=0,1,2$, of the populations $g_i^{\lambda}$. A discussion is in order.

\begin{itemize}
	
\item	
A rationale for using Grad's projection for regularized transformation is to essentially impose a moment hierarchy in the new reference frame: the low-order moments retained in Grad's projection ($\bm{M}_k$, $k=0,1,2,3$ in the case of Eq.\ \eqref{eq:f_transformed})  are independent and can be identified as "slow" moments. The remaining higher-order moments are considered as "fast" moments, enslaved by the slow ones and given by Grad's closure \cite{GorbanKarlin}.
\item Grad's projection is not a unique regularization strategy. For instance, another possibility is to set the higher-order moments to  equilibrium following the construction proposed in \cite{Multiscale2021}: Let us denote $M_q=\{M_0,\bm{M}_1,\bm{M}_2,\bm{M}_3\}$ the subset of slow moments, where $q$ is the dimension of the subspace, $q=10$ for $D=2$ and $q=20$ for $D=3$, while $M_{Q-q}$  stands for the fast moments.
Moment vector of the $f$-populations $M_Q$ can be considered as an element of a direct sum,
	\begin{align}	
		{M}_{Q}={M}_q \oplus{M}_{Q-q}.
	\end{align}
In the new reference frame $\lambda'$, consider a moment vector ${M}_Q^{\lambda\to\lambda'}$ using the $\lambda$-frame to evaluate the slow moments, $M_q^{\lambda'}=M_q^{\lambda}$, as before, while equilibrating the fast moments in the $\lambda'$ reference: 
\begin{align}
	{M}_Q^{\lambda\to\lambda'}={M}^{\lambda}_q \oplus \left({M}_{Q-q}^{\lambda'}\right)^{\rm eq}.
\end{align}
The regularized transformed populations are obtained by moment inversion,
\begin{align}
	f^{\lambda'}={\mathcal{M}_{\lambda'}}^{-1} {M}_Q^{\lambda\to\lambda'},\label{eq:liftingF}
\end{align}
where $\mathcal{M}_{\lambda'}$ is the $Q\times Q$ matrix of the populations-to-moments transform in the $\lambda'$ reference frame. The difference between the two regularization methods is that, with Grad's projection, the fast moments are slaved to the nonequilibrium slow moments, $M_{Q-q}^{\rm Grad}=M_{Q-q}(M_q)$, while in the equilibration case they are $M_{Q-q}^{\rm eq}=M_{Q-q}^{\rm eq}(M_0,\bm{M}_1)$. While both approaches are on equal footing, Grad's projection appears to be more economic as it does not need a matrix inversion \eqref{eq:liftingF} and shall be used below in this paper.

\item In two dimensions, the $D2Q16$ is the minimal product-form lattice based on Hermite roots that enables the third-order Grad's projection \eqref{eq:f_transformed}. While some previous realizations of PonD method for compressible flows used the standard $D2Q9$ lattice \cite{Pond,Ehsan2020,Multiscale2021},  inclusion of the third-order moment tensor $\bm{M}_3$ into the list of slow moments appears to be necessary in order to simulate hypersonic flows with very strong discontinuities and near-vacuum components such as those presented in Sec.\ \ref{sec:1Dstrong}.

\end{itemize}

\section{Semi-Lagrangian and finite volume realizations}
\label{Numerical_discretization}

In this section, we present two methods for the numerical discretization of the proposed PonD model. First, we review the semi-Lagrangian scheme, as formulated in \cite{Pond,Ehsan2020,Ehsan2021}. A finite volume method, based on the DUGKS numerical scheme \cite{Guo_DUGKS_IsoT}, is subsequently presented.

\subsection{Semi-Lagrangian realization}
\label{sec:SemiLag}
The semi-Lagrangian realization follows the spirit of LBM, where the governing continuous equations \eqref{eq:f_equation},\eqref{eq:g_equation} are integrated along the characteristics and a variable transformation is used to eliminate the implicitness of the scheme \cite{LBM_VarTransform1,LBM_VarTransform2}, 
\begin{align}
    \Tilde{f}_{i} &= {f}_{i}-\frac{\delta t}{2\tau}(f_i^{\rm eq}-f_i),\\  
\Tilde{g}_{i} &= {g}_{i}-\frac{\delta t}{2\tau}(g_i^{\rm eq}-g_i).    
\end{align}
The final equations, which describe the propagation and collision steps, can be expressed only through the $\Tilde{f}$- and $\Tilde{g}$-populations. For simplicity, we lift in this section the tilde notation. For the reconstruction of the populations at any point $\bm{x}$ and time $t$, we use the following formula \cite{Pond},

\begin{align}
 f_i(\bm{x},t) &=\sum_{s=1}^{m} \Lambda(\bm{x}-\bm{x}_s)f_i^{\lambda}(\bm{x}_s,t),\\
g_i(\bm{x},t) &=\sum_{s=1}^{m} \Lambda(\bm{x}-\bm{x}_s)g_i^{\lambda}(\bm{x}_s,t), 
\end{align}
where $\bm{x}_s$ are the collocation points, $\Lambda$ is the interpolation kernel, and it is assumed that regularized populations at the collocation points, $f_i^{\lambda}(\bm{x}_s,t)$ and  $g_i^{\lambda}(\bm{x}_s,t)$, are transformed into the same reference frame $\lambda$, as explained in Sec.\ \ref{sec:transformation}. Below, we use a 4-point stencil ($m=4$) with a B-spline interpolation kernel in a combination with limiters, as detailed in \cite{EhsanThesis}.

We consider the propagation step at a monitoring point $\bm{x}$ and time $t$. Semi-Lagrangian advection is performed at the departure points of characteristic lines $\bm{x}-\bm{v}_i^{\lambda_0} \delta t$,

\begin{align}
\label{advectionPond}
   f_i^{\lambda_0} &={f}_i(\bm{x}-\bm{v}_i^{\lambda_0} \delta t,t-\delta t),\\
    \label{advectionPondg}
g_i^{\lambda_0} &={g}_i(\bm{x}-\bm{v}_i^{\lambda_0} \delta t,t-\delta t).
\end{align}
The reference frame $\lambda_0=\{ \bm{u}_0,T_0 \}$ is initialized using the local flow velocity and the local temperature, which are available from the previous time step, $\bm{u}_0=\bm{u}(\bm{x},t-\delta t), {T}_0={T}(\bm{x},t-\delta t) $. Eqs.\eqref{advectionPond}-\eqref{advectionPondg} constitute the predictor propagation step. The density, momentum and temperature are consequently computed by 
\begin{align}
\label{correctorfieldsrho}
    \rho_1 &=\sum_{i=0}^{Q-1} f_i^{\lambda_0} , \\
 \label{correctorfieldsu}   
    \rho_1 \bm{u}_1 &=\sum_{i=0}^{Q-1} \bm{v}_i^{\lambda_0} f_i^{\lambda_0}, \\
 \label{correctorfieldT}     
     \rho_1 {E}_1 &=\sum_{i=0}^{Q-1} \frac{{(\bm{v}_i^{\lambda_0})}^2}{2} f_i^{\lambda_0} +\sum_{i=0}^{Q-1}g_i^{\lambda_0}.
\end{align}
The computed velocity  (\ref{correctorfieldsu}) and temperature (\ref{correctorfieldT})  define the corrector reference frame  $\lambda_1=\{ \bm{u}_1,T_1 \}$ at the monitoring point and the propagation step \eqref{advectionPond}-\eqref{advectionPondg} is repeated with the updated reference frame. The predictor-corrector process is iterated until convergence with the limit values, 
$$\rho (\bm{x},t),\bm{u} (\bm{x},t),T (\bm{x},t), f_i^{\lambda(\bm{x},t)} =\lim_{n \to \infty} \rho_n,\bm{u}_n, T_n, f_i^{\lambda_n}, $$
defining the density, velocity, temperature and the pre-collision populations at the monitoring point $\bm{x}$ at time $t$. The predictor-corrector iteration loop ensures that the propagation and the collision steps are performed at the co-moving reference frame, in which the local equilibrium populations \eqref{feqPond}-\eqref{geqPond} are exact. 

The collision step follows the BGK collision model,
\begin{align}
    f_i(\bm{x},t) &=f_i^{\lambda(\bm{x},t)}+2\beta \left[\rho(\bm{x},t)W_i-f_i^{\lambda(\bm{x},t)} \right],\\
    g_i(\bm{x},t) &=g_i^{\lambda(\bm{x},t)}+2\beta \left[\rho\left(C_v-\frac{D}{2}\right)T(\bm{x},t)W_i-g_i^{\lambda(\bm{x},t)} \right],
\end{align}
where the relaxation parameter $\beta$ is related to the kinematic viscosity by $\nu=T(\frac{1}{2\beta}-\frac{1}{2})\delta t$. The thermal conductivity is $\kappa=C_p(\frac{1}{2\beta}-\frac{1}{2})\rho T \delta t$, which yields Prandtl equal to one. 

\subsection{Finite-volume realization}
\label{sec:FV}
The semi-Lagrangian propagation, coupled with a local collision step, is a simple and efficient numerical scheme for the realization of PonD. It should be noted however that this method is not strictly conservative. While there exist strategies to partially alleviate this problem \cite{ConservSemiLag,ConserSemiLag2}, we propose a finite-volume discretization scheme which naturally restores the conservation. Specifically, we reformulate the DUGKS algorithm \cite{Guo_DUGKS_IsoT,Guo_DUGKS_Compres,Guo_DUGKS_Rev} in a co-moving reference frame. Here we outline the main points of the discretization procedure.

\subsubsection{Updating rule}

The evolution of the populations is governed by the following equations in the DUGKS framework \cite{Guo_DUGKS_IsoT},
\begin{align}
    \label{eq:FinalUpdateDUGKS}
    \begin{split}
    \Tilde{f}_{i}(\bm{x}_j,t_{n+1})&=\left( \frac{2\tau-\delta t}{2\tau+\delta t} \right) \Tilde{f}_{i}(\bm{x}_j,t_{n})+ \\&
    \left(\frac{2\delta t}{2\tau+\delta t} \right){f}_{i}^{{\rm eq}}(\bm{x}_j,t_{n})
-\frac{\delta t}{V_j} F_{f,i}(\bm{x}_j,t_{n+1/2}),
    \end{split} \\
     \begin{split}
    \Tilde{g}_{i}(\bm{x}_j,t_{n+1})&=\left( \frac{2\tau-\delta t}{2\tau+\delta t} \right) \Tilde{g}_{i}(\bm{x}_j,t_{n})+\\&
    \left(\frac{2\delta t}{2\tau+\delta t} \right){g}_{i}^{{\rm eq}}(\bm{x}_j,t_{n})-\frac{\delta t}{V_j} F_{g,i}(\bm{x}_j,t_{n+1/2}). 
     \end{split}
\end{align}
The update equations are derived from the integration of the continuous equations \eqref{eq:f_equation},\eqref{eq:g_equation} in a control volume centered at $\bm{x}_j$, with volume $V_j$, from time $t_n$ to $t_{n+1}=t_n+\delta t$, using the midpoint rule for the convection term and the trapezoidal rule for the collision term \cite{Guo_DUGKS_IsoT}. To remove the implicitness, the DUGKS scheme adopts the variable transformation from the standard LBM practice \cite{LBM_VarTransform1,LBM_VarTransform2}

\begin{equation}
    \label{eq:variableTransform}
    \Tilde{\phi}_{i}={\phi}_{i}-\frac{\delta t}{2}\Omega_{\phi,i}={\phi}_{i}-\frac{\delta t}{2\tau}(\phi_i^{\rm eq}-\phi_i),  
\end{equation}
where $\phi$ stands for the $f$- and $g$-populations. The fluxes of the populations $F_{\phi,i}(\bm{x}_j,t_{n+1/2})$ across the surface of the control volume are defined as,

\begin{equation}\label{eq:fluxesDUGKS}
    F_{\phi,i}(\bm{x}_j,t_{n+1/2})=\int_{\partial V_j}(\bm{v}_i \cdot \bm{n})\phi_{i}(\bm{x},t_{n+1/2})d\bm{S},
\end{equation}
where $\bm{n}$ is the outward unit vector normal to the surface. Finally, we remark that within the finite volume context, the populations and the collision terms are cell-averaged quantities,

\begin{equation}
     \phi_{i}(\bm{x}_j,t_{n})=\frac{1}{V_j}\int_{V_j}{\phi_{i}(\bm{x},t_n)d\bm{x}}.
\end{equation}

\subsubsection{Evolution in the co-moving reference frame}

In this section, we present the implementation of DUGKS with an adaptive reference frame formulation. We consider a cell with center $\bm{x}_j$, at time $t_n$. The populations $f_{i}^{\lambda}(\bm{x}_j,t_{n}), g_{i}^{\lambda}(\bm{x}_j,t_{n})$ are known from the previous time step (or initial conditions) and they are expressed in the local reference frame $\lambda=\{\bm{u}(\bm{x}_j,t_n), T(\bm{x}_j,t_n)\}$. With the exact equilibria \eqref{feqPond} and \eqref{geqPond}, the update equations become,

\begin{widetext}
\begin{align}
    \label{eq:FinalUpdateDUGKS}
    \Tilde{f}_{i}^{\lambda}(\bm{x}_j,t_{n+1})&=\left( \frac{2\tau-\delta t}{2\tau+\delta t} \right) \Tilde{f}_{i}^{\lambda}(\bm{x}_j,t_{n})+\left(\frac{2\delta t}{2\tau+\delta t} \right)\rho(\bm{x}_j,t_{n}) W_i-\frac{\delta t}{V_j} F_{f,i}^{\lambda}(\bm{x}_j,t_{n+1/2}), \\ \label{eq:FinalUpdateDUGKS_g}
    \Tilde{g}_{i}^{\lambda}(\bm{x}_j,t_{n+1})&=\left( \frac{2\tau-\delta t}{2\tau+\delta t} \right) \Tilde{g}_{i}^{\lambda}(\bm{x}_j,t_{n})+\left(\frac{2\delta t}{2\tau+\delta t} \right)\left(C_v-\frac{D}{2}\right)\rho(\bm{x}_j,t_{n})T(\bm{x}_j,t_{n}) W_i-\frac{\delta t}{V_j} F_{g,i}^{\lambda}(\bm{x}_j,t_{n+1/2}).    
\end{align}
\end{widetext}
The fluxes \eqref{eq:fluxesDUGKS} featured in the update equations \eqref{eq:FinalUpdateDUGKS} and \eqref{eq:FinalUpdateDUGKS_g} are evaluated in the next section. 

\subsubsection{Flux evaluation in the co-moving reference frame}
\label{Comoving Flux}
The key element of the update equations \eqref{eq:FinalUpdateDUGKS},\eqref{eq:FinalUpdateDUGKS_g} is the evaluation of the flux term, $F_{\phi,i}(\bm{x}_j,t_{n+1/2})$, which contains the unknown populations ${\phi}_{i}(\bm{x}_b,t_{n+1/2})$ at the cell interface $\bm{x}_b$ and time $t_{n+1/2}$. Integrating eqs.(\ref{eq:f_equation},\ref{eq:g_equation}) along the characteristics for half-time step shows that the requested populations are connected with the known populations at time $t_{n}$ through the following equation \cite{Guo_DUGKS_IsoT},

\begin{equation}
    \label{eq:InterfacePop}
    \bar{\phi}_{i}(\bm{x}_b,t_{n+1/2})=\bar{\phi}_{i}^{+}(\bm{x}_b-\bm{v}_i \delta t/2,t_{n}),
\end{equation}
where,
\begin{align}\label{eq:FluxTransf}
    \bar{\phi}_{i} &={\phi}_{i}-\frac{\delta t/2}{2}\Omega_{\phi,i},\\ \label{eq:Fbarplus}
    \bar{\phi}_{i}^{+} &={\phi}_{i}+\frac{\delta t/2}{2}\Omega_{\phi,i}.
\end{align}
Eq.\eqref{eq:InterfacePop} is essentially a half-time step semi-Lagrangian advection step, as in LBM, with the final point located at the interface $\bm{x}_b$, at $t_{n+1/2}$. Following the spirit of PonD, we realize this step in the co-moving reference frame, with the following iterating procedure.

The reference frame $\lambda_0=\{ \bm{u}_0,T_0 \}$ (predictor reference frame) at the cell interface $\bm{x}_b$ is initialized with the fluid velocity and temperature from the previous step, $\bm{u}_0=\bm{u}(\bm{x}_b,t_{n-1/2}), {T}_0={T}(\bm{x}_b,t_{n-1/2})$. The populations $\bar{\phi}_{i}^{+, \lambda_0}$ and the spatial gradients $\bm{\sigma}_{i}^{\lambda_0} =\nabla \bar{\phi}_{i}^{+,\lambda_0} $ are subsequently evaluated in the neighbouring cells of the interface, at time $t_n$. In this work, Van Leer and minmod slope limiters were used for the computation of the spatial derivatives \cite{VanLeerLimiter,RoeLimiter}. We also note that the regularized reference frame transformation is applied, to express the required populations from their original reference frame to the target reference frame $\lambda_0$. The populations are reconstructed at the departure point $\bm{x}'=\bm{x}_b-\bm{v}_i^{\lambda_0}\delta t/2$, with the MUSCL scheme \cite{VanLeerMuscl},

\begin{equation}
    \bar{\phi}_{i}^{+,\lambda_0}(\bm{x}',t_n)=\bar{\phi}_{i}^{+,\lambda_0}(\bm{x}_j,t_n)+(\bm{x}'-\bm{x}_j)\cdot \bm{\sigma}_{i}^{\lambda_0}(\bm{x}_j,t_n).
\end{equation}

According to eq. \eqref{eq:InterfacePop}, we obtain the $\bar{\phi}_{i}^{\lambda_0}$ populations at the interface $\bm{x}_b$ and time $t_{n+1/2}$ by

\begin{equation}
    \bar{\phi}_{i}^{\lambda_0}(\bm{x}_b,t_{n+1/2})=\bar{\phi}_{i}^{+,\lambda_0}(\bm{x}',t_n).
\end{equation}

The density, momentum and temperature are finally computed by 
\begin{align}
\label{correctorfieldsrhoDUGKS}
    \rho_1 &=\sum_{i=0}^{Q-1} \bar{f}_{i}^{\lambda_0}(\bm{x}_b,t_{n+1/2}) , \\
 \label{correctorfieldsuDUGKS}   
    \rho_1 \bm{u}_1 &=\sum_{i=0}^{Q-1} \bm{v}_i^{\lambda_0} \bar{f}_{i}^{\lambda_0}(\bm{x}_b,t_{n+1/2}), \\
    \label{correctorfieldsTDUGKS}  
     \rho_1 {E}_1 &=\sum_{i=0}^{Q-1} \frac{{(\bm{v}_i^{\lambda_0})}^2}{2} \bar{f}_{i}^{\lambda_0}(\bm{x}_b,t_{n+1/2}) +\sum_{i=0}^{Q-1}\bar{g}_{i}^{\lambda_0}(\bm{x}_b,t_{n+1/2}).
\end{align}

The computed moments define the corrector reference frame $\lambda_1=\{ \bm{u}_1,T_1 \}$. We repeat the above half-time step semi-Lagrangian advection step with the updated $\lambda_1$  and the predictor-corrector loop is continued upon reference frame convergence $\lambda_{b}=\lim_{k \to \infty}{\lambda_k}(\bm{x}_b,t_{n+1/2})$. With this procedure we enforce the execution of the advection at the optimal co-moving reference frame. With the completion of the advection, eq. \eqref{eq:FluxTransf} is used to obtain the populations ${\phi}_{i}^{\lambda_b}(\bm{x}_b,t_{n+1/2})$, where the equilibria that are needed are the exact co-moving equilibria \eqref{feqPond}, \eqref{geqPond}. The flux of the populations across the interface of the cell $\bm{x}_j$, in the local reference frame $\lambda$, can then be computed as,

\begin{equation} \label{fluxEq}
    F_{\phi,i}^{\lambda}(\bm{x}_j,t_{n+1/2})=\sum_{c}(\bm{v}_i^{\lambda} \cdot \bm{n}_c) \phi_{i}^{\lambda}(\bm{x}_{b,c},t_{n+1/2}),
\end{equation}
where $\bm{x}_{b,c}$ designates the center of the $c$-th face of the cell and $\bm{n}_c$ is the outwards normal vector.

\subsubsection{Summary of the algorithm}

Based on the previous steps, we summarize the evolution procedure from time $t_n$ to $t_{n+1}$:
\begin{enumerate}
    \item Initialization of the populations
    \begin{itemize}
        \item Loop over cell centers $\bm{x}_j$: The populations $\Tilde{\phi}_{i}(\bm{x}_j,t_{n})$ are expressed according to local reference frame, $\lambda=\{\bm{u}(\bm{x}_j,t_{n}),T(\bm{x}_j,t_{n})\}$.
    \end{itemize}
    
    \item Calculation of the fluxes
    
    \begin{itemize}
        \item Loop over cell centers $\bm{x}_j$: Calculation of the $\bar{\phi}_{i}^{+}(\bm{x}_j,t_{n})$ populations, according to eq. \eqref{eq:Fbarplus}. 
        \item Loop over cell faces $\bm{x}_b$: Calculation of the populations ${\phi}_{i}(\bm{x}_b,t_{n+1/2})$ at the co-moving reference frame, according to the iteration procedure in Sec. \ref{Comoving Flux}.
    \end{itemize}
    \item Population update
    \begin{itemize}
        \item  Loop over cell centers $\bm{x}_j$: Computation of the fluxes to the local reference frame of the cell eq. \eqref{fluxEq} and update the populations through eqs. \eqref{eq:FinalUpdateDUGKS}, \eqref{eq:FinalUpdateDUGKS_g}. 
    \end{itemize}
\end{enumerate}

It is interesting to underline the differences between the proposed formulation in the optimal reference frame and the original DUGKS scheme \cite{Guo_DUGKS_IsoT}. First we note that in the proposed scheme, the reference frame is adaptive in space and time and the regularized transformation (Sec.\ref{sec:transformation}) is applied when it is necessary to connect different reference frames. The key point of the current scheme is the computation of the fluxes in the co-moving reference frame. This construction, enforced by the predictor-corrector iterative procedure, ensures Galilean invariance and avoids any errors originating from truncanted equilibria. While the computational cost is increased relative to the original DUGKS, the operational range of a given lattice is extended greatly without increasing the number of discrete speeds. As shown in the results, extreme compressible flows can be accurately and robustly captured with the $D2Q16$ lattice in a co-moving reference frame, which is not feasible if a uniform reference frame is imposed.

\section{Results and discussion}
\label{Results}

In this section we validate our model through 1D and 2D benchmarks. First we test the model with 1D Riemann problems, involving up to moderate discontinuities. Flows with low density-near vacuum regions and very strong discontinuities are investigated subsequently with the finite volume scheme (Sec.\ref{sec:FV}). In these flows, where the mass conservation is of high importance, we compare the performance of the two discretization schemes. We conclude the results with classical high Mach 2D problems. Unless stated otherwise, the numerical parameters of the simulations are the following. The time step $\delta t$ is such that the Courant–Friedrichs–Lewy (CFL) number is $\text{CFL}= \max|v_{i\alpha}|(\delta t /\delta x) =0.2$, where $\delta x$ is the grid resolution. The adiabatic exponent is $\gamma=1.4$. Finally, the viscosity is low enough such that the results remain invariant (typically $\mu \sim \mathcal{O}(10^{-3}-10^{-2})$).

\subsection{1D gas dynamic problems with weak to moderate discontinuities}

\subsubsection{Sod's shock tube}

\begin{figure}
    \centering
   \includegraphics[width=0.4\textwidth]{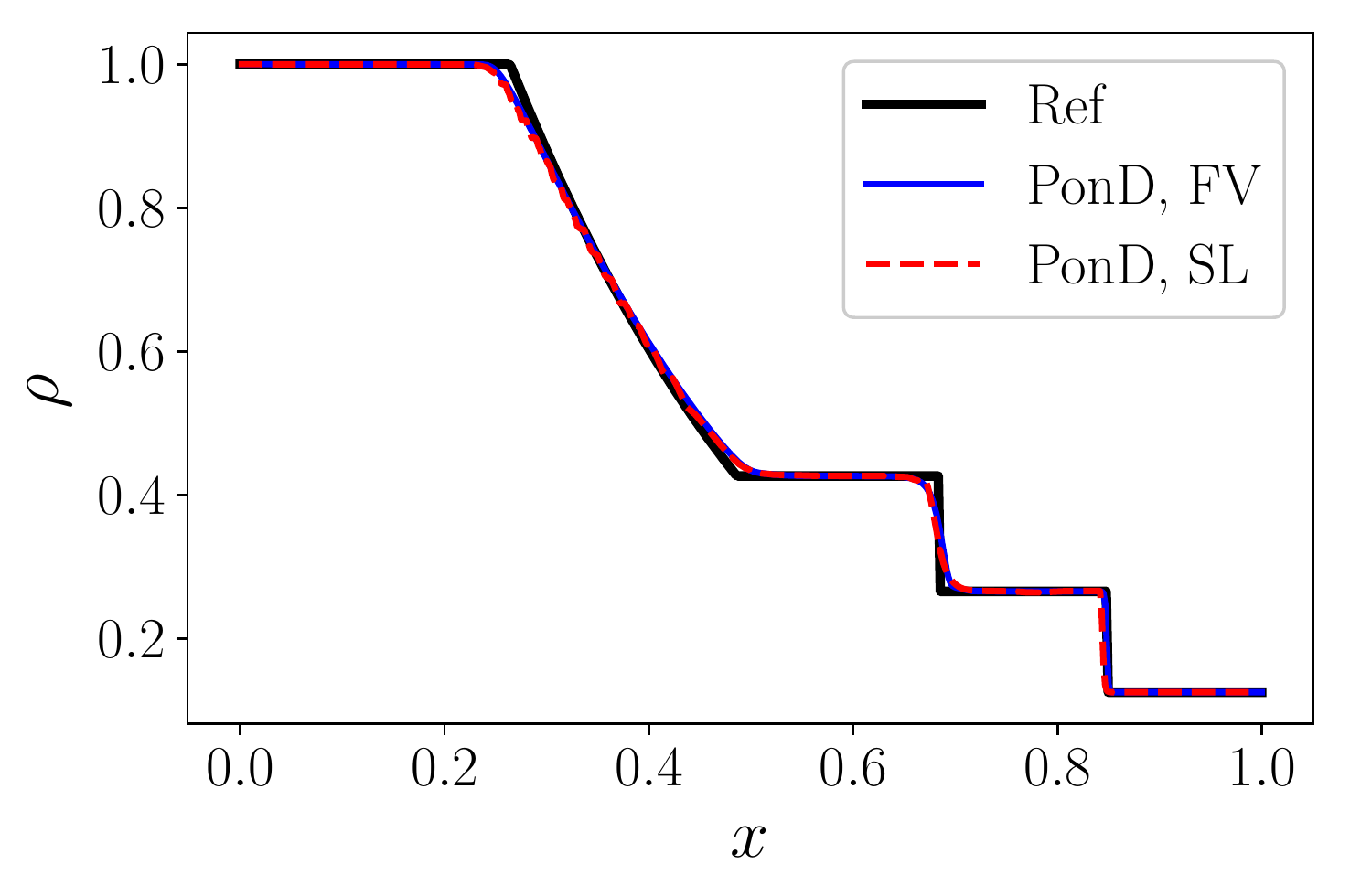}
    \includegraphics[width=0.4\textwidth]{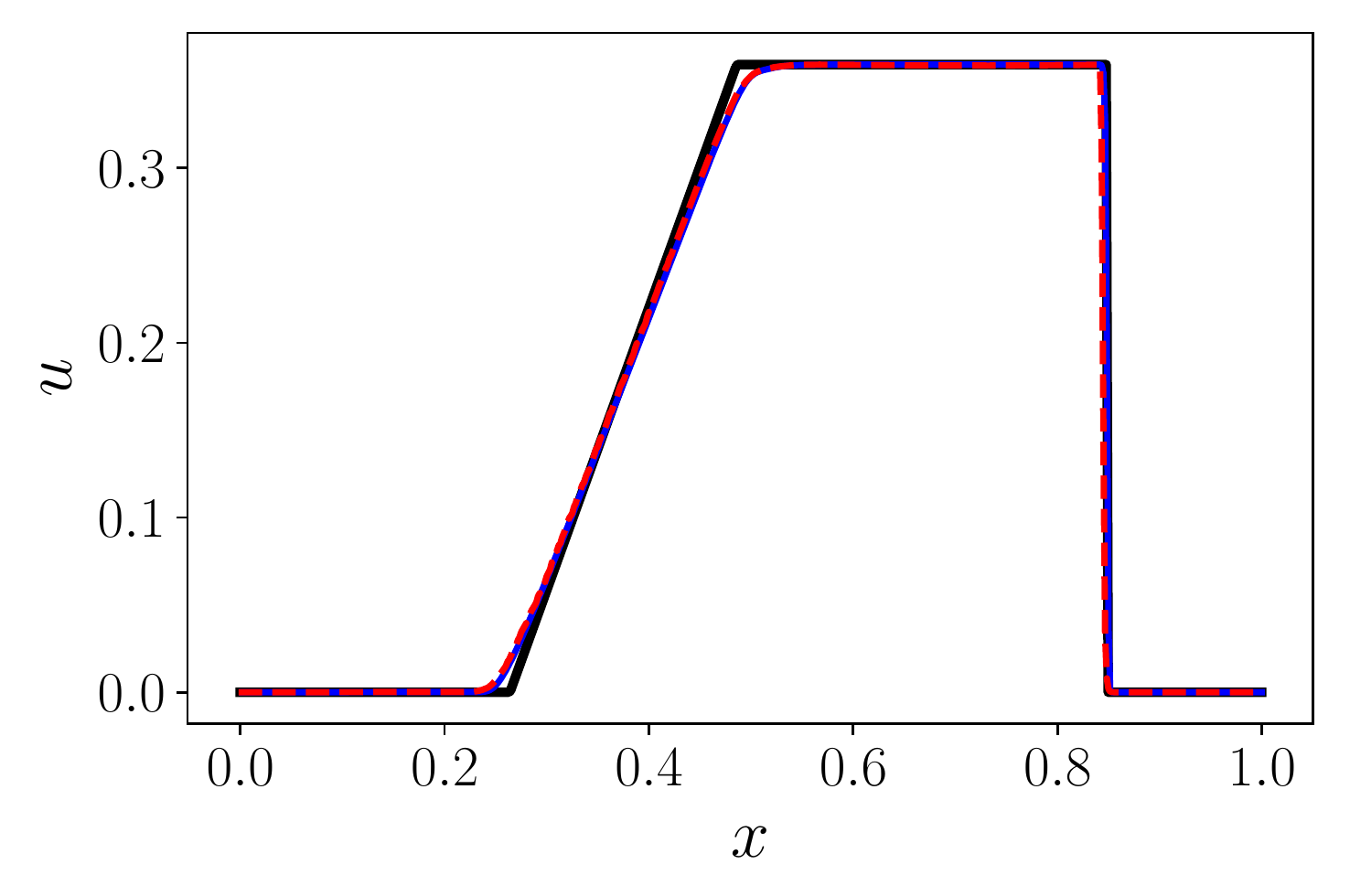}
     \includegraphics[width=0.4\textwidth]{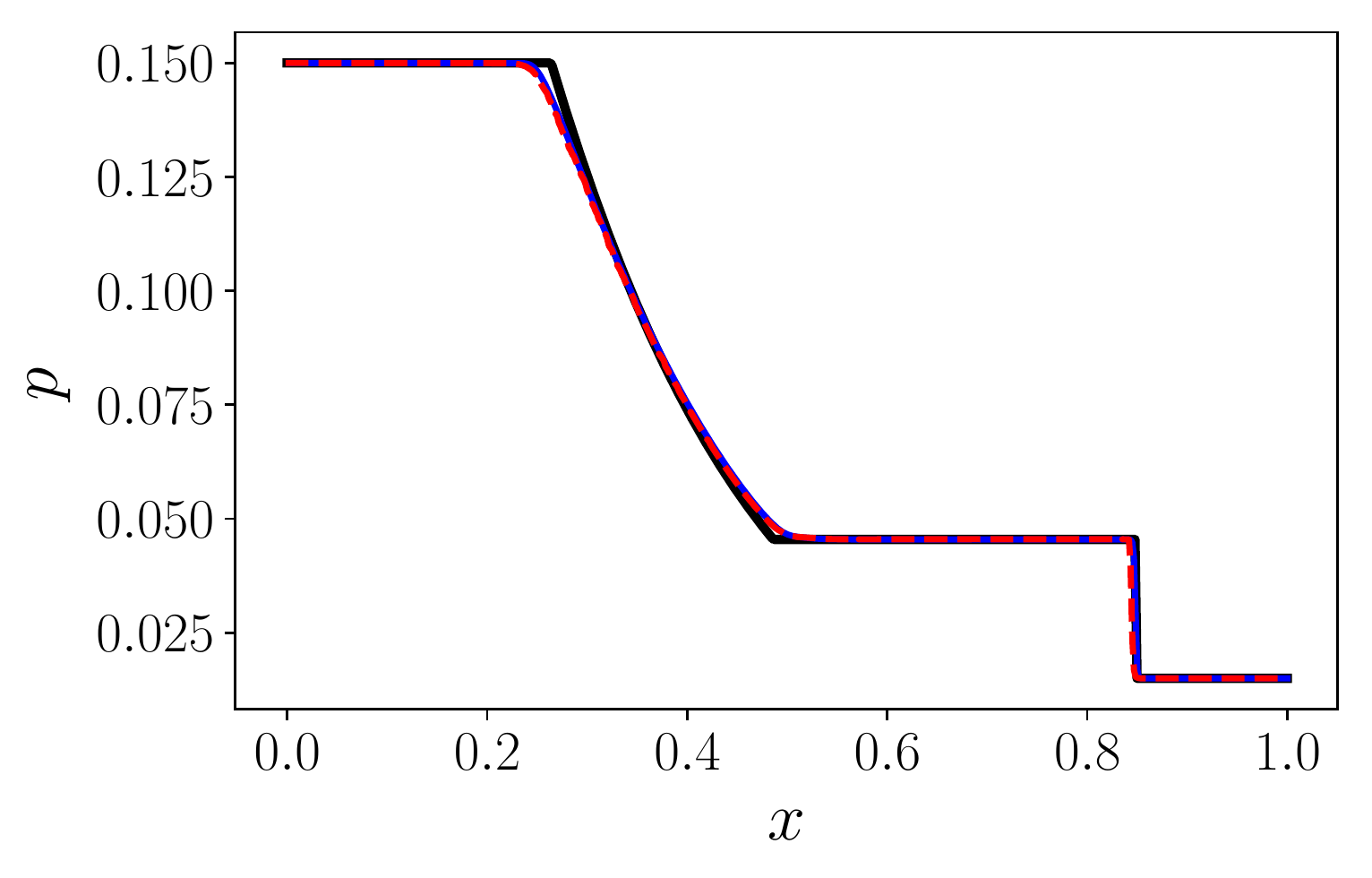}
    \caption{Density (top), velocity (middle) and pressure (bottom) profiles for Sod's shock tube, at $t=0.2$. Dashed line: semi-Lagrangian (SL) scheme. Solid line: finite-volume (FV) scheme. Thick solid line: Reference from an exact Riemann solver.}
    \label{fig:Sod_Tube}
\end{figure}

In the first case we simulate Sod's shock tube \cite{SodTube}, which is a typical benchmark Riemann problem for a compressible flow solver. The initial conditions are,

\begin{equation}
    (\rho,u,p)=\begin{cases}
    (1, 0, 0.15),& 0 \leq x<0.5, \\
    (0.125, 0, 0.015),& 0.5 \leq x \leq 1. \\
    \end{cases}
\end{equation}
The resolution of the computational domain is $L=600$. The results for the density, velocity and pressure profiles at  time $t=0.2$, are shown in Fig.\ \ref{fig:Sod_Tube}, indicating very good match with the exact solution.

\subsubsection{Lax problem}

\begin{figure}
    \centering
   \includegraphics[width=0.4\textwidth]{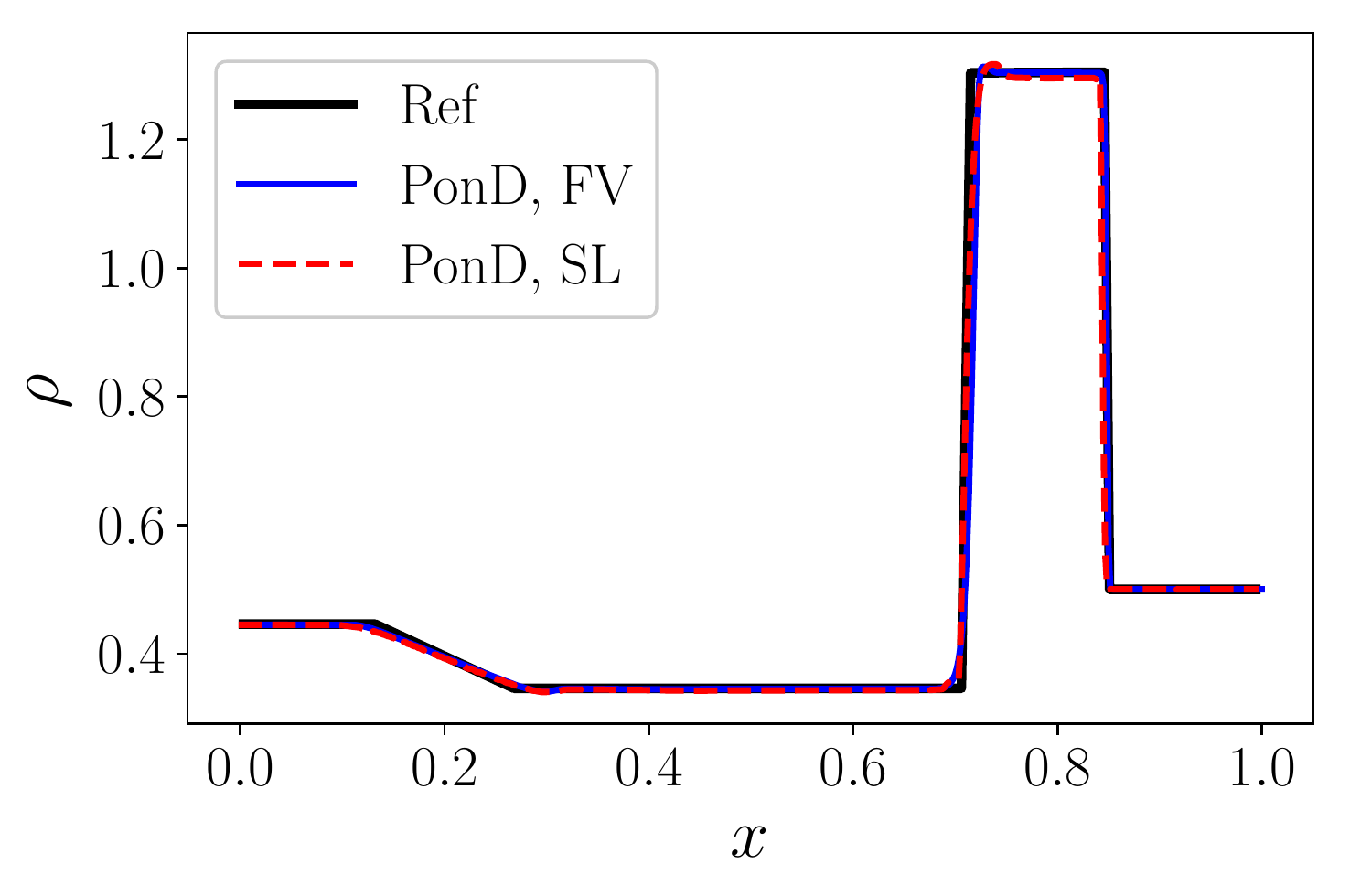}
    \includegraphics[width=0.4\textwidth]{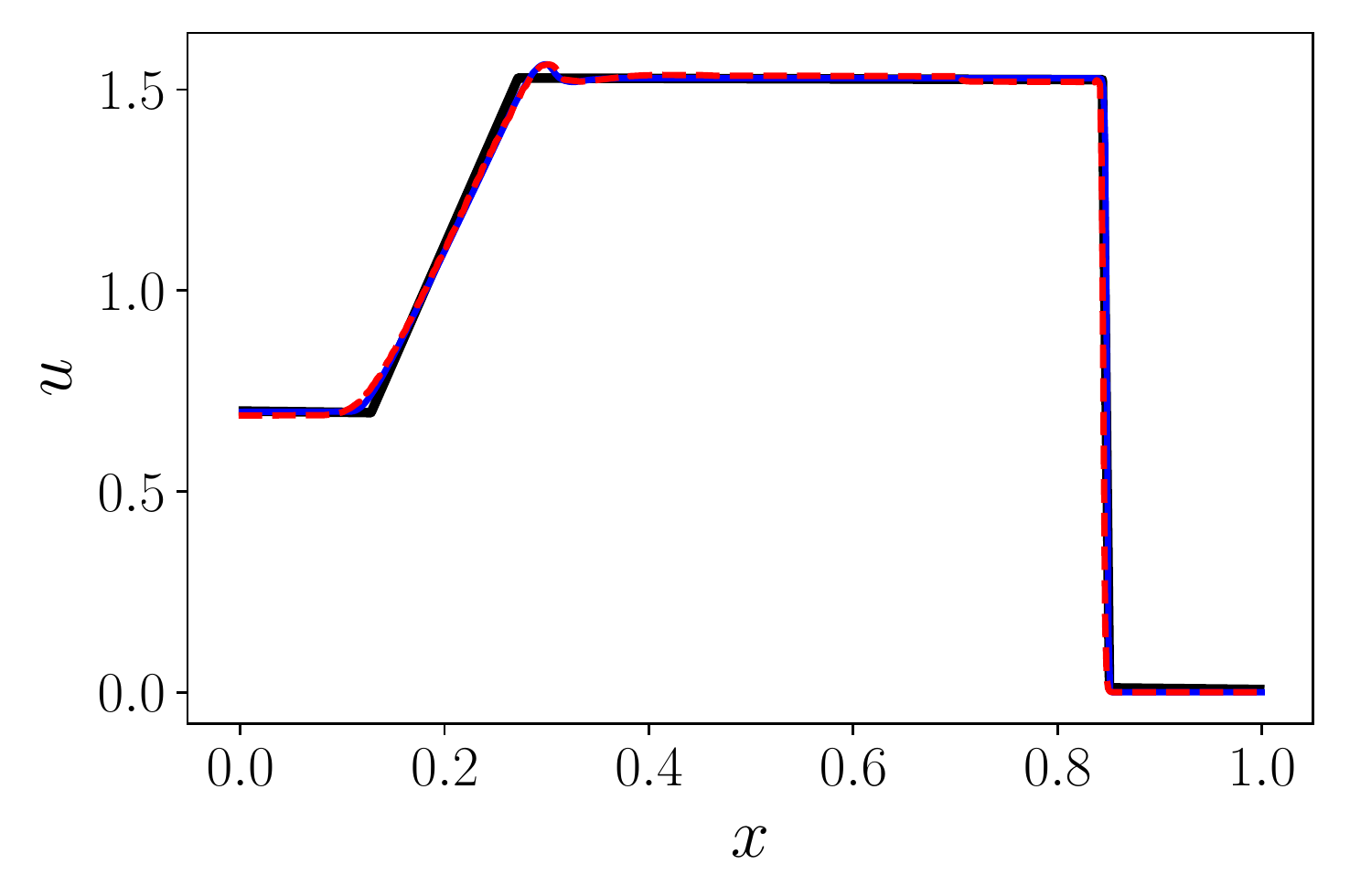}
     \includegraphics[width=0.4\textwidth]{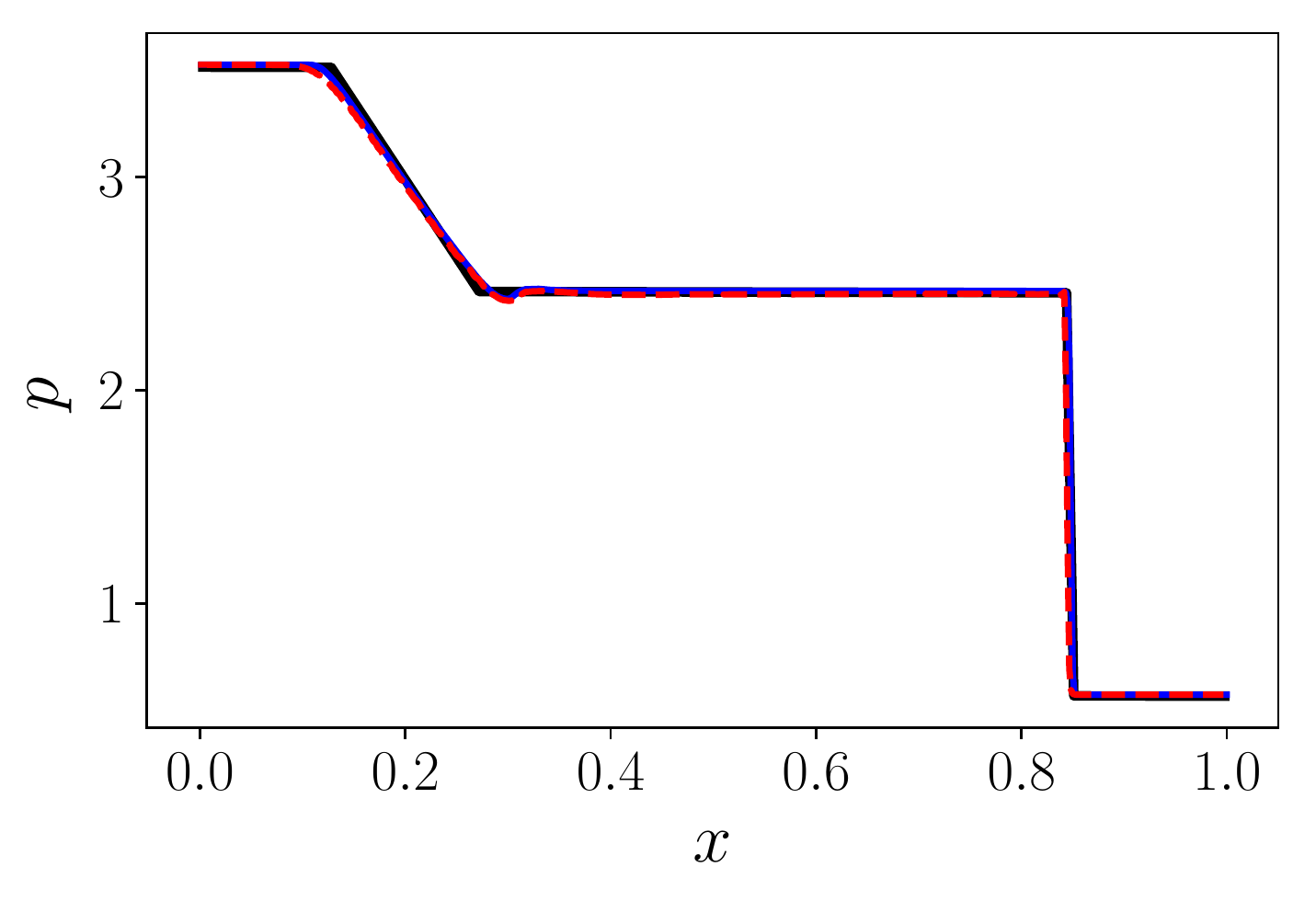}
    \caption{Density (top), velocity (middle) and pressure (bottom) profiles for Lax problem, at $t=0.14$. Dashed line: semi-Lagrangian (SL) scheme. Solid line: finite-volume (FV) scheme. Thick solid line: Reference from an exact Riemann solver.}
    \label{fig:Lax_Problem}
\end{figure}

We continue with the Lax problem \cite{LaxTube}, with the following initial conditions,

\begin{equation}
    (\rho,u,p)=\begin{cases}
    (0.445, 0.698, 3.528),& 0 \leq x<0.5, \\
    (0.5, 0, 0.571),& 0.5 \leq x \leq 1. \\
    \end{cases}
\end{equation}
The simulation is performed with $L=600$, until $t=0.14$. The results, shown in Fig. \ref{fig:Lax_Problem}, compare very well with the exact solution, with the exception of minor oscillations.

\subsubsection{Shock density-wave interaction}
\begin{figure}
    \centering
   \includegraphics[width=0.4\textwidth]{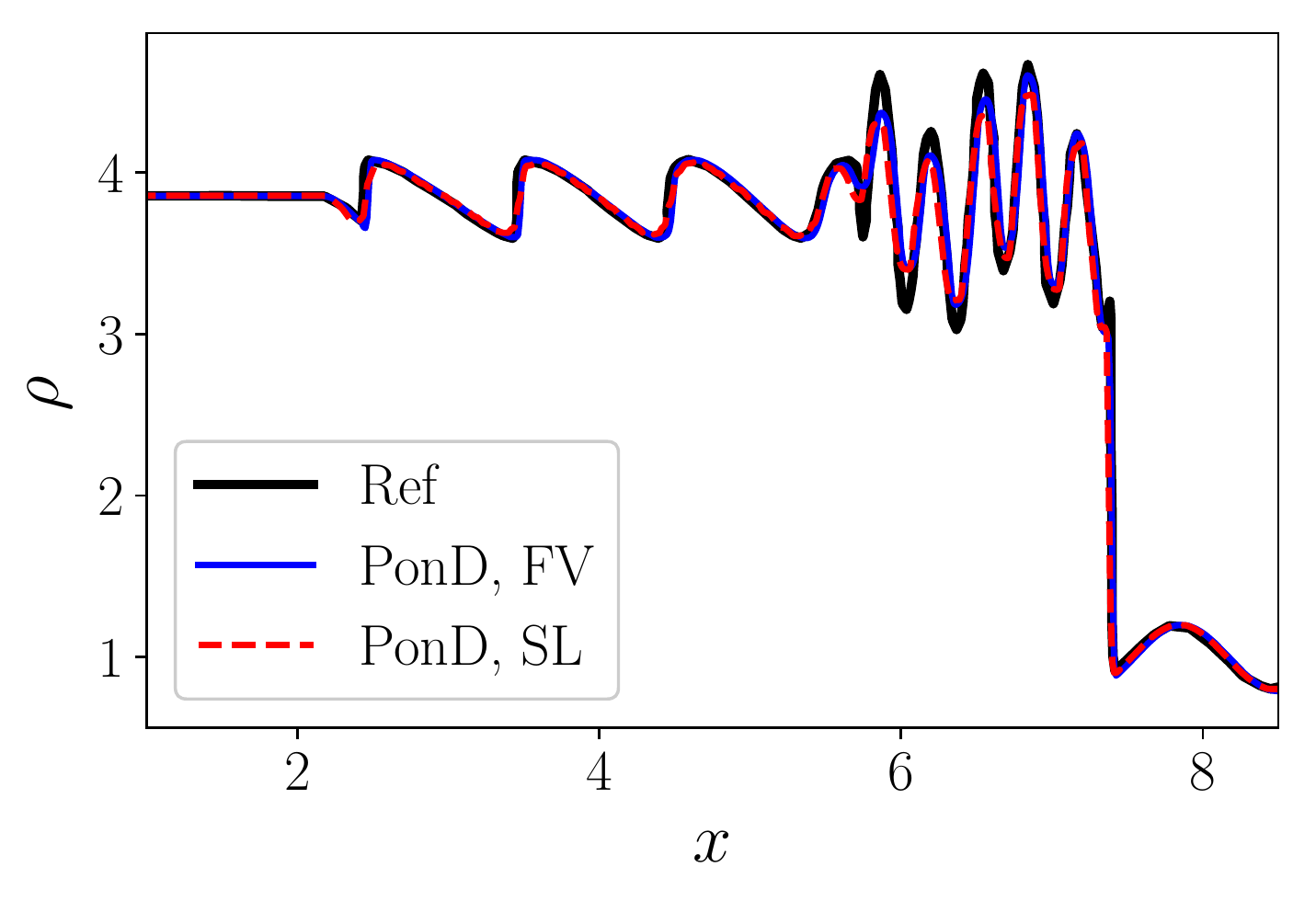}
   \includegraphics[width=0.4\textwidth]{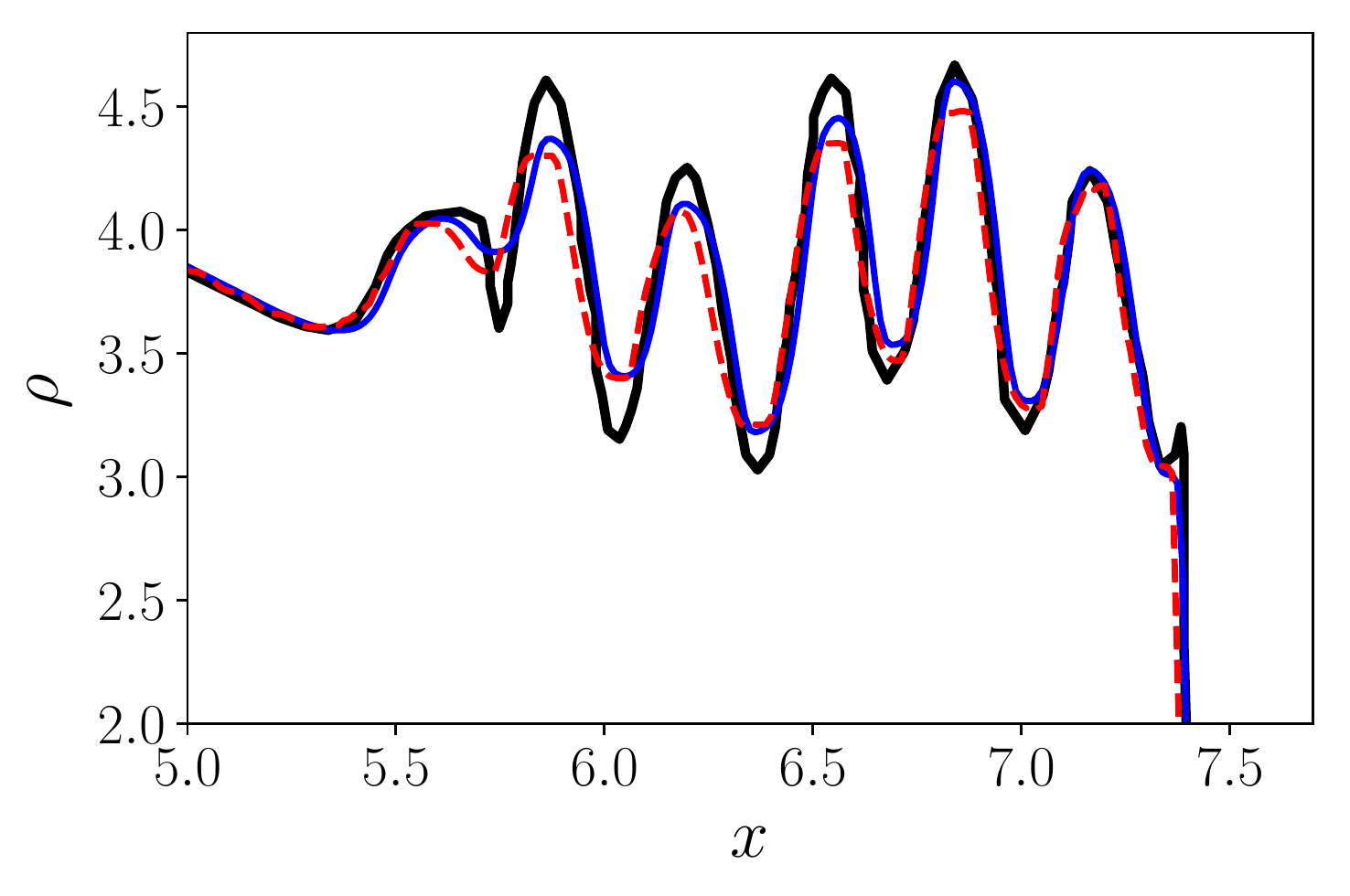}
   
    \caption{Density profile for the Shu-Osher problem, at $t=1.8$. Dashed line: semi-Lagrangian (SL) scheme. Solid line: finite-volume (FV) scheme. Thick solid line: Reference solution \cite{ShuOsherRefSol}. Bottom: a zoom into the high frequency wave region.}
    \label{fig:ShuOsher_Problem}
\end{figure}

In this case, also known as Shu-Osher problem \cite{ShuOsherProblem}, a Mach 3 shock wave interacts with a perturbed density field. The interaction leads to discontinuities and the formation of small structures. The initial conditions are,
\begin{equation}
    (\rho,u,p)=\begin{cases}
    (3.857, 2.629, 10.333),& 0 \leq x<1, \\
    (1+0.2 \sin (5(x-5)), 0, 1),& 1 \leq x \leq 10. \\
    \end{cases}
\end{equation}
The results for the density profile, at $t=1.8$ and $L=800$, are shown in Fig. \ref{fig:ShuOsher_Problem}. It is clear that the shock location and the high frequency waves are captured very well, apart from a small underestimation of the amplitudes of the post-shock waves.

\subsection{1D gas dynamic problems with very strong discontinuities  }
\label{sec:1Dstrong}

In this section we validate the model against flows with very strong discontinuities. We continue with the finite volume discretization, the conservative properties of which are important for this regime. A comparison between the semi-Lagrangian and the finite volume scheme is discussed in the following section (\ref{sec:FVSLcomparison}).

\subsubsection{Strong shock tube}
 We consider the case of a strong shock tube \cite{StrongShock}, where the ratio between the temperature of the left and right side is $10^5$. The initial conditions for this problem are,
\begin{equation}
    (\rho,u,p)=\begin{cases}
    (1, 0, 1000),& 0 \leq x<0.5, \\
    (1, 0, 0.01),& 0.5 \leq x \leq 1. \\
    \end{cases}
\end{equation}

\begin{figure}
    \centering
   \includegraphics[width=0.4\textwidth]{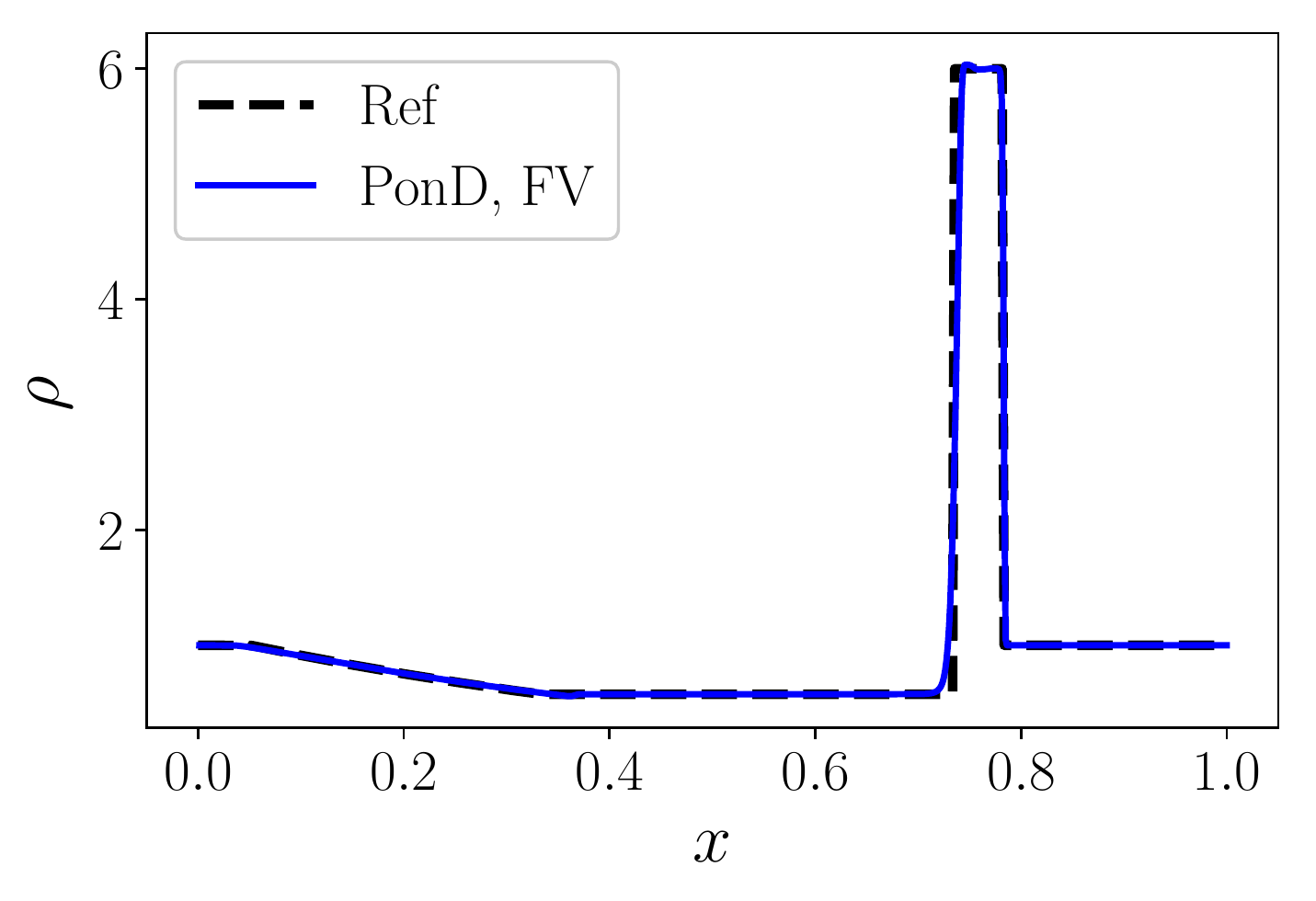}
    \includegraphics[width=0.4\textwidth]{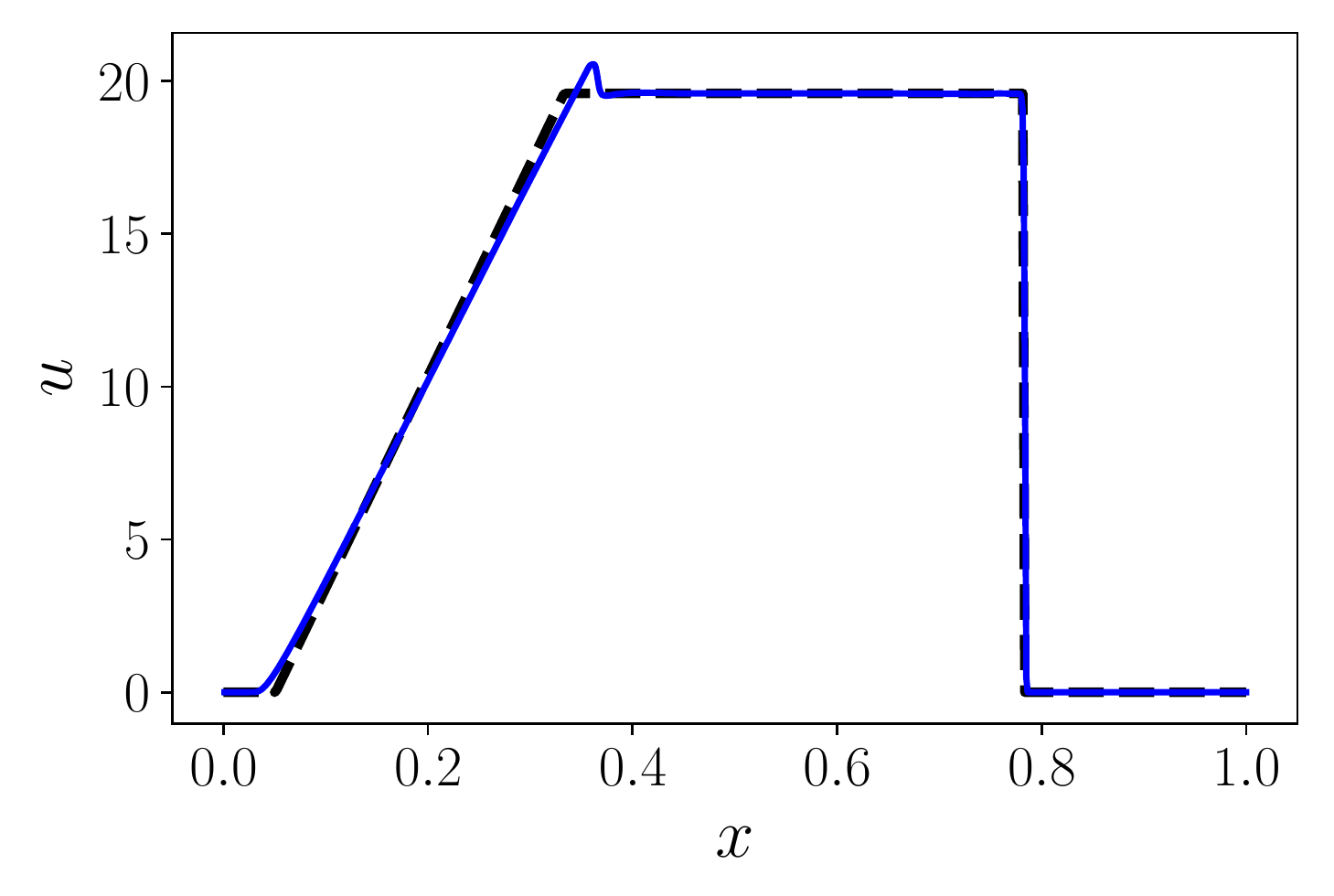}
     \includegraphics[width=0.4\textwidth]{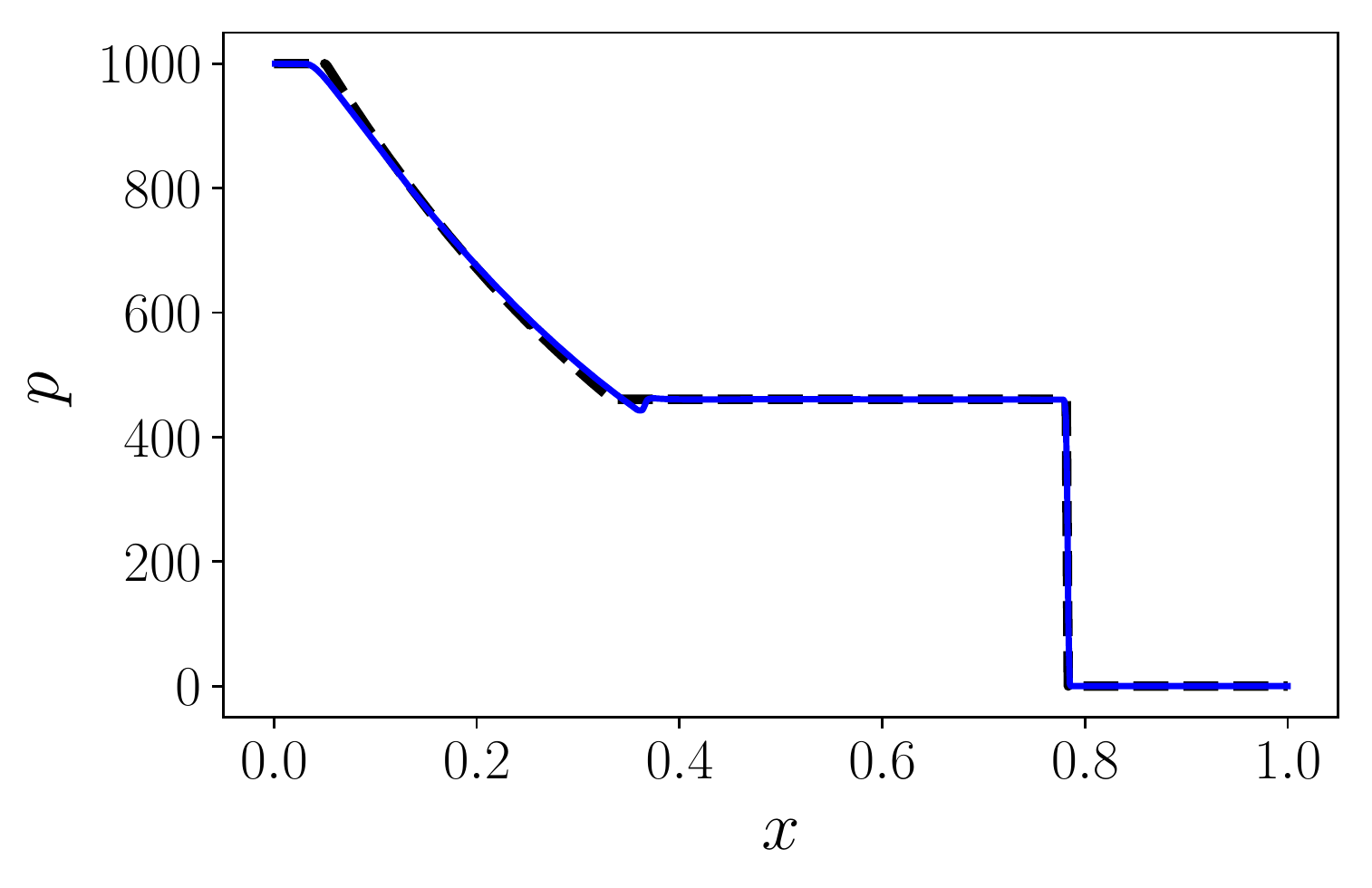}
    \caption{Density (top), velocity (middle) and pressure (bottom) profiles for the strong shock tube problem, at $t=0.012$. Solid line: finite-volume (FV) scheme. Dashed line: Reference from an exact Riemann solver. }
    \label{fig:Strong_ShockTube_Problem}
\end{figure}

This problem, characterized by the strong temperature discontinuity and a high Mach number of 198, probes the robustness and accuracy of the numerical methods. The results of the simulation, at $t=0.012$ and $L=800$, are shown in Fig. \ref{fig:Strong_ShockTube_Problem}. Overall, a very good agreement with the exact solution is noted.

\subsubsection{Two blast waves interaction problem}
The next test case is the two-blast-wave interaction problem, proposed by Woodward and Colella \cite{WoodwardCollela_DoubleMachReflection}. The following initial conditions are imposed for this problem,

\begin{equation}
    (\rho,u,p)=\begin{cases}
    (1, 0, 1000),& 0\leq x <0.1, \\
    (1, 0, 0.01),& 0.1\leq x <0.9, \\
     (1, 0, 100),& 0.9\leq x <1. \\
    \end{cases}
\end{equation}

\begin{figure}
    \centering
   \includegraphics[width=0.4\textwidth]{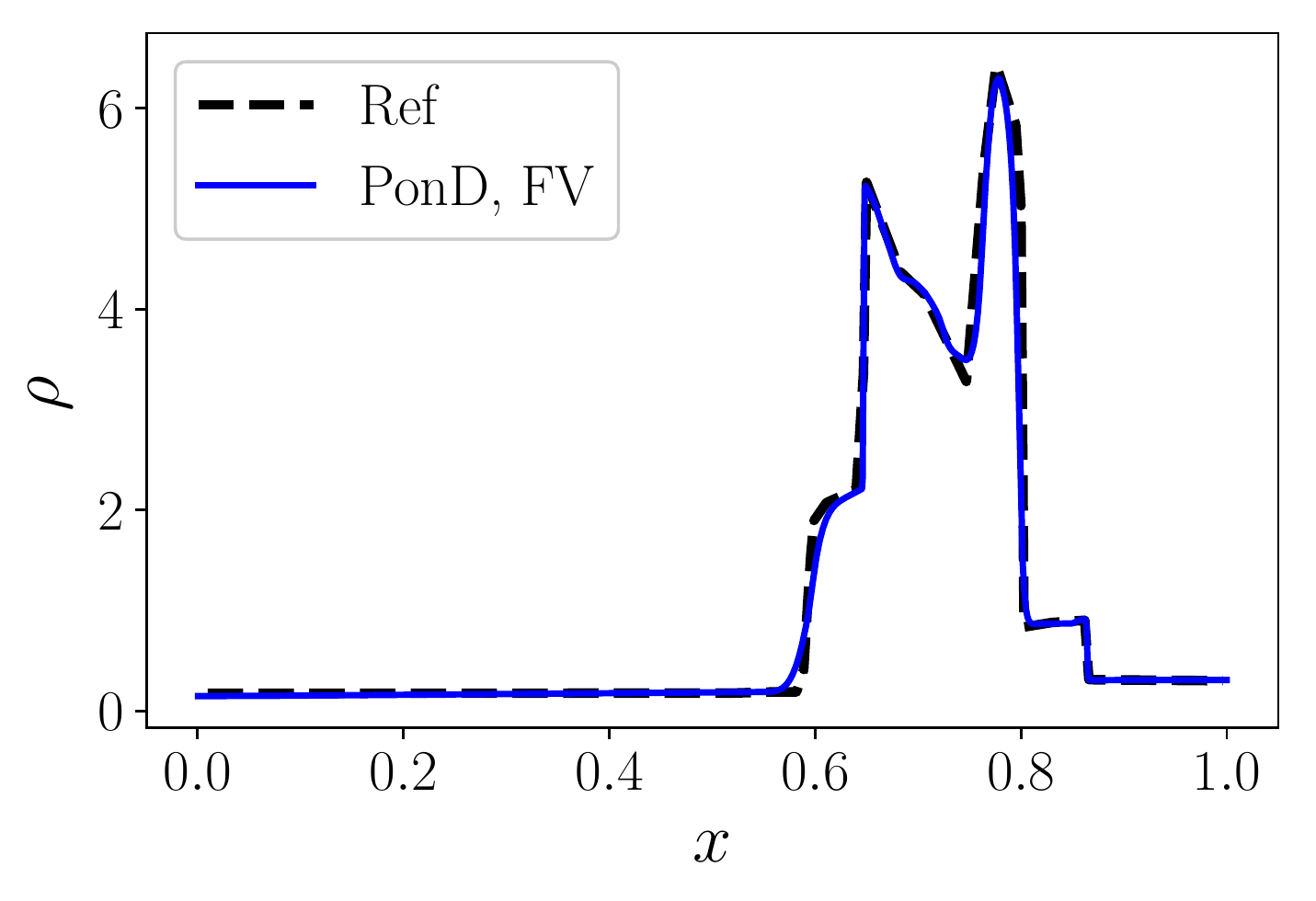}
    \includegraphics[width=0.4\textwidth]{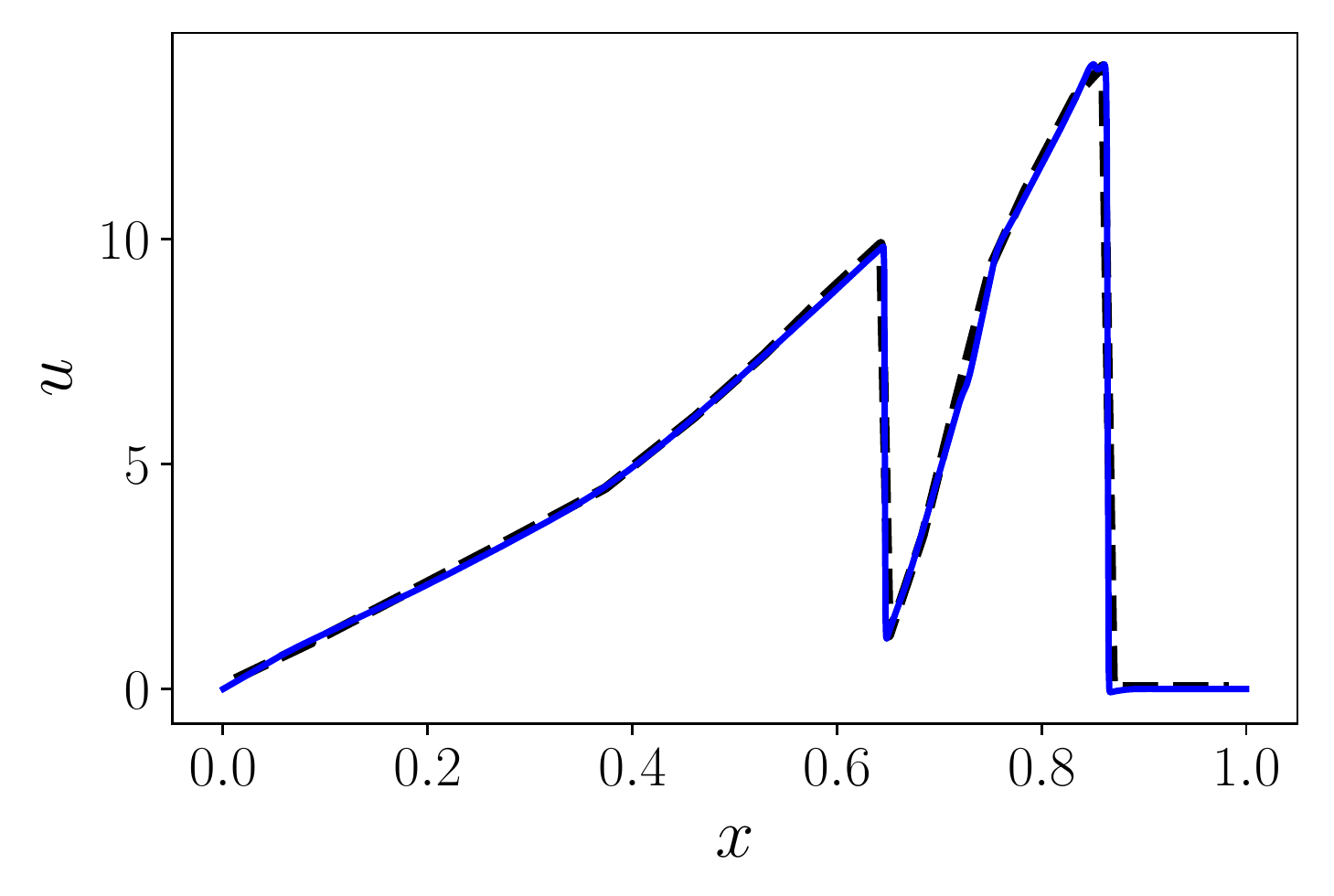}
     \includegraphics[width=0.4\textwidth]{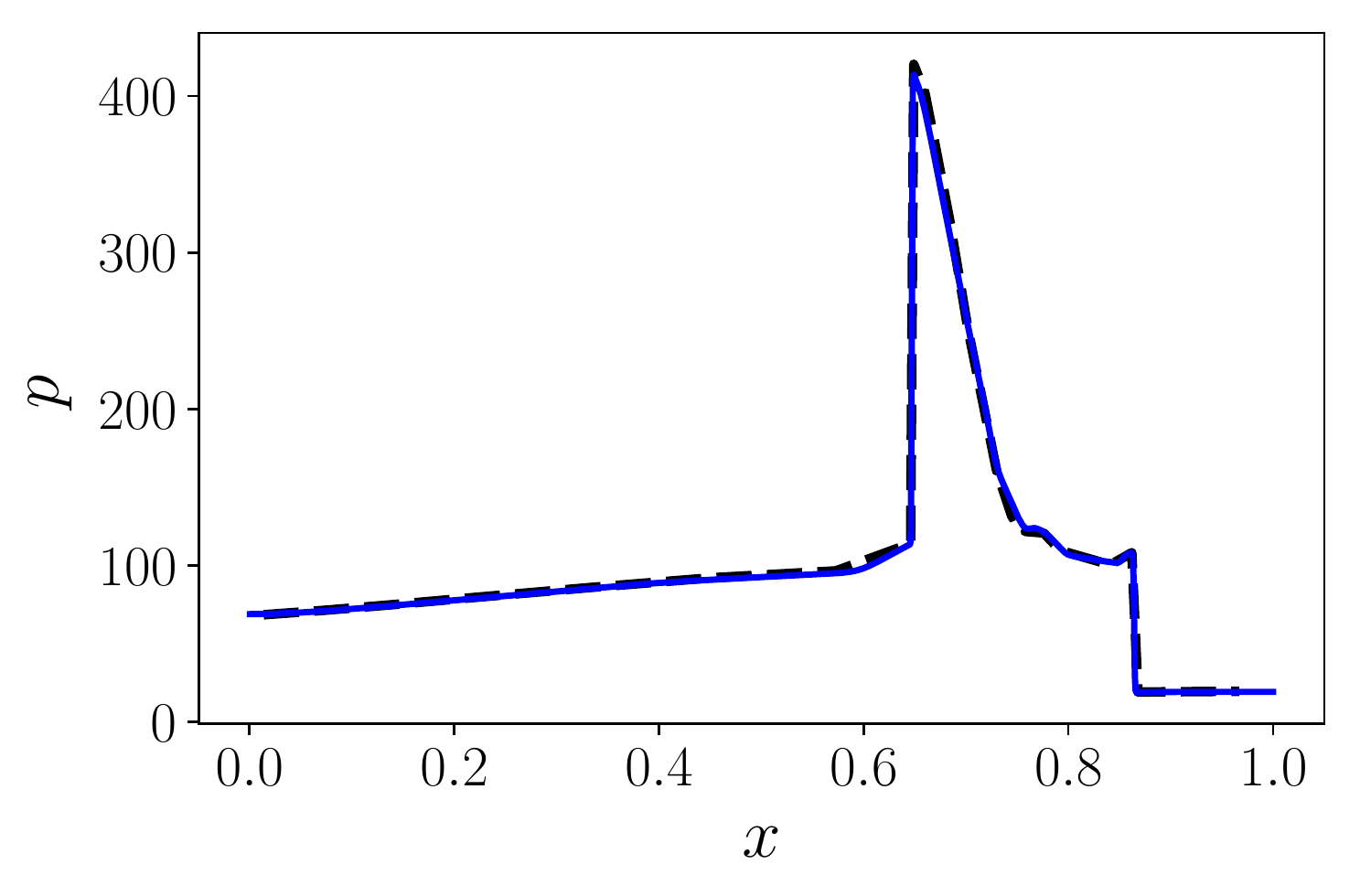}
    \caption{Density (top), velocity (middle) and pressure (bottom) profiles for the two blast wave problem, at $t=0.038$. Solid line: finite-volume (FV) scheme. Dashed line: Reference solution \cite{LinFuAllSpeed}.}
    \label{fig:Blast_Wave_Interaction_Problem}
\end{figure}

The resolution is $L=1600$ and reflective boundary conditions (BCs) are applied at $x=0$ and $x=1$. 
The results at $t=0.038$, shown in Fig. \ref{fig:Blast_Wave_Interaction_Problem}, are in very good  agreement with the reference solution from \cite{LinFuAllSpeed}.

\subsubsection{Double rarefaction problem}

We continue with a near-vacuum test case, which is known as the double rarefaction problem \cite{DoubleRaref}. The initial conditions are as follows:

\begin{equation}
    (\rho,u,p)=\begin{cases}
    (1, -2, 0.1),& 0 \leq x<0.5, \\
    (1, 2, 0.1),& 0.5 \leq x \leq 1. \\
    \end{cases}
\end{equation}

\begin{figure}
    \centering
   \includegraphics[width=0.4\textwidth]{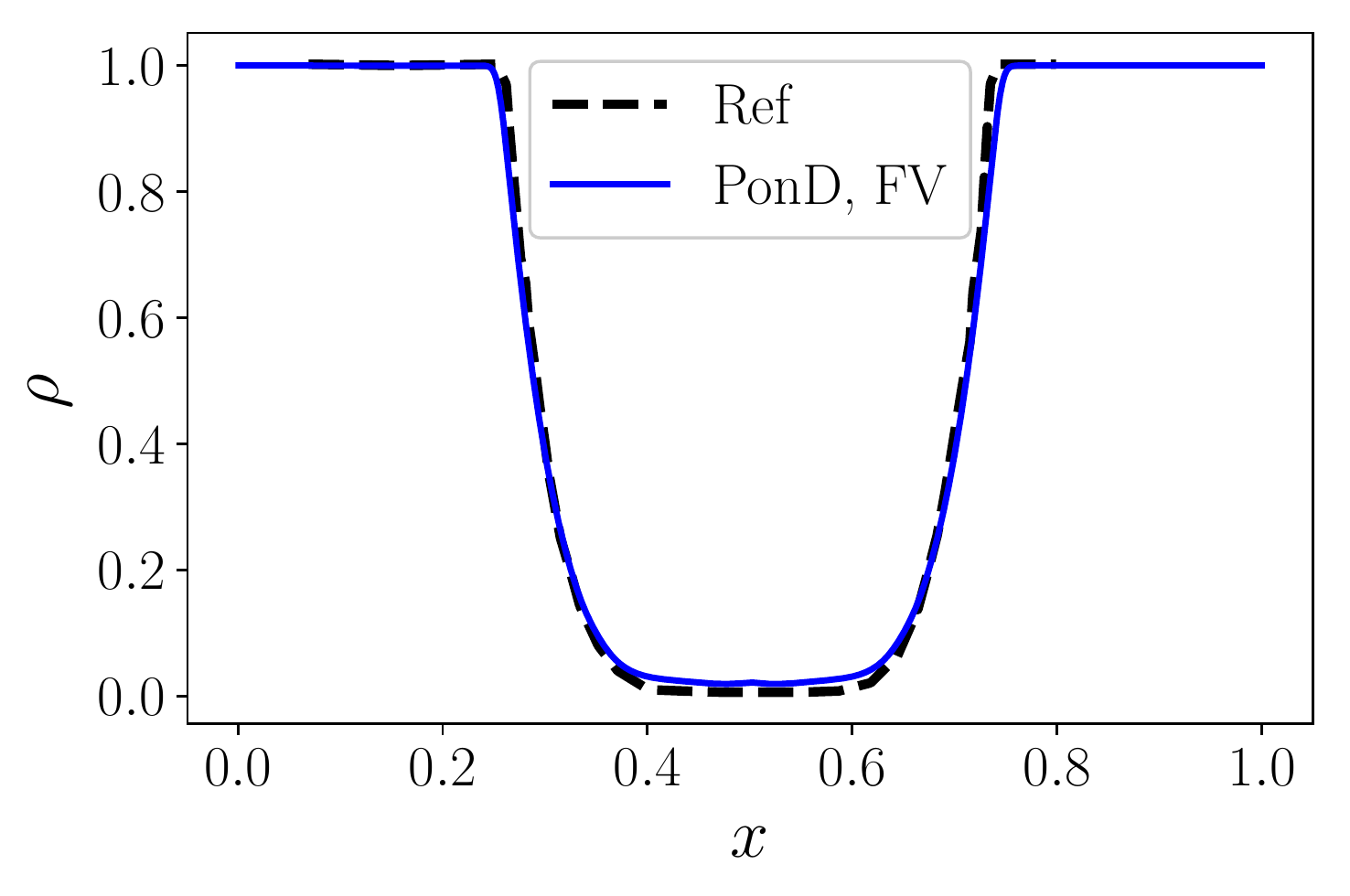}
    \includegraphics[width=0.4\textwidth]{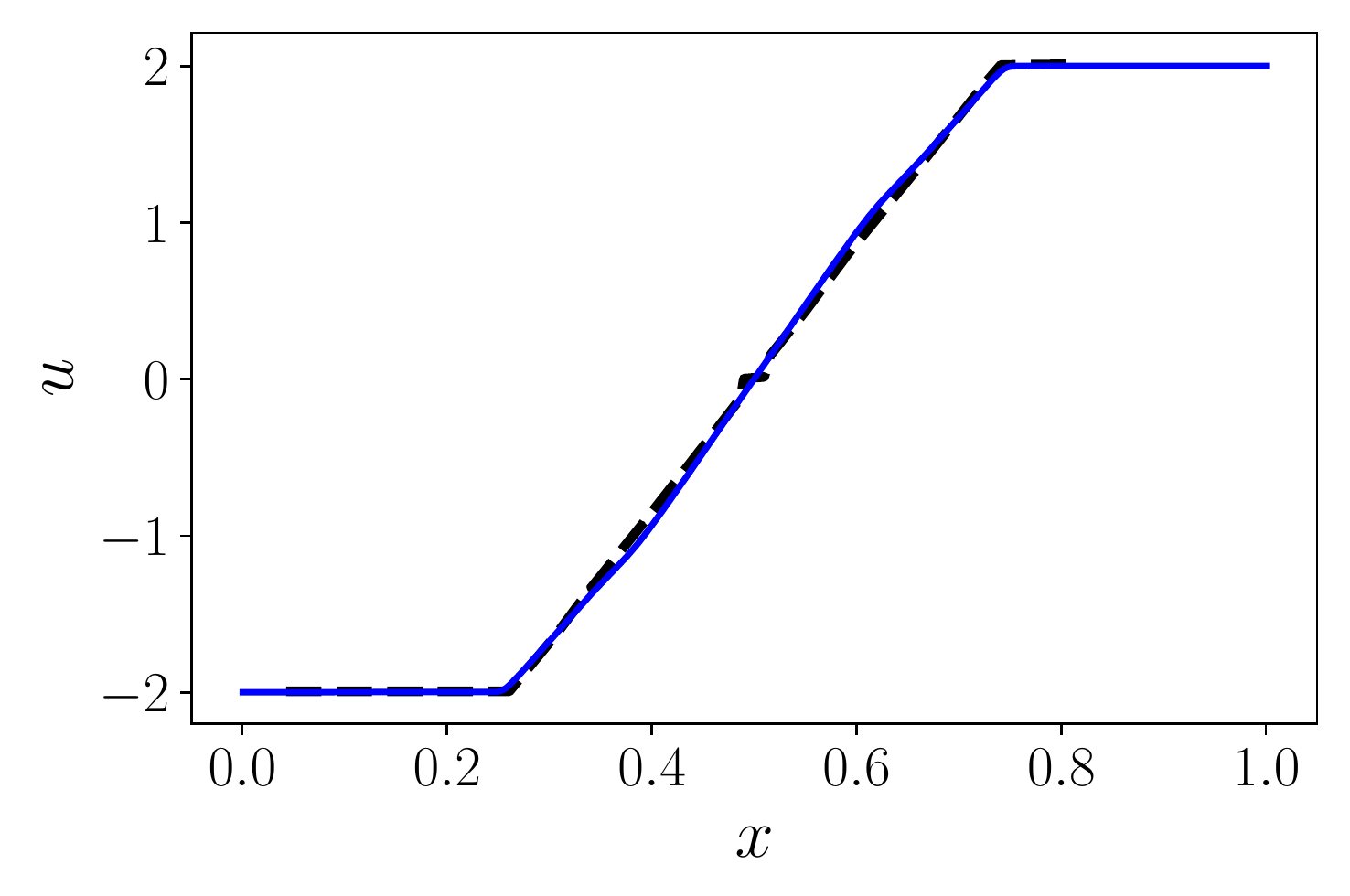}
     \includegraphics[width=0.4\textwidth]{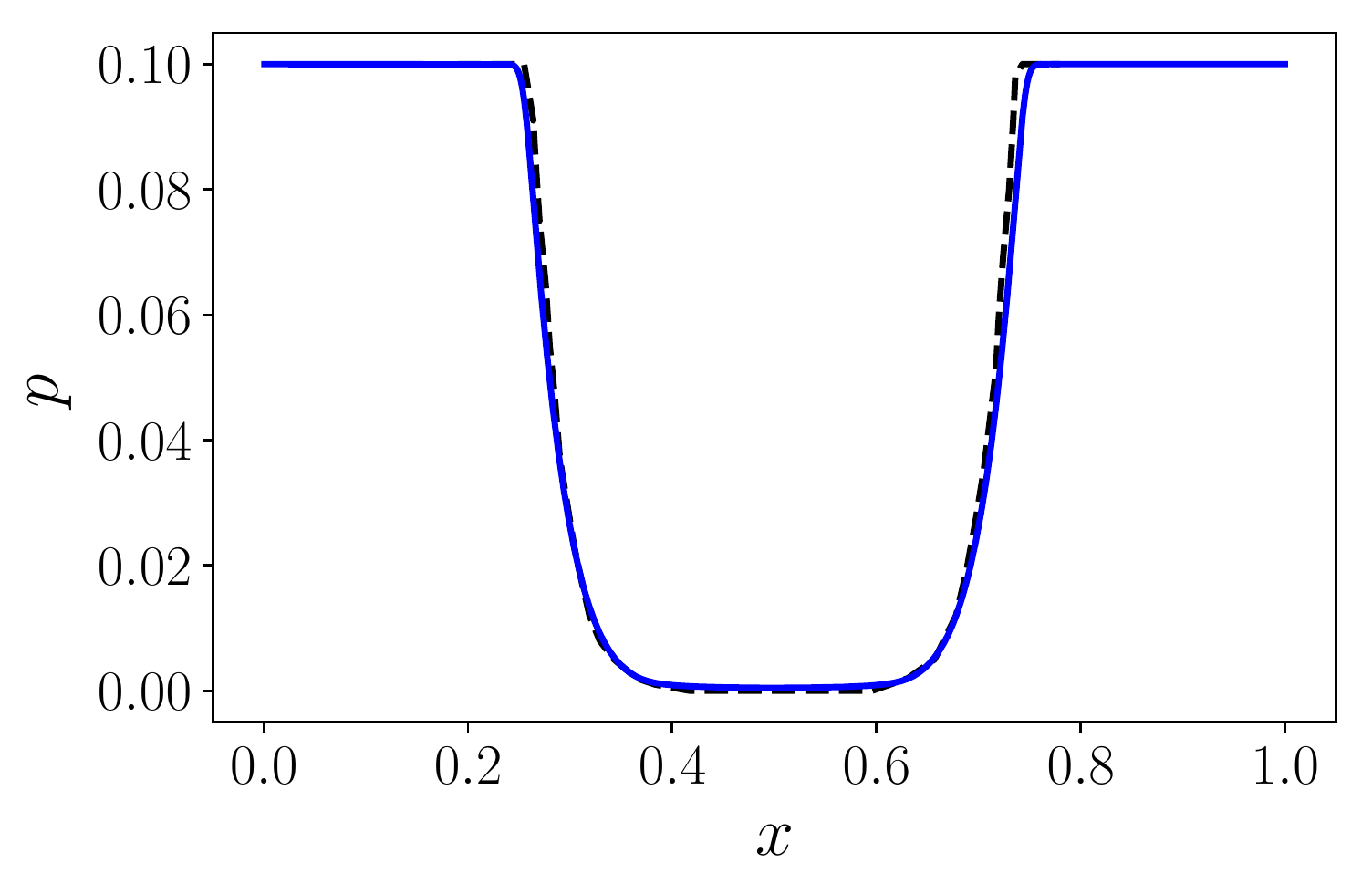}
    \caption{Density (top), velocity (middle) and pressure (bottom) profiles for the double rarefaction problem, at $t=0.1$. Solid line: finite-volume (FV) scheme. Dashed line: Reference from an exact Riemann solver.}
    \label{fig:Double_Rar_Problem}
\end{figure}

The results are compared with the exact solution at $t=0.1$ and $L=800$, as shown in Fig. \ref{fig:Double_Rar_Problem}. It can be seen that as the two rarefaction waves propagate towards opposite directions, a near-vacuum is formed in the center of the domain. Nonetheless, a good agreement to the reference solution can be observed.

\subsubsection{Le Blanc problem}

The Le Blanc problem is considered next \cite{LeBlanc}, which involves very strong discontinuities and is initialized with the following conditions,

\begin{equation}
    (\rho,u,p)=\begin{cases}
    (1, 0, 2/3 \times 10^{-1}),& 0\leq x <3, \\
    (10^{-3}, 0, 2/3 \times 10^{-10}),& 3\leq x \leq 9. \\
    \end{cases}
\end{equation}

\begin{figure}
    \centering
   \includegraphics[width=0.4\textwidth]{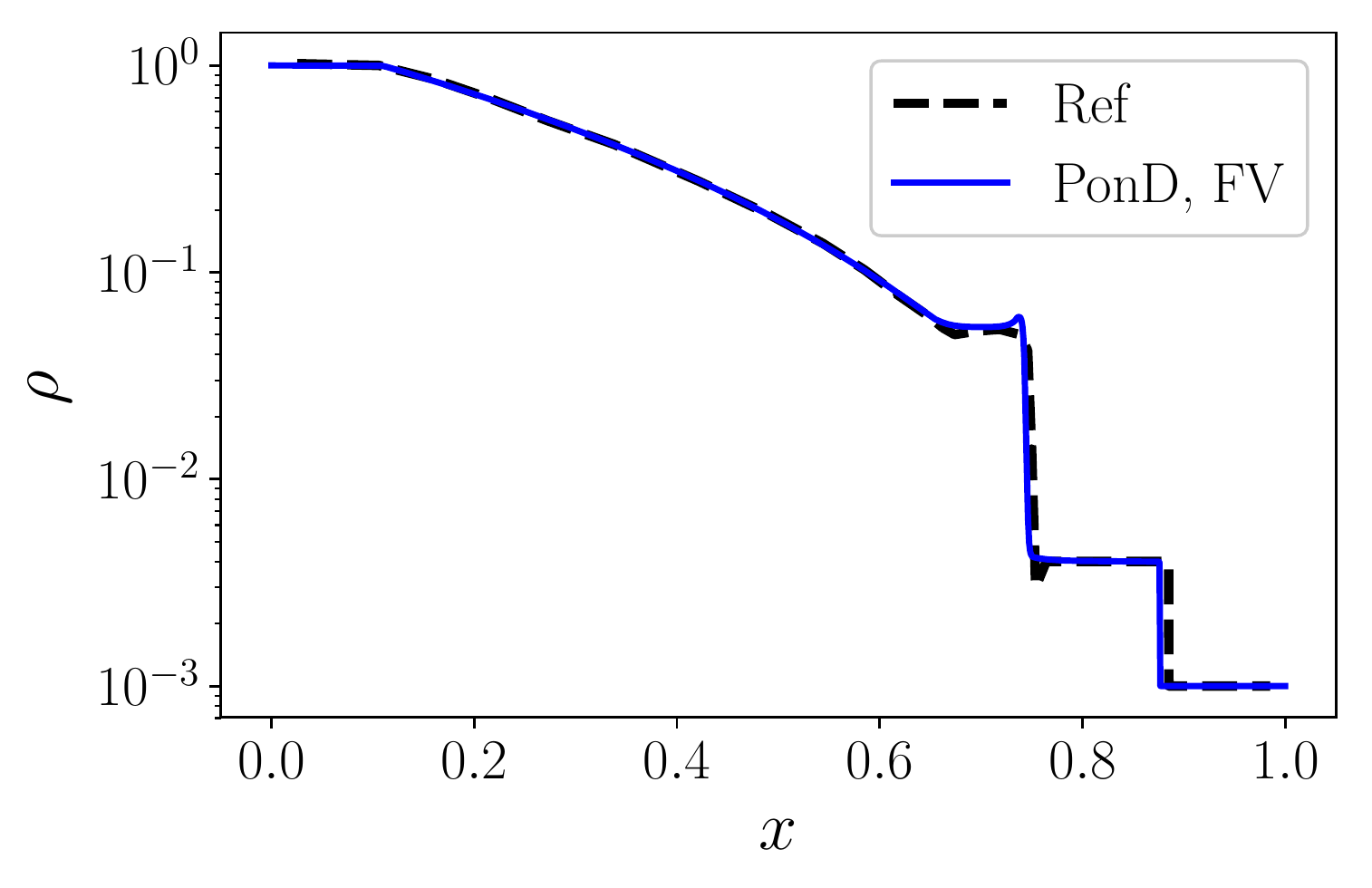}
    \includegraphics[width=0.4\textwidth]{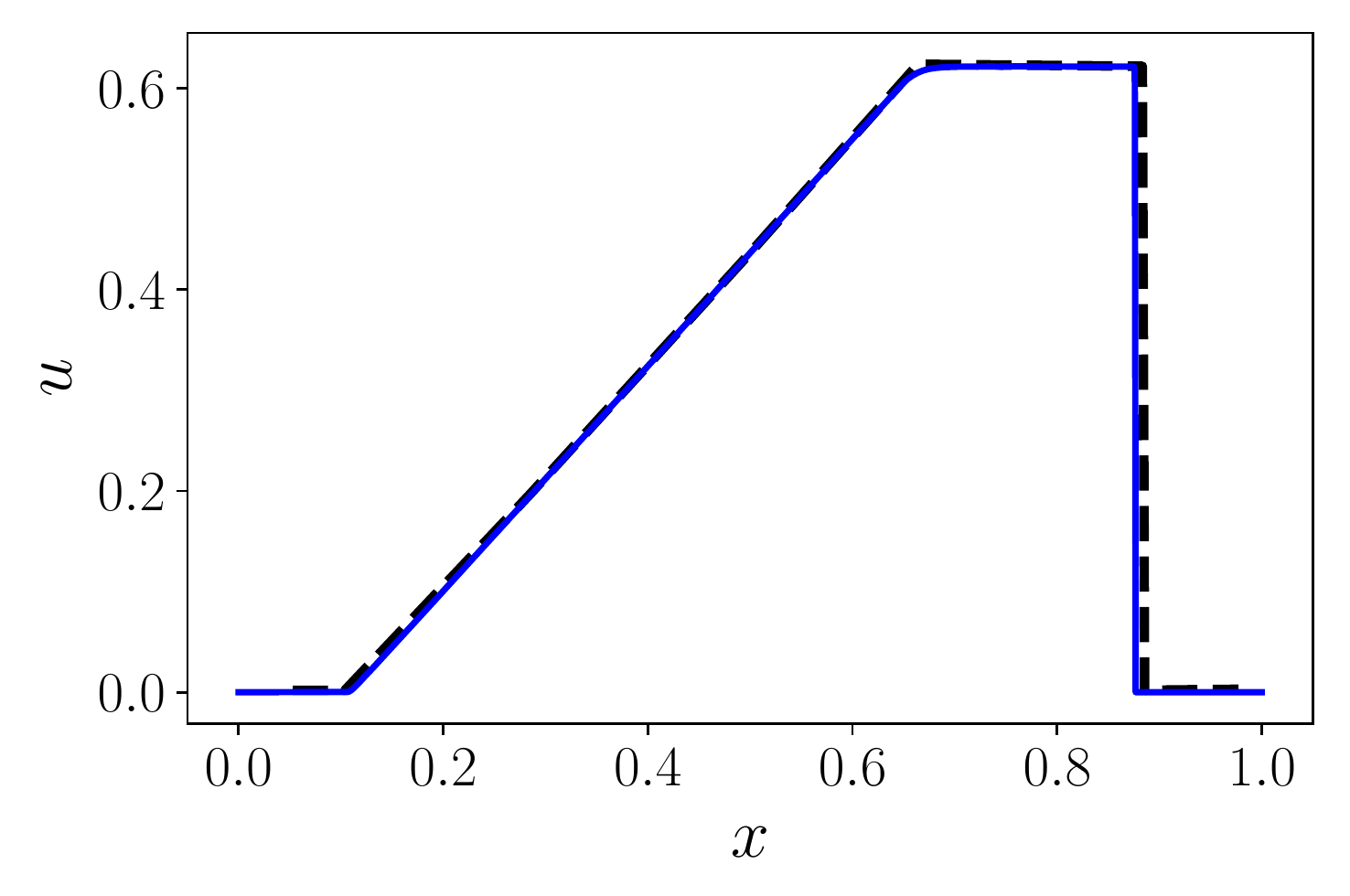}
     \includegraphics[width=0.4\textwidth]{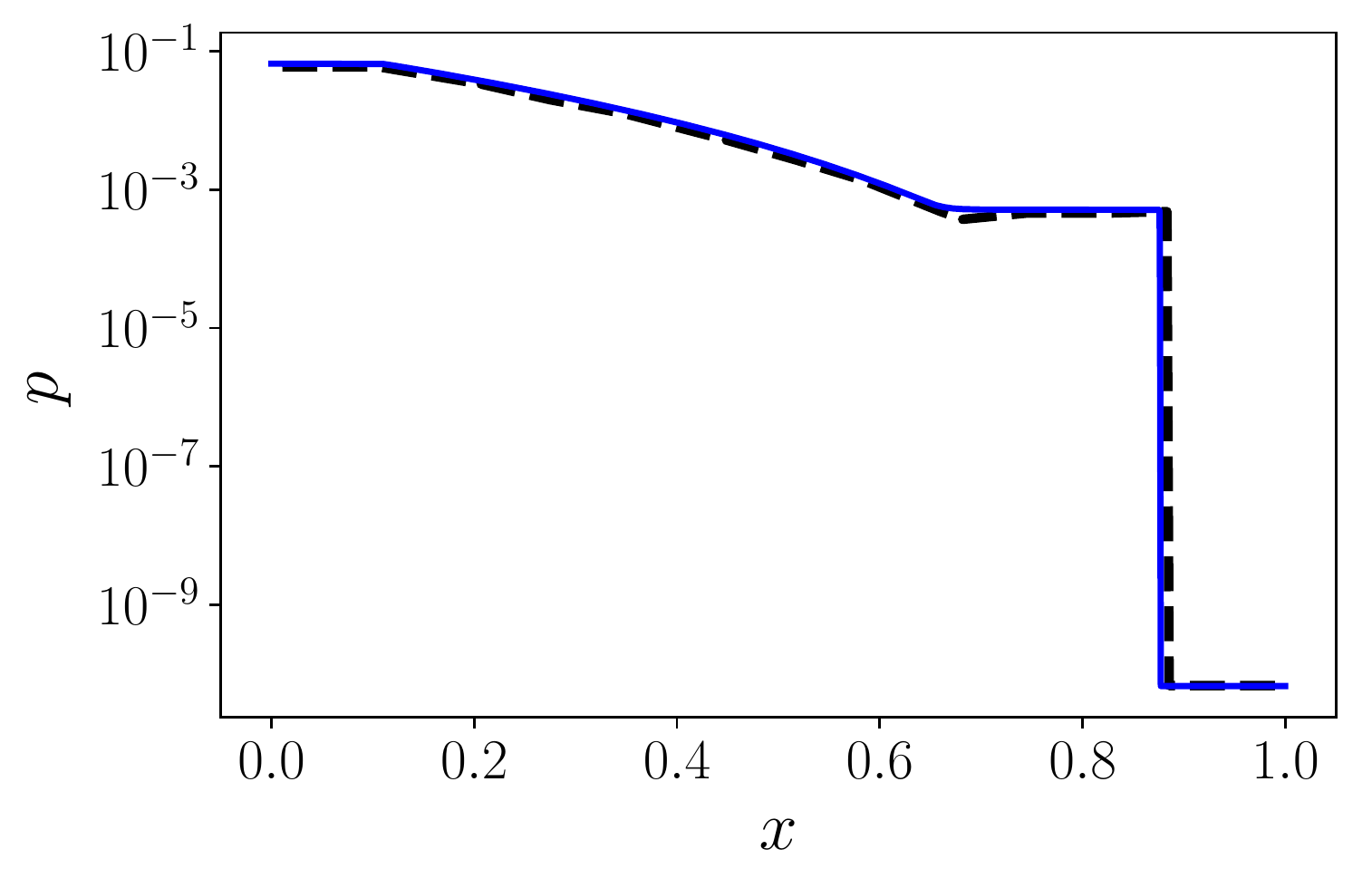}
    \caption{Density (top), velocity (middle) and pressure (bottom) profiles for the Le Blanc problem, at $t=6$. Solid line: finite-volume (FV) scheme. Dashed line: Reference solution \cite{LinFuAllSpeed}.}
    \label{fig:Le Blanc_Problem}
\end{figure}

In this problem, the adiabatic exponent is fixed to $\gamma=5/3$. Fig. \ref{fig:Le Blanc_Problem} shows the results at $t=6$ and $L=4000$. With the exception of minor oscillations, a very good agreement of the present scheme with the reference solution \cite{LinFuAllSpeed} is observed.

\subsubsection{Planar Sedov blast-wave problem}

\begin{figure}
    \centering
   \includegraphics[width=0.4\textwidth]{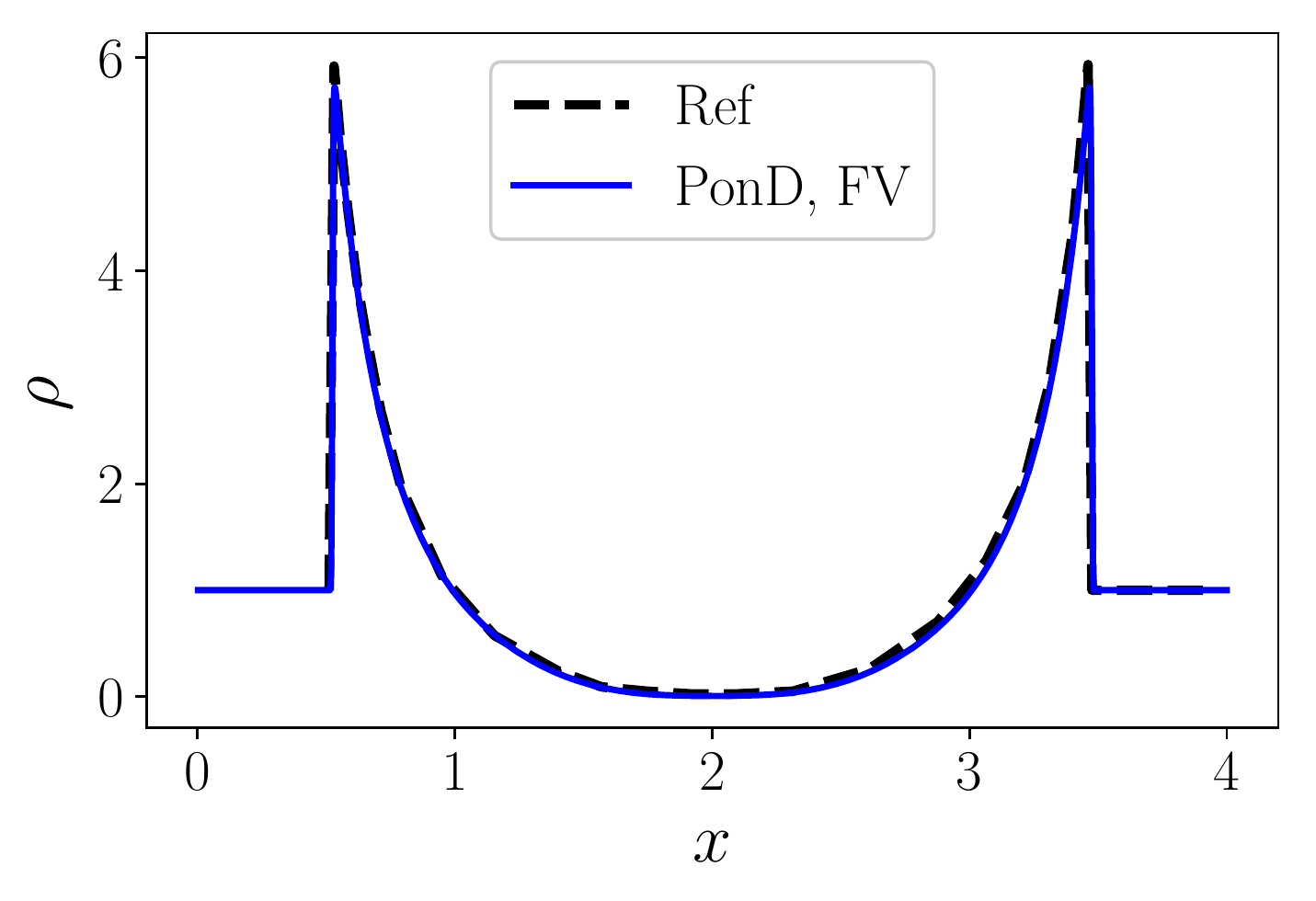}
    \includegraphics[width=0.4\textwidth]{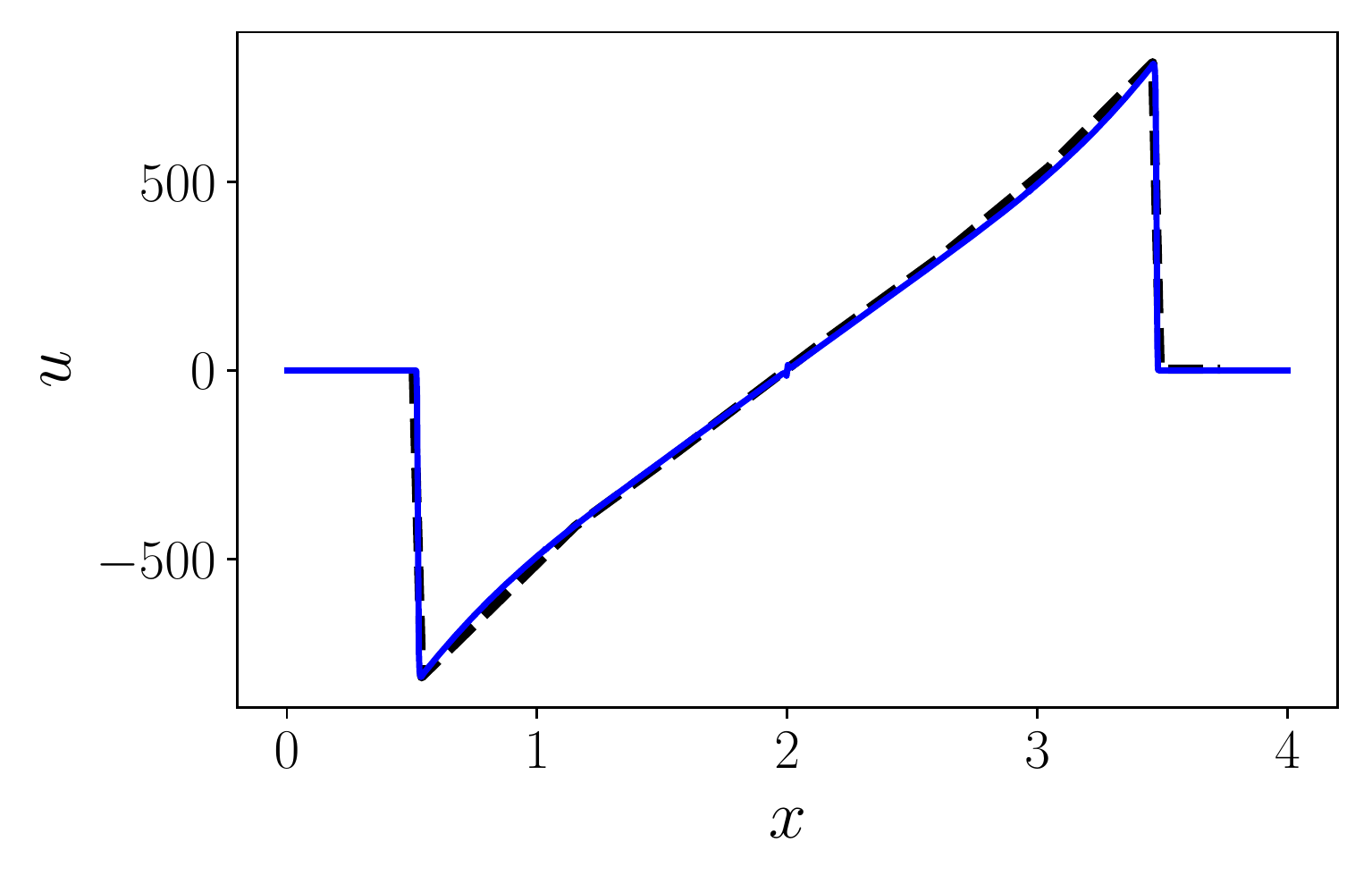}
     \includegraphics[width=0.4\textwidth]{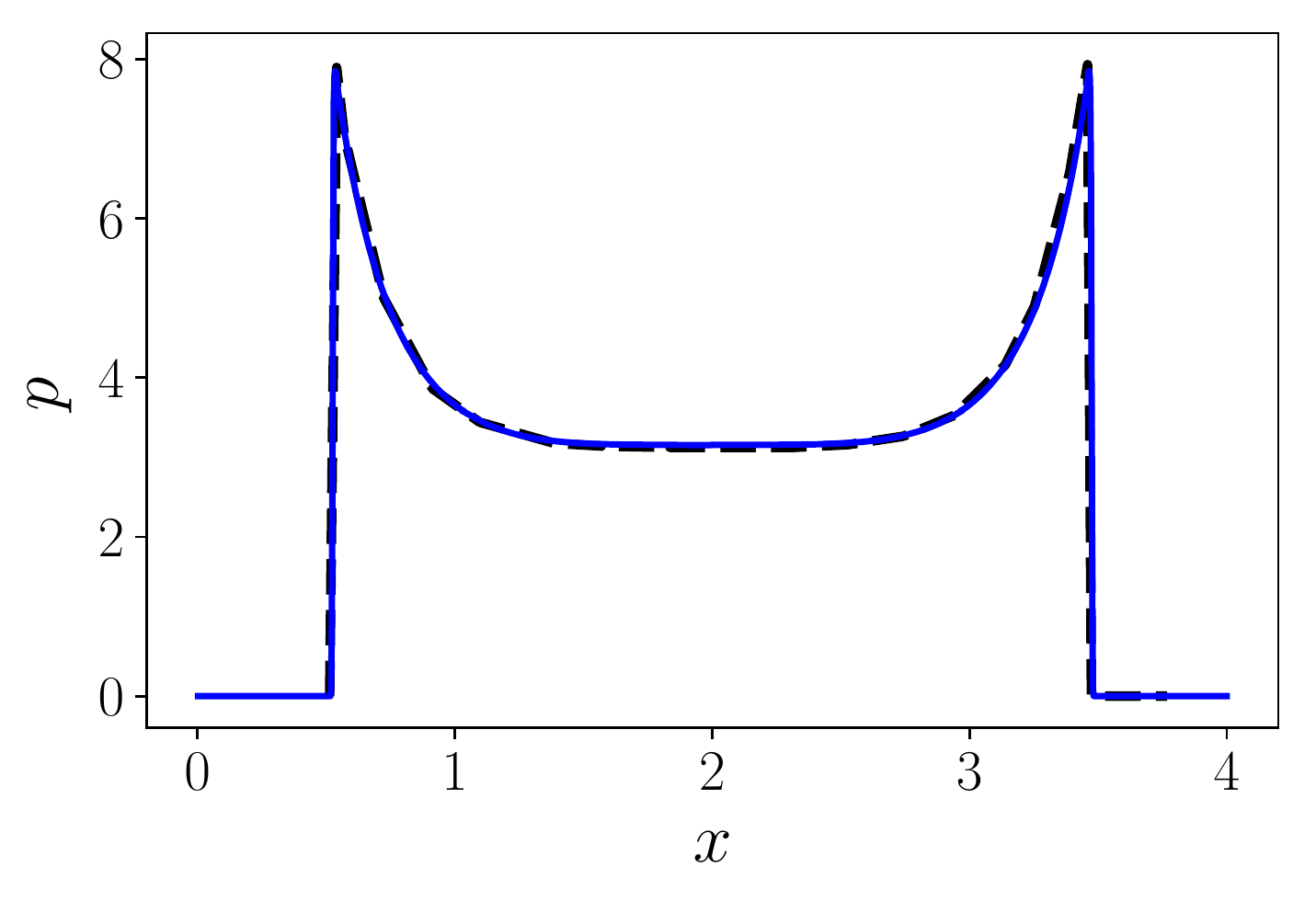}
    \caption{Density (top), velocity (middle) and pressure (bottom) profiles for the Sedov blast wave problem, at $t=0.001$. Solid line: finite-volume (FV) scheme. Dashed line: Reference solution \cite{LinFuAllSpeed}.}
    \label{fig:Sedov_Problem}
\end{figure}

The final 1D test case is the planar Sedov blast-wave problem \cite{Sedov}. The initial conditions for this case are the following,
\begin{equation}
\begin{split}
(\rho,u,p)=\begin{cases}
    (1, 0, 2.56 \times 10^{8}),\  2-0.5\delta x\leq x \leq 2+0.5\delta x, \\
    (10^{-3}, 0, 4 \times 10^{-13}),\ \mathrm{otherwise}.
    \end{cases}
    \end{split}
\end{equation}
The initial conditions of this problem approximate a delta function of pressure, concentrated at the center of the domain and almost vanishing everywhere else. The blast wave emanating from the center propagates outwards, creating a post-shock region of very low density. It is noted that the initial ratio of pressure between the center and the surroundings is 21 orders of magnitude. The results are shown in Fig. \ref{fig:Sedov_Problem}, at $t=0.001$ and $L=1600$, demonstrating very good agreement with the reference solution \cite{LinFuAllSpeed}.

\subsubsection{Comparison of semi-Lagrangian and finite volume schemes}
\label{sec:FVSLcomparison}

\begin{figure}
     \centering
   \includegraphics[width=0.4\textwidth]{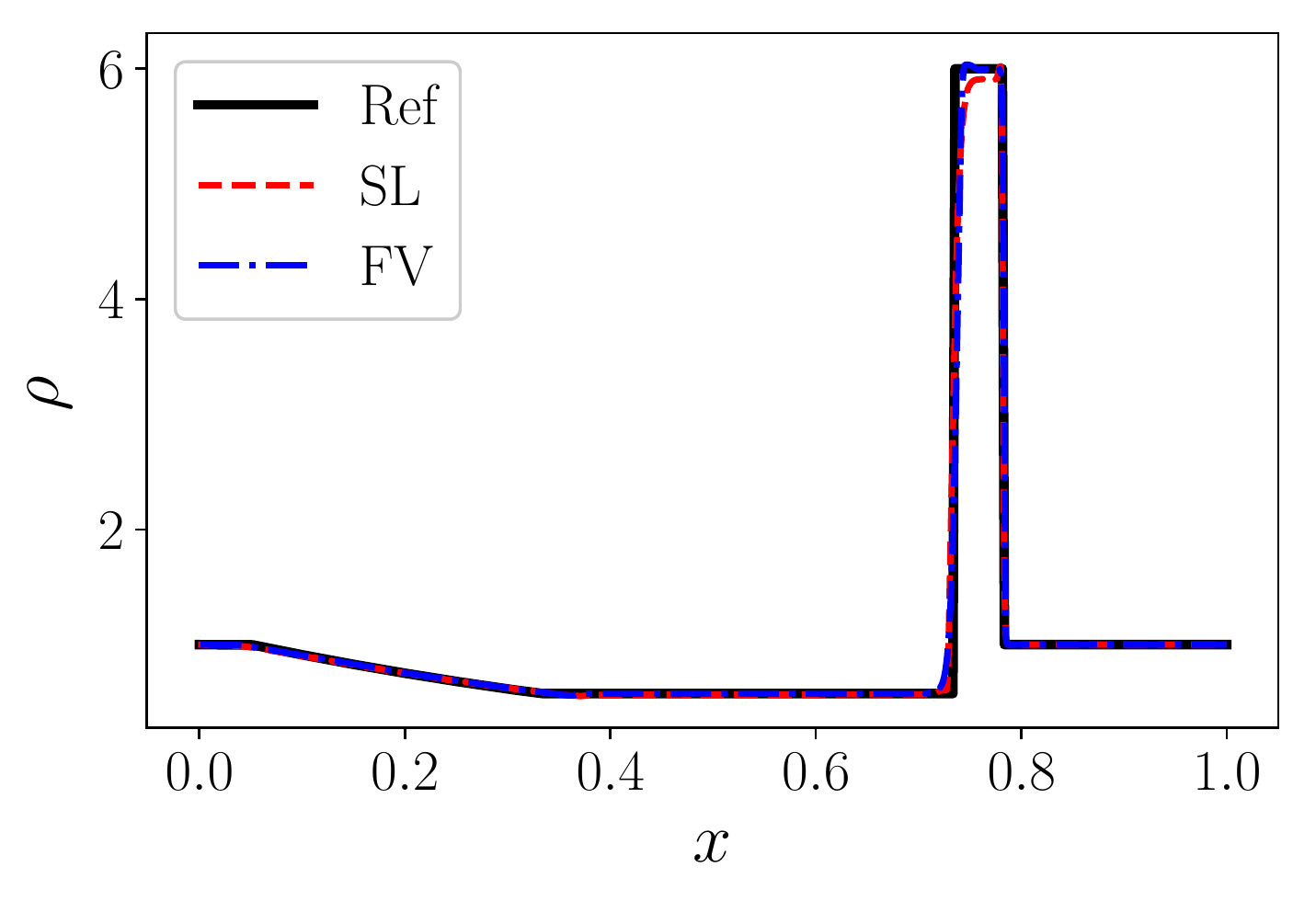}
     \includegraphics[width=0.4\textwidth]{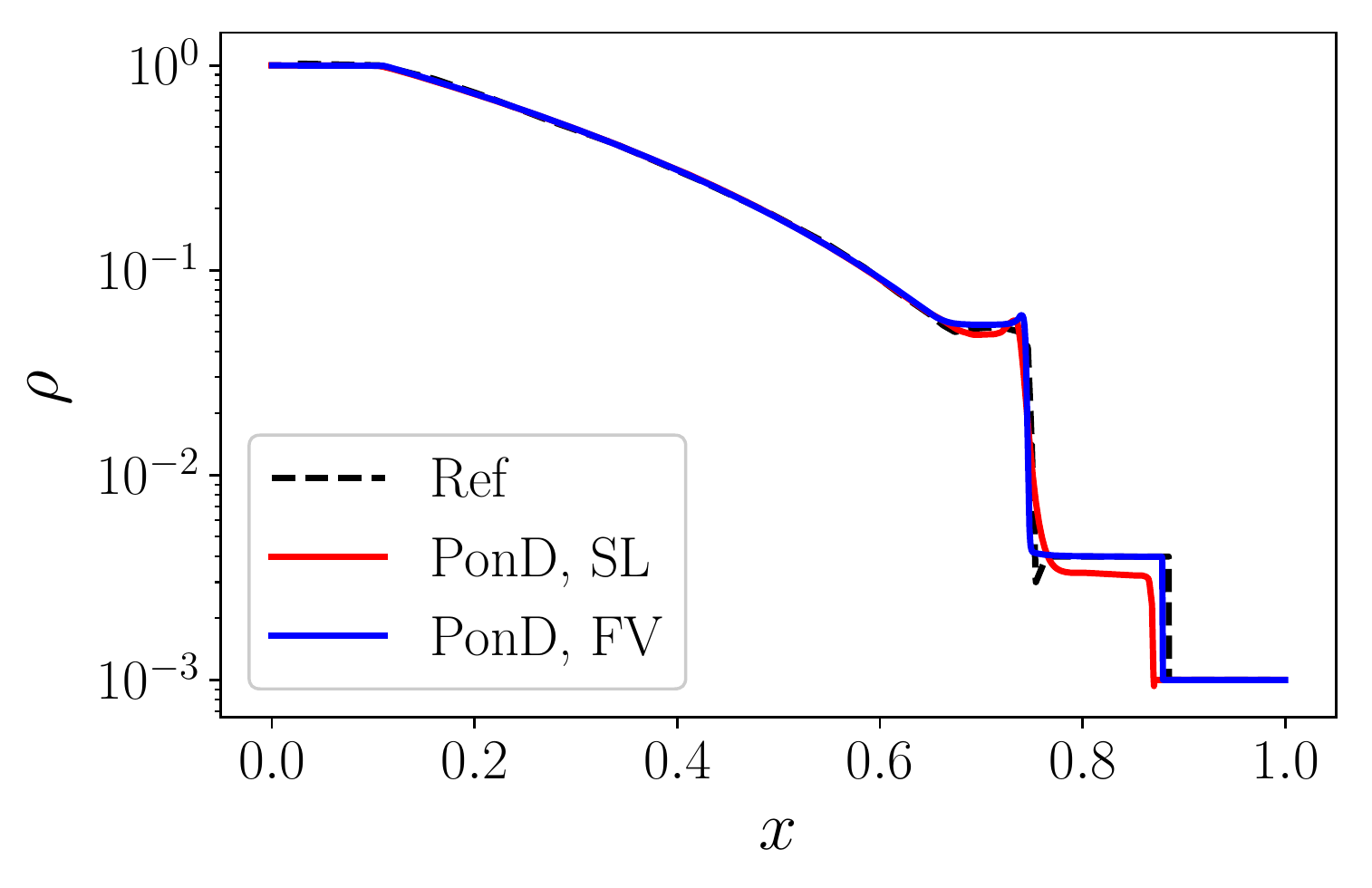}
     \includegraphics[width=0.4\textwidth]{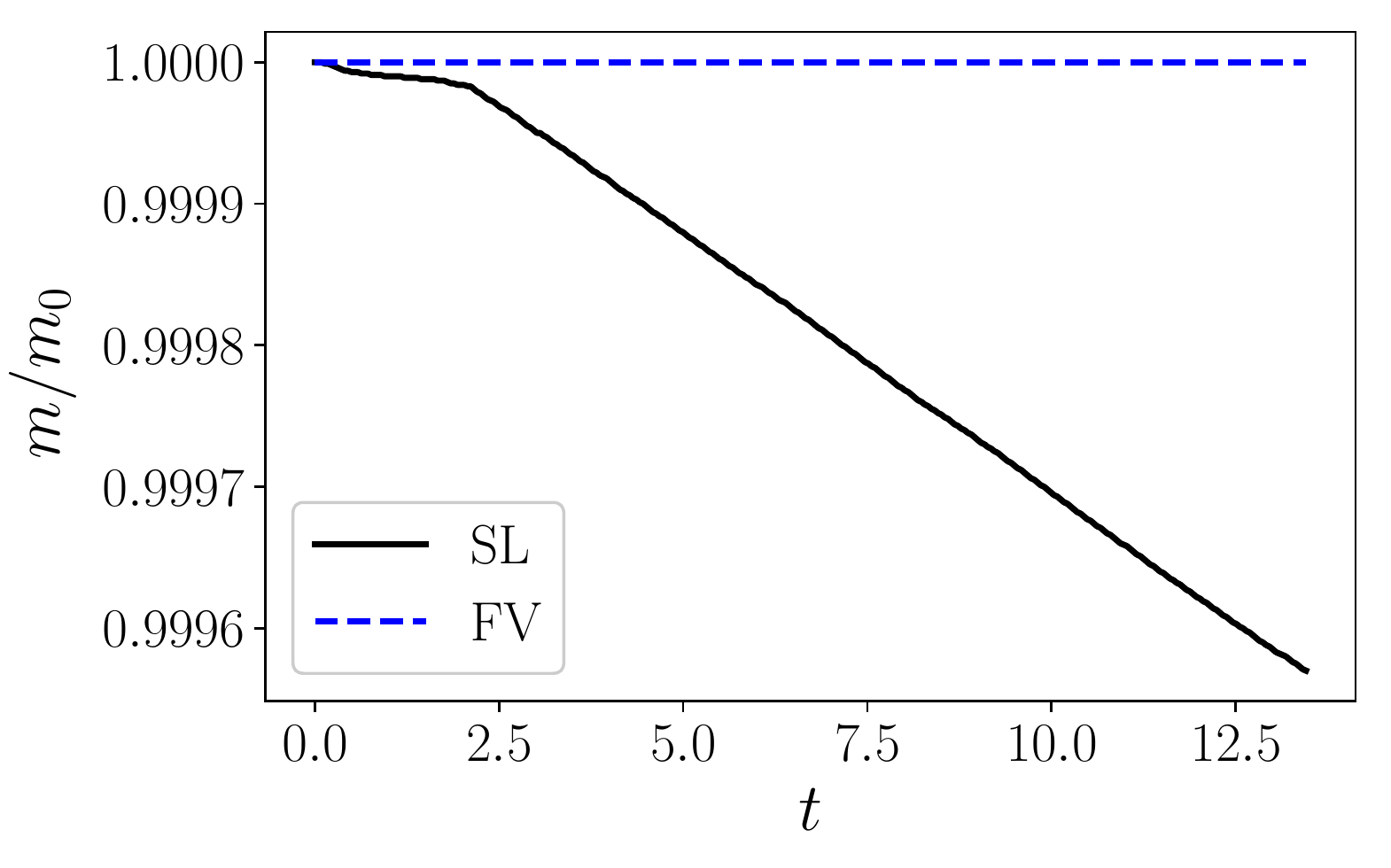}
    
     \caption{Comparison of semi-Lagrangian (SL) and finite volume schemes (FV). Top: strong shock tube. Middle: Le Blanc problem. Bottom: Evolution of mass for the Le Blanc problem.}
     \label{fig:Comparison}
 \end{figure}

Finally, we compare the two numerical schemes that were used in this work. 
As mentioned above, the key distinction between the two methods is that the finite volume discretization is strictly conservative, whereas the semi-Lagrangian method is not. While there exist conservative formulations \cite{ConservSemiLag}, they have not been the focus of the current work. To highlight the effect trough our test cases, we compare the strong shock tube and the Le Blanc problem. The former is characterised by strong discontinuities in pressure and temperature but the density is of the same order. In contrast, the Le Blanc problem involves large variations in the density. Fig.\ \ref{fig:Comparison} compares the results of the two numerical schemes for these problems. For the case of the strong shock tube, an almost identical profile is attained, with minor oscillations being more pronounced for the semi-Lagrangian scheme. On the other hand, the performance of the two schemes deviates for the case of the Le Blanc problem. Here we notice that the finite volume discretization leads to a very accurate comparison with the reference solution. However, a discrepancy in density is observed for the semi-Lagrangian scheme. In particular, an under prediction in the density as well as a mismatch of the shock location are present. This behaviour is expected for problems involving low densities, since the effect of the conservation error is more pronounced. 

We conclude the comparison with a comment regarding the computational efficiency of the two schemes. 
Comparing the two algorithms, one can notice that per time-step one semi-Lagrangian step is included in the flux calculation for the finite volume realization. Therefore, the computational cost of the finite volume scheme, per time step, is in general higher than the semi-Lagrangian scheme. However, the added cost to ensure strict conservation remains reasonable. In particular, we compared the runtimes of the two schemes, for a given level of accuracy (same $L_2$ error) for the Shu-Osher problem and observed only $20 \%$ increase for the finite volume implementation. We need to stress however that a comprehensive study of the computational performance of the two schemes, including efficiency, stability and numerical dissipation, requires further in-depth investigations and is left for future work.

\subsection{2D cases}

\subsubsection{2D Riemann problem}

\begin{figure*}
    \centering
   \includegraphics[width=0.9\textwidth]{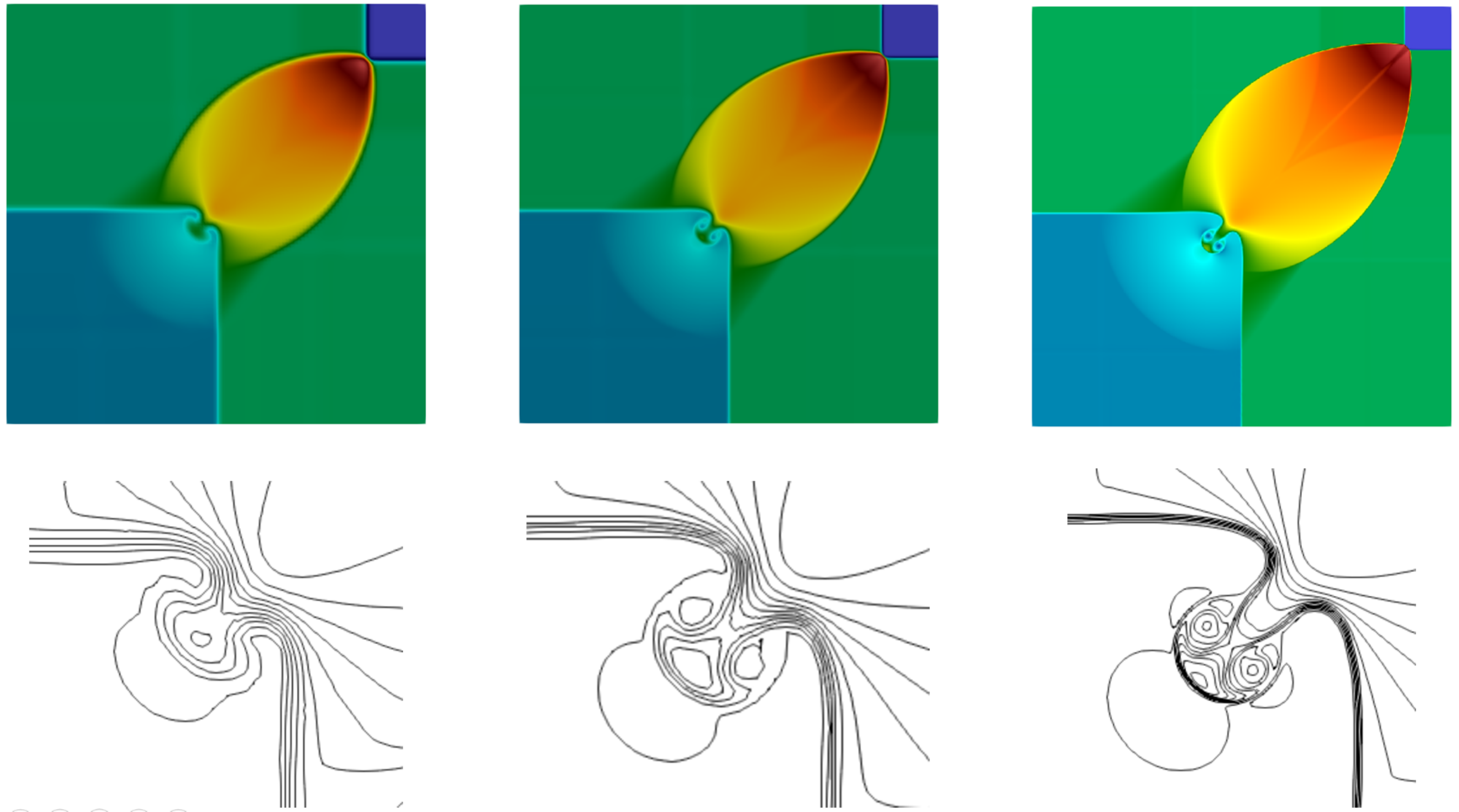}

    \caption{2D Riemann problem with three different resolutions, $[125,125]$ (left), $[250,250]$ (middle), $[750,750]$ (right). The top row shows plots of the density fields, at $t=0.25$. The bottom row represent 30 equidistant density contours.}
    \label{fig:2D_Riemann}
\end{figure*}

As a first validation in two dimensions we simulate a 2D Riemann problem, which is a classical benchmark for compressible flow solvers \cite{Lax_RiemannReference}. A square domain $(x,y) \in [0,1]\times [0,1]$ is divided into four quadrants, each of which is initialized with constant values of density, velocity and pressure as follows:
\begin{equation}
     (\rho,u_x,u_y,p) = \begin{cases}
    (0.5313, 0, 0, 0.4),& x>0.5, y>0.5, \\
    (1, 0.7276, 0, 1),& x\leq 0.5, y>0.5, \\
    (0.8, 0, 0, 1),& x\leq 0.5, y\leq 0.5, \\
    (1, 0, 0.7276, 1),& x>0.5, y\leq 0.5. \\
    \end{cases}
\end{equation}
At the boundaries, a zero-gradient BC was imposed $\partial_{\bm{n}} f=0$, where $\bm{n}$ is the outwards unit normal vector. Three simulations with increasing resolution, $L=125,250,750$, were performed. The results of the density field, as well as  density contours near the center of the domain, are depicted in Fig. \ref{fig:2D_Riemann}. The specified initialization of the Riemann problem leads to shock waves interacting and propagating towards the upper right quadrant, while a complex pattern is formed in the opposite direction. The results show a very good agreement with the reference solutions in \cite{Lax_RiemannReference,2DRiem_ref}. Moreover, the refinement of the mesh leads to an increasing resolution of the finer structures near the origin.

\subsubsection{2D explosion in a box}

\begin{figure}
    \centering
   \includegraphics[width=0.4\textwidth]{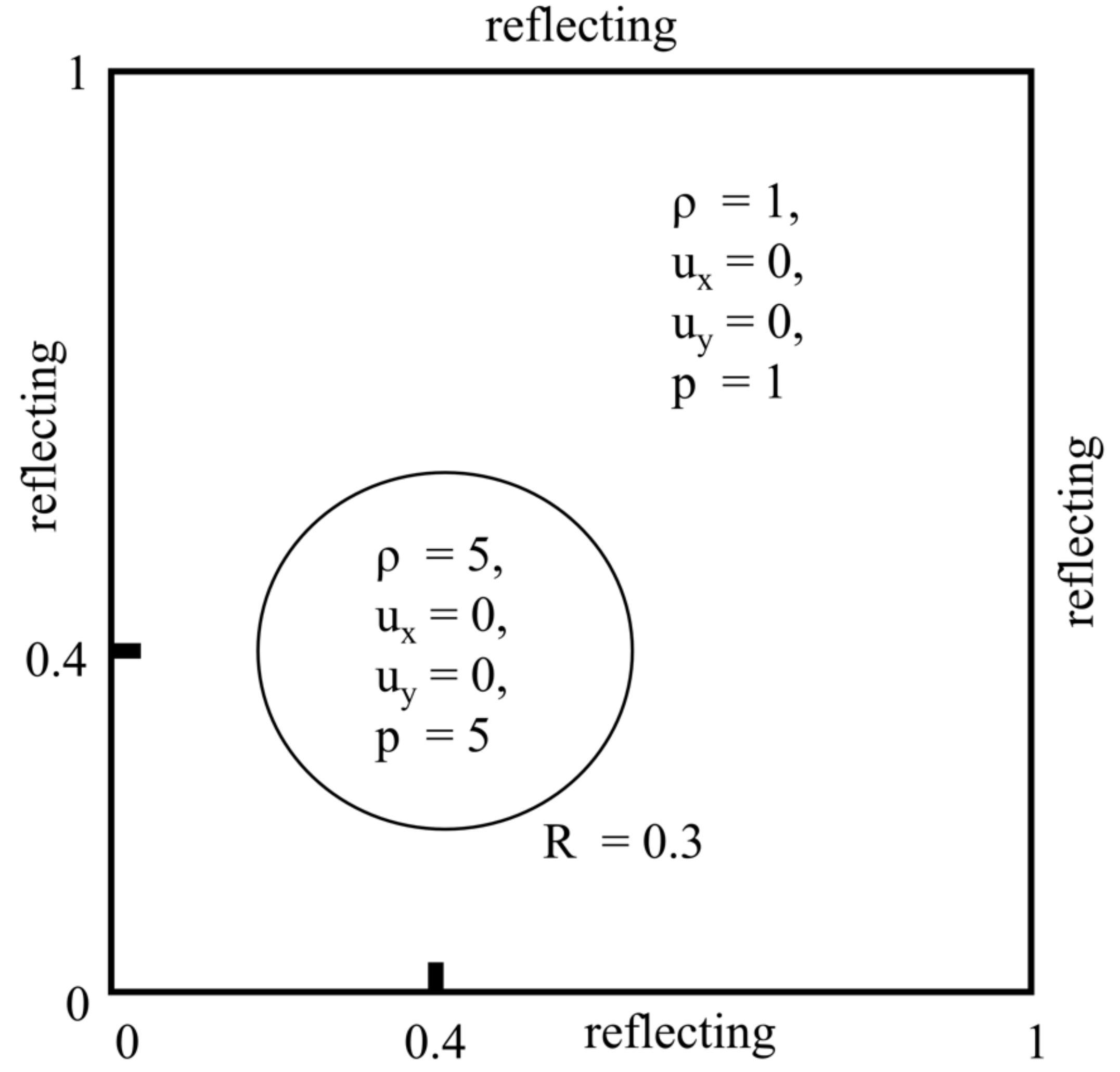}
   \includegraphics[width=0.4\textwidth]{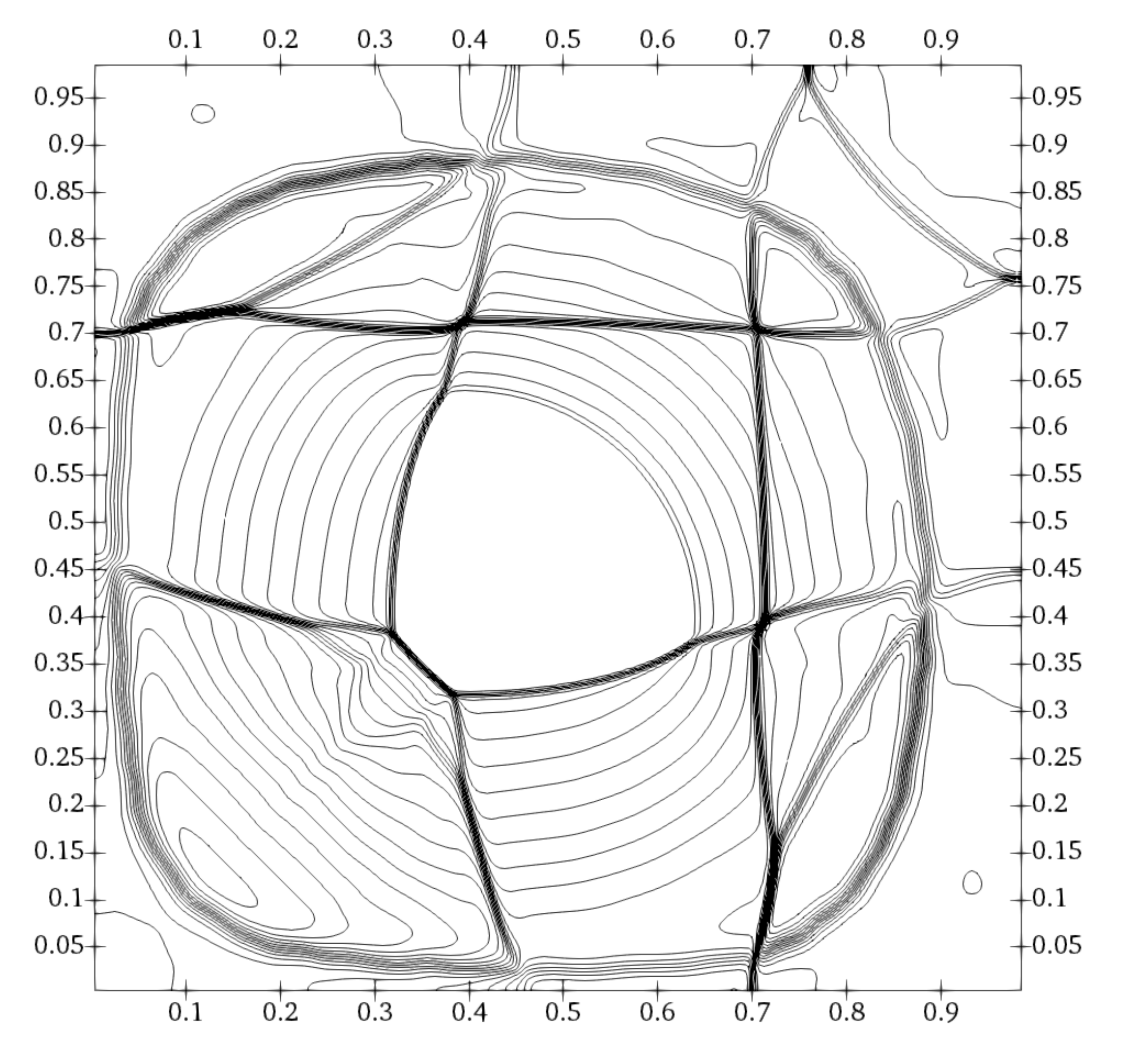}

    \caption{Top: The configuration of the 2D explosion in a box. Bottom:  $30$ equally spaced density contours in the range $\rho \in [0.52,3.8]$ and at $t=0.5$ are shown.}
    \label{fig:2D_Box_Init}
\end{figure}

We consider here an unsteady explosion enclosed in a 2D box. The configuration of this case is shown schematically in Fig. \ref{fig:2D_Box_Init}. The computational domain $[0,1]\times [0,1]$ is initialized with the following conditions,
\begin{equation}
    (\rho,u_x,u_y,p)=\begin{cases}
    (5, 0, 0, 5),& |(x,y)-(0.4,0.4)|<0.3, \\
    (1, 0, 0, 1),& \mathrm{otherwise}. \\
    \end{cases}
\end{equation}
The domain was discretized with 256 points per direction and reflective BCs were imposed on the walls of the box. With this setup, the circular shock waves expand towards the boundaries of the box and the reflected waves interact in a complicated manner. A snapshot from the evolution at $t=0.5$ is shown in Fig. \ref{fig:2D_Box_Init}, which depicts 30 density contours. A comparison with the results obtained from a block-structured adaptive mesh refinement solver in \cite{ExplosionBoxRef} demonstrates an excellent agreement between the observed patterns in the density field.

\subsubsection{Richtmyer--Meshkov instability}

We proceed further with the validation of our model and consider the simulation of the Richtmyer--Meshkov instability (RMI) \cite{RMI}. In the RMI problem, a shock wave collides with the interface of two fluids with different densities. In the following, we will compare our results with a numerical study of the RMI from the reference \cite{RMI_Ref}.

\begin{figure}
    \centering
   \includegraphics[width=0.4\textwidth]{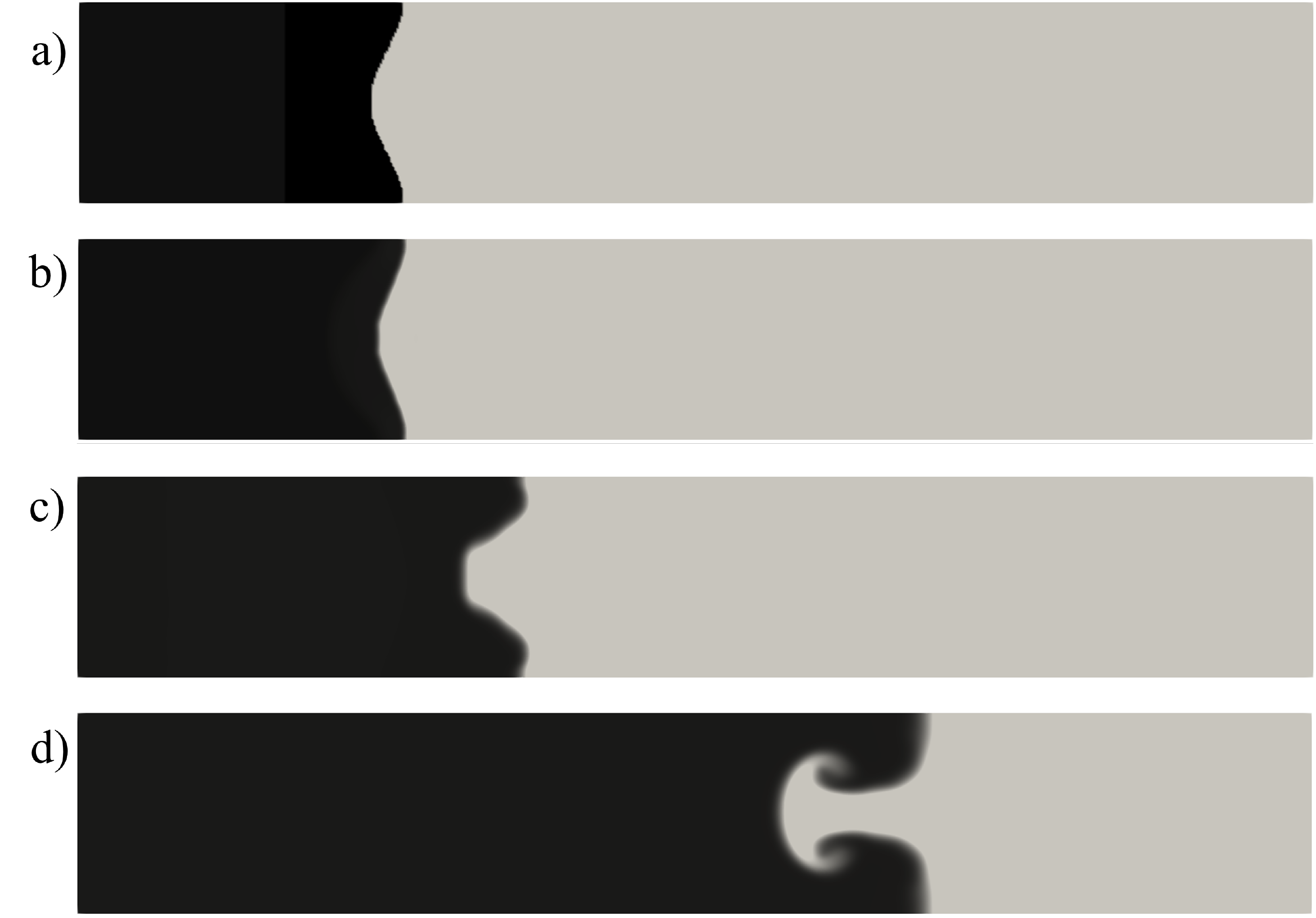}
    \includegraphics[width=0.4\textwidth]{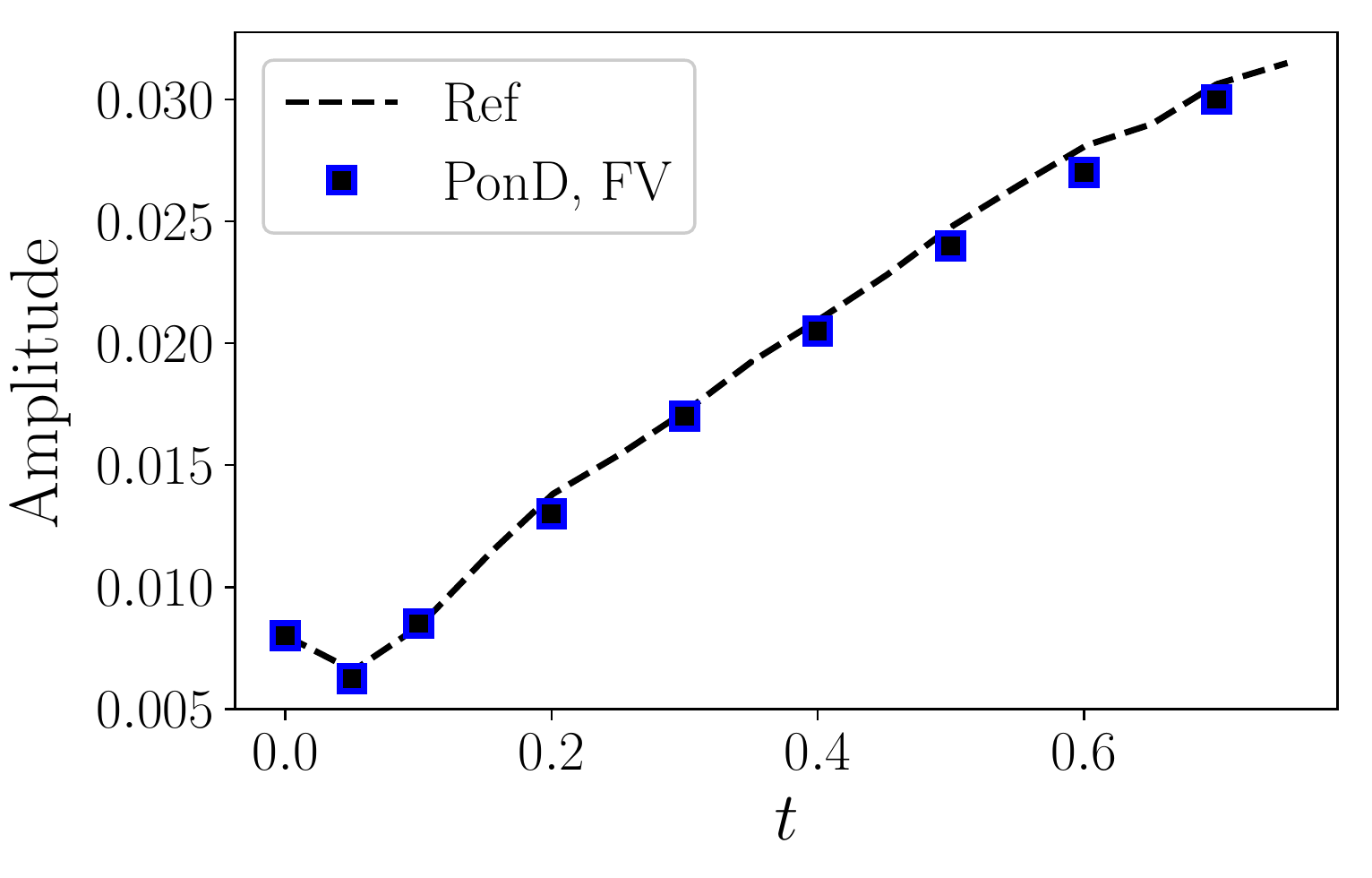}
    \caption{RMI problem, with the shock wave travelling from the light medium towards the heavy one. Top: Density field at times: $t=0,0.06,0.3,1.15$. Bottom: Amplitude growth of the instability and comparison with the reference \cite{RMI_Ref}. }
    \label{fig:RMI_type1}
\end{figure}

The first type of RMI problem to consider is the case of a shock wave, with Mach number of 1.2, travelling from the light medium to the heavy one. The computational domain $[0,0.6]\times [0,0.1]$ is initialized with the following conditions,

\begin{equation}
\begin{split}
    &(\rho,u_x,u_y,p) \\& =\begin{cases}
    (1.34161, 0.361538,0,0.151332),& 0\leq x <1/6, \\
    (1, 0, 0,1),& 1/6\leq x <1/4, \\
     (5.04, 0, 0,1),& \mathrm{otherwise}. \\
    \end{cases}
    \end{split}
\end{equation}
Additionally, a sinusoidal perturbation with amplitude of 0.008 is imposed on the interface. The simulation was performed with a grid $[600,100]$. The numerical setup is concluded with inflow BC on the left boundary, outflow BC on the right boundary and zero-gradient BC at the bottom and top boundaries. The temporal evolution of the instability is captured in Fig. \ref{fig:RMI_type1}, which shows the density field at different times. The simulation compares very well with the corresponding results from \cite{RMI_Ref}. A quantitative comparison is demonstrated in Fig. \ref{fig:RMI_type1}, which shows the change of the perturbation amplitude with time.

\begin{figure}
    \centering
   \includegraphics[width=0.4\textwidth]{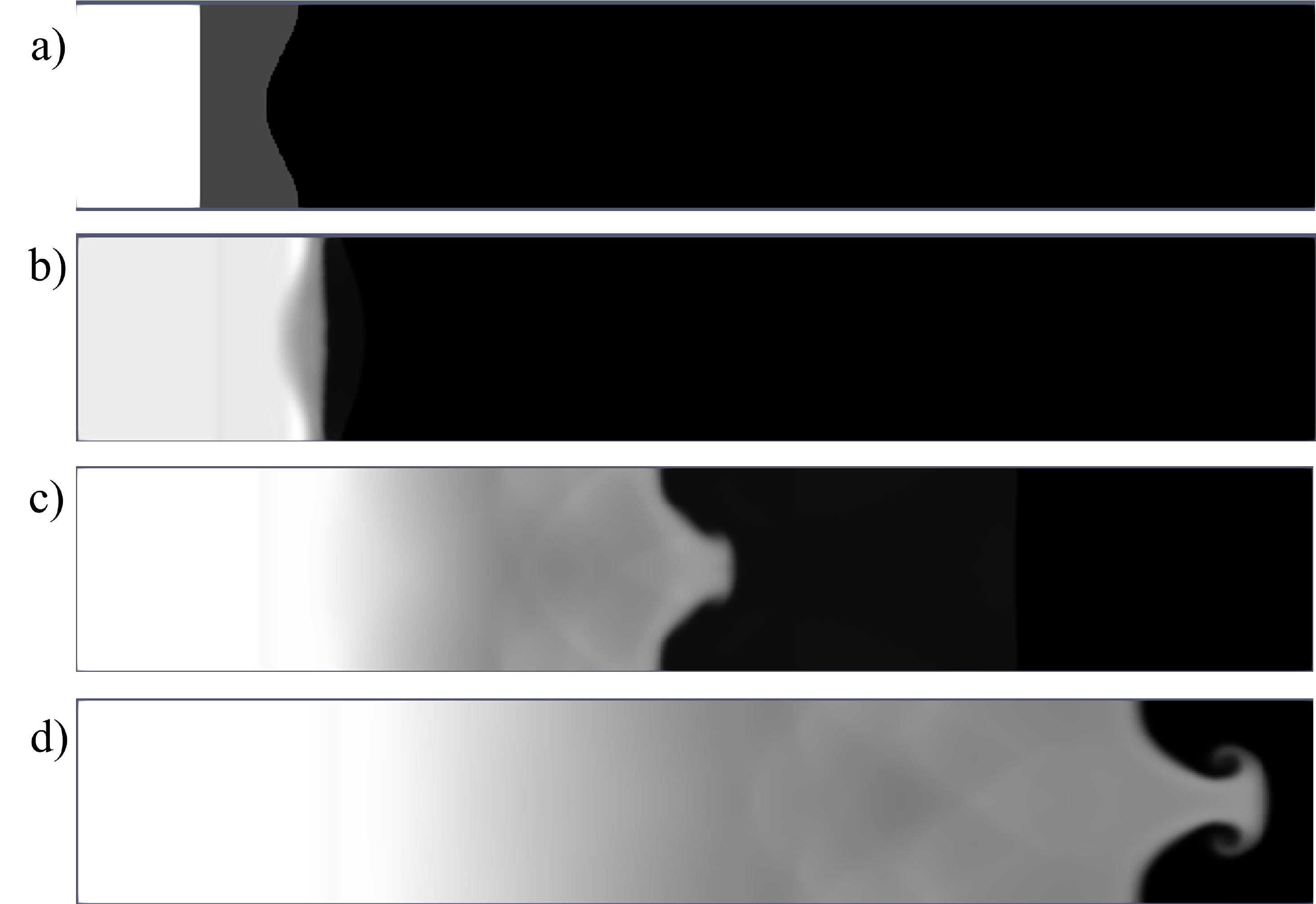}
    \includegraphics[width=0.4\textwidth]{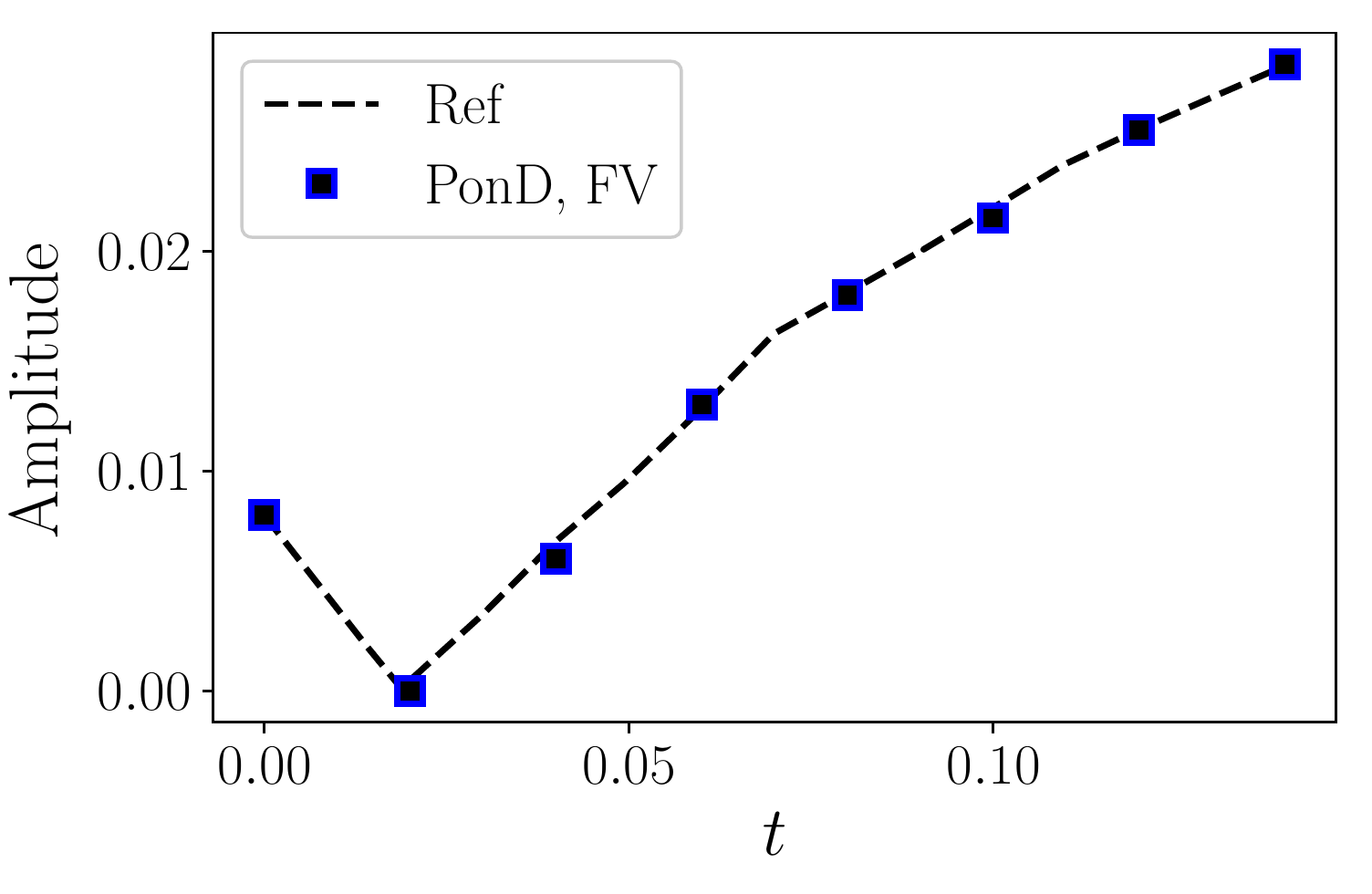}
    \caption{RMI problem, with the shock wave travelling from the heavy medium towards the light one. Top: Density field at times: $t=0,0.02,0.08,0.16$. Bottom: Amplitude growth of the instability and comparison with the reference \cite{RMI_Ref}. }
    \label{fig:RMI_type2}
\end{figure}

In the second type of the RMI problem, a shock wave with 2.5 Mach number, travels from the heavy medium to the light one. This configuration is achieved with the following initial conditions,
\begin{equation}
    (\rho,u_x,u_y,p)=\begin{cases}
    (3.33, 2.07,0,7.125),& 0\leq x <1/12, \\
    (1, 0, 0,1),& 1/12\leq x <1/6, \\
     (0.138, 0, 0,1),& \mathrm{otherwise}. \\
    \end{cases}
\end{equation}
Apart from the initial conditions, the numerical setup is the same as in the previous RMI simulation. The results, shown in Fig. \ref{fig:RMI_type2}, are in very good agreement with the reference \cite{RMI_Ref}.

\subsubsection{Double Mach reflection}

The double Mach reflection (DMR) problem is an important benchmark for compressible solvers, which has been studied extensively experimentally, theoretically and numerically \cite{DoubleMachRef_Exp,DoubleMachRef_Theory,WoodwardCollela_DoubleMachReflection}. In this setup, a Mach 10 shock wave collides with a reflecting wall, which is inclined $30^{\circ}$ counter-clockwise with respect to the shock propagation direction. The computational domain is a $[0,4]\times [0,1]$ rectangle discretized with a resolution $[1000,250]$.

Following the conventional configuration \cite{DoubleMachRef_Conf}, the simulation is initialized with a shock inclined $60^{\circ}$ to the horizontal, intersecting the bottom boundary at $x=1/6$. The undisturbed state of the gas is $(\rho,u_x,u_y,p)=(1.4, 0, 0, 1)$ and the post-shock state $(\rho,u_x,u_y,p)=(8, 4.125\sqrt{3}, -4.125, 116.5)$. At the bottom boundary, the fixed post-shock conditions are imposed along $x \in [0,1/6]$ and reflecting BCs along $x \in [1/6,4]$. At the left boundary, the fixed post-shock conditions are also imposed and on the right boundary zero gradient BCs. At the top boundary, time-dependent BC are specified, which track the motion of the initial Mach $10$ shock wave \cite{WoodwardCollela_DoubleMachReflection}.

\begin{figure*}
    \centering
   \includegraphics[width=0.8\textwidth]{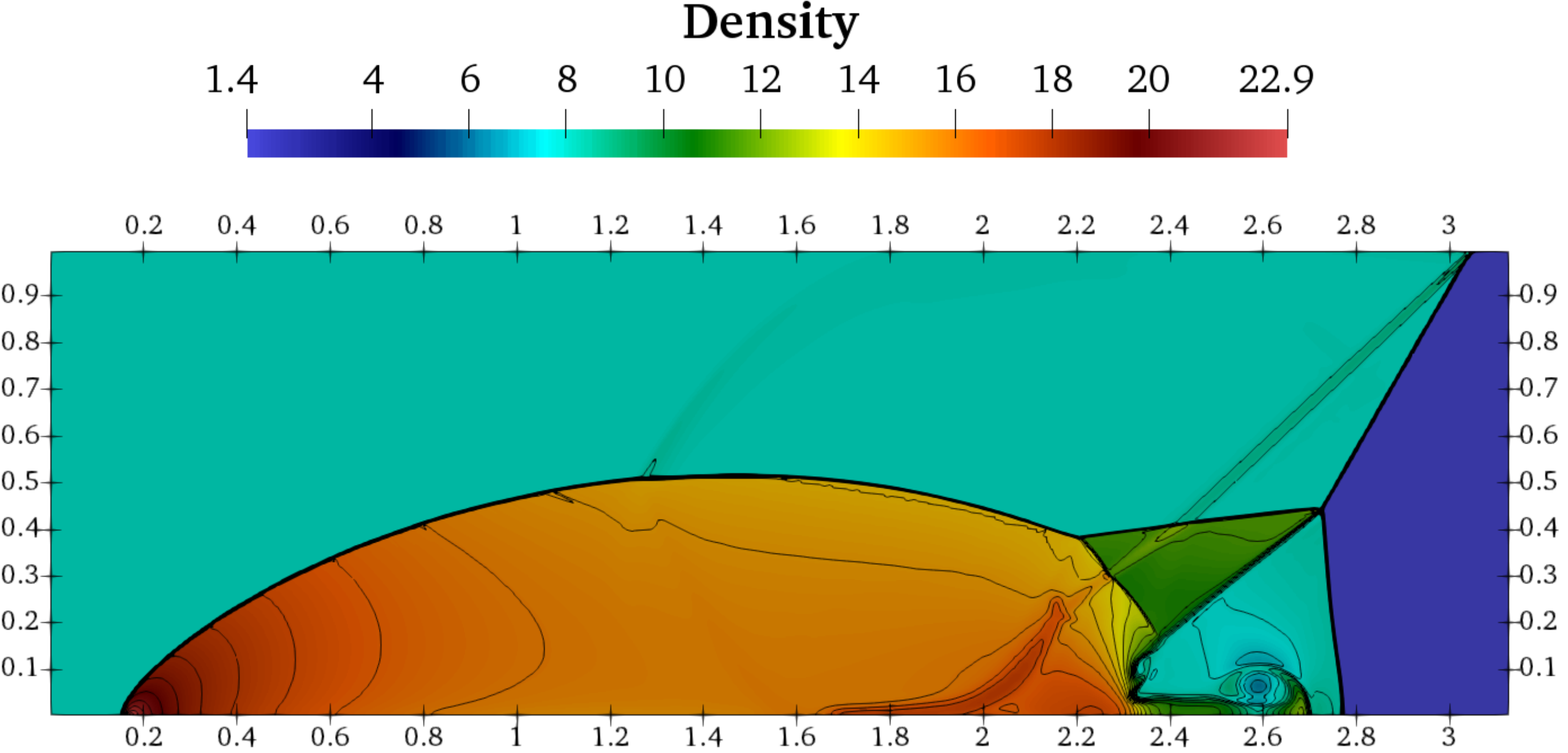}
 \includegraphics[width=0.8\textwidth]{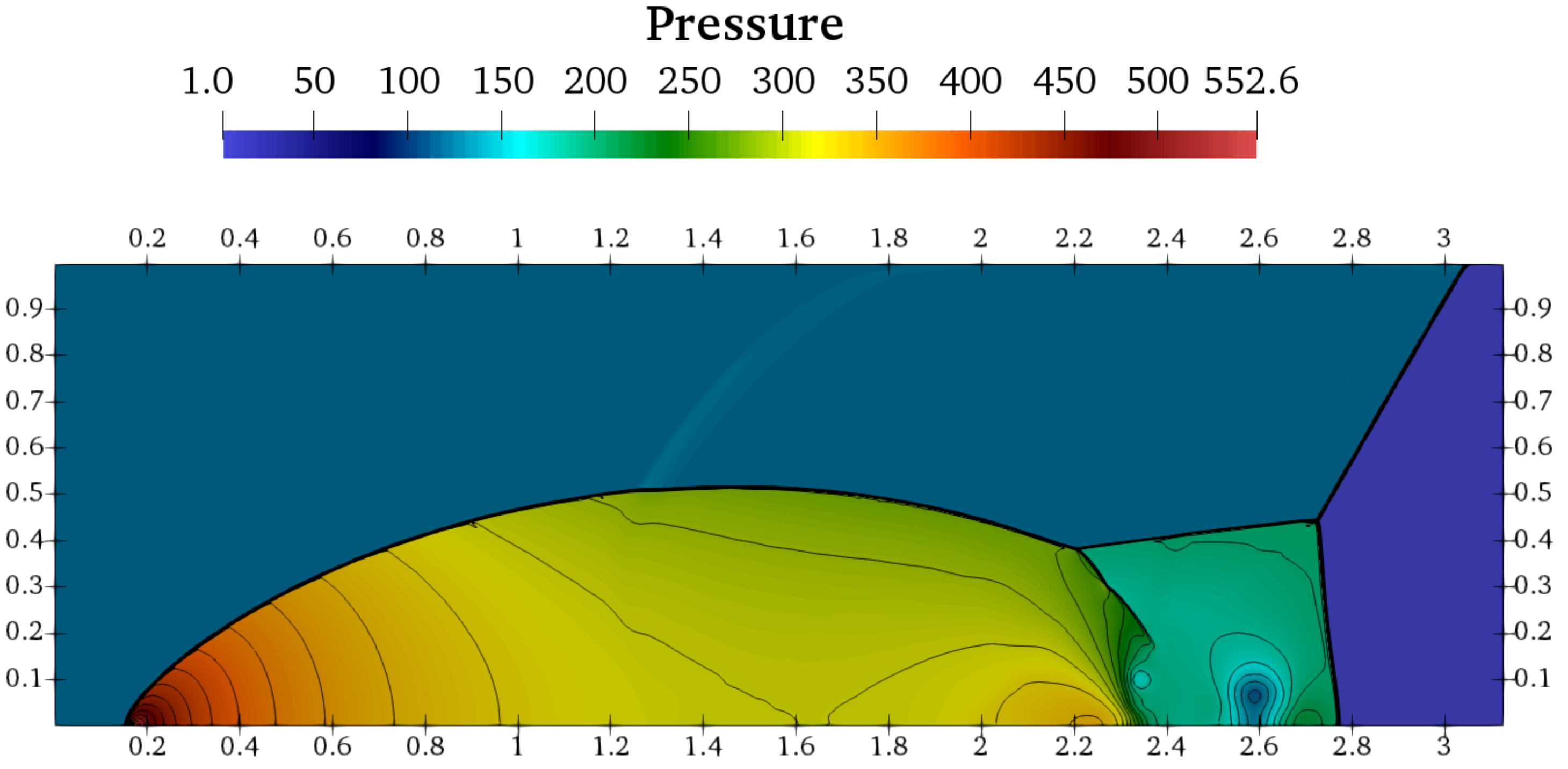}
    \caption{The double Mach reflection problem. Top: density field. Bottom: pressure field.}
    \label{fig:DoubleMachReflection}
\end{figure*}

The results for the density and pressure fields at $t=0.2$ are shown in Fig. \ref{fig:DoubleMachReflection}, with the flow characteristics being in very good agreement with corresponding results from the literature. Following the impact of the shock wave on the reflecting wall, a self-similar structure is formed and growing along the propagation of the shock. The key features of the flow are distinguished in the results, including the two Mach stems, two triple points, a prime slip line and a fainted secondary slip line as well as the jet formation near the wall. A comparison of the density and pressure fields with references from the literature \cite{WoodwardCollela_DoubleMachReflection,DoubleMachRef_Conf} demonstrate very good match in terms of the feature locations and the magnitude of the hydrodynamic fields.

 \subsubsection{Astrophysical jet}

\begin{figure*}
    \centering
   \includegraphics[width=0.32\textwidth]{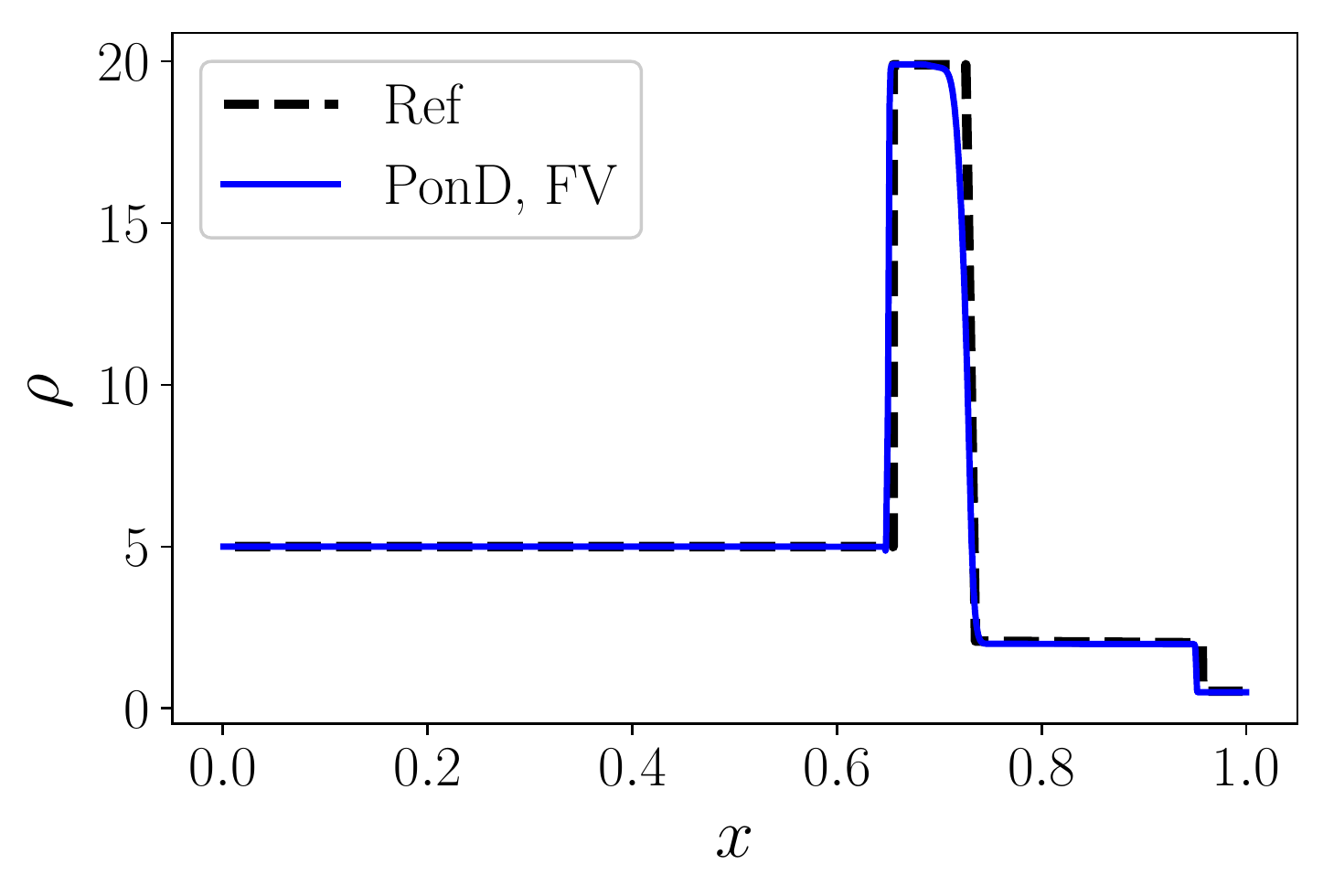}
    \includegraphics[width=0.32\textwidth]{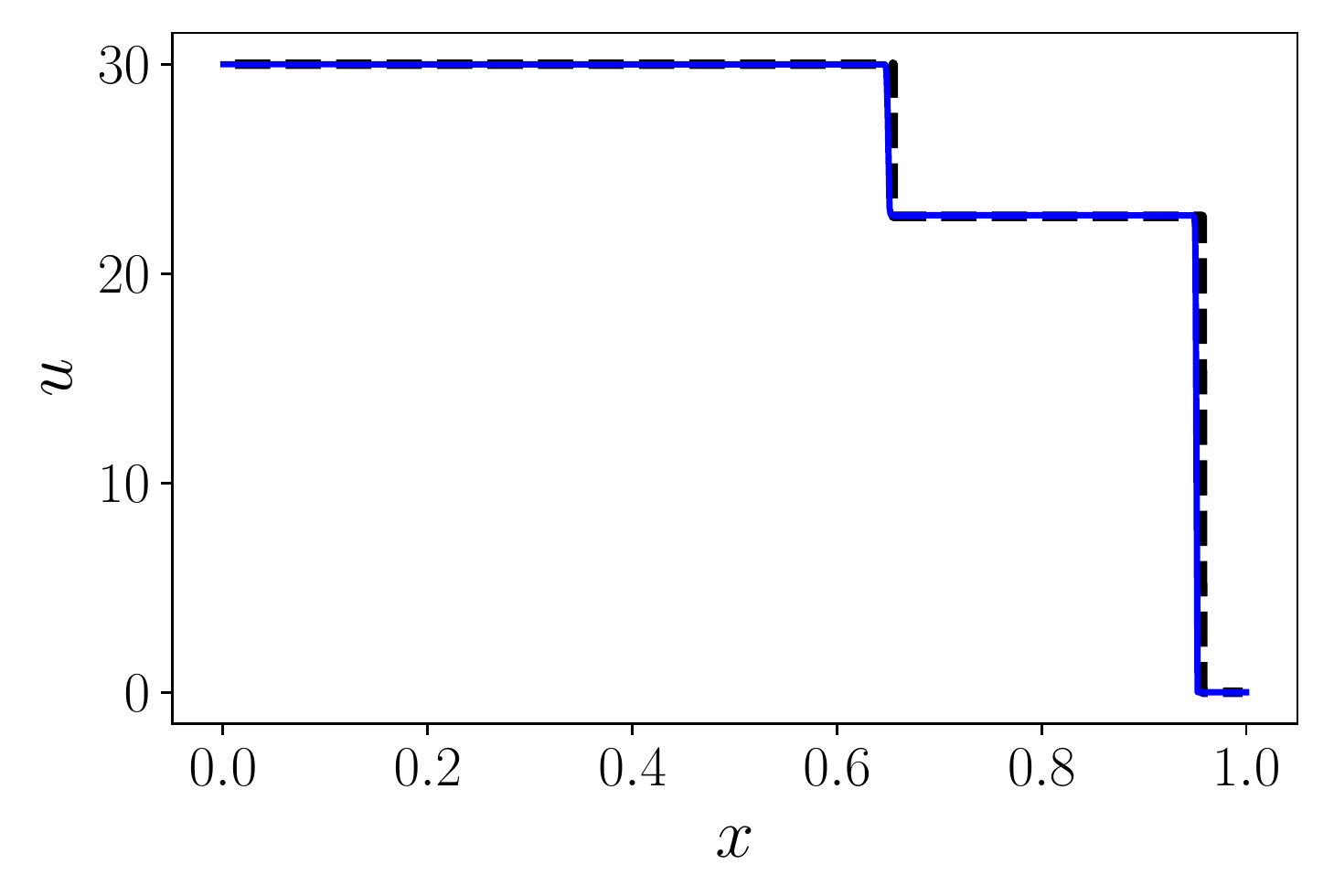}
     \includegraphics[width=0.32\textwidth]{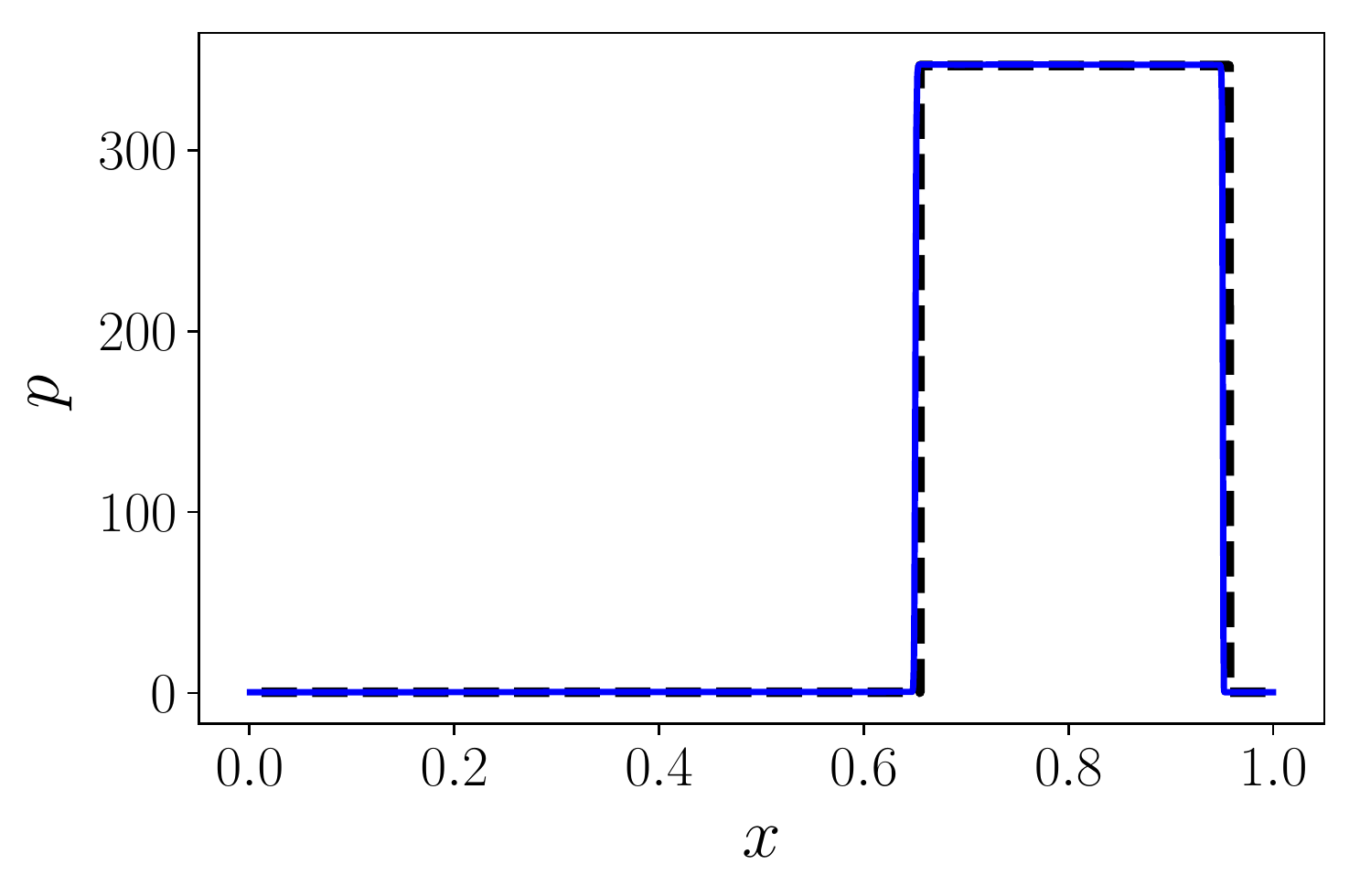}
    \caption{Density (left), velocity (middle) and pressure (right) profiles for the 1D Riemann "jet", at $t=0.06$. Solid line: finite-volume (FV) scheme. Dashed line: Reference from exact Riemann solver.}
    \label{fig:1D Jet}
\end{figure*}

\begin{figure*}
     \centering
   \includegraphics[width=0.48\textwidth]{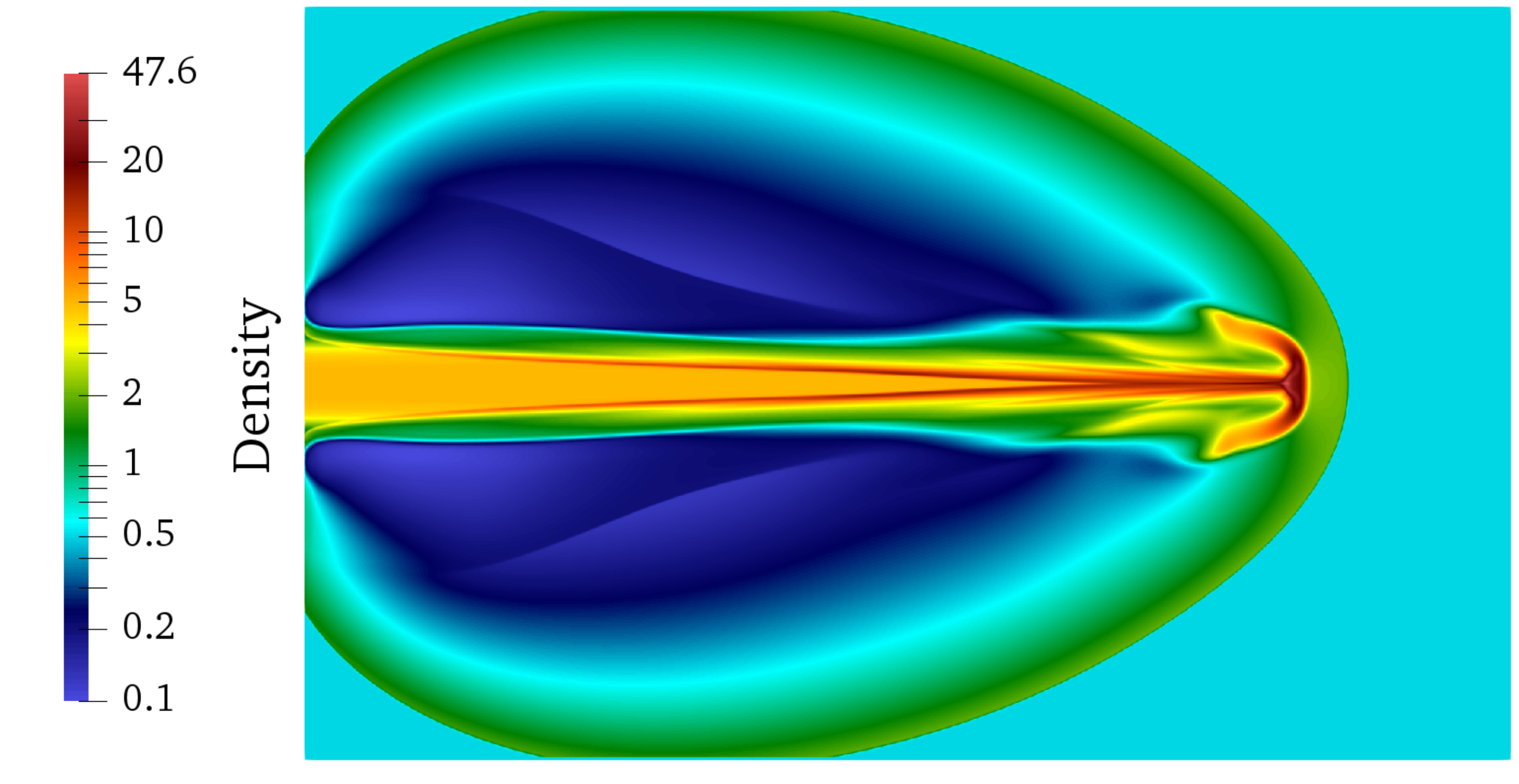}
     \includegraphics[width=0.48\textwidth]{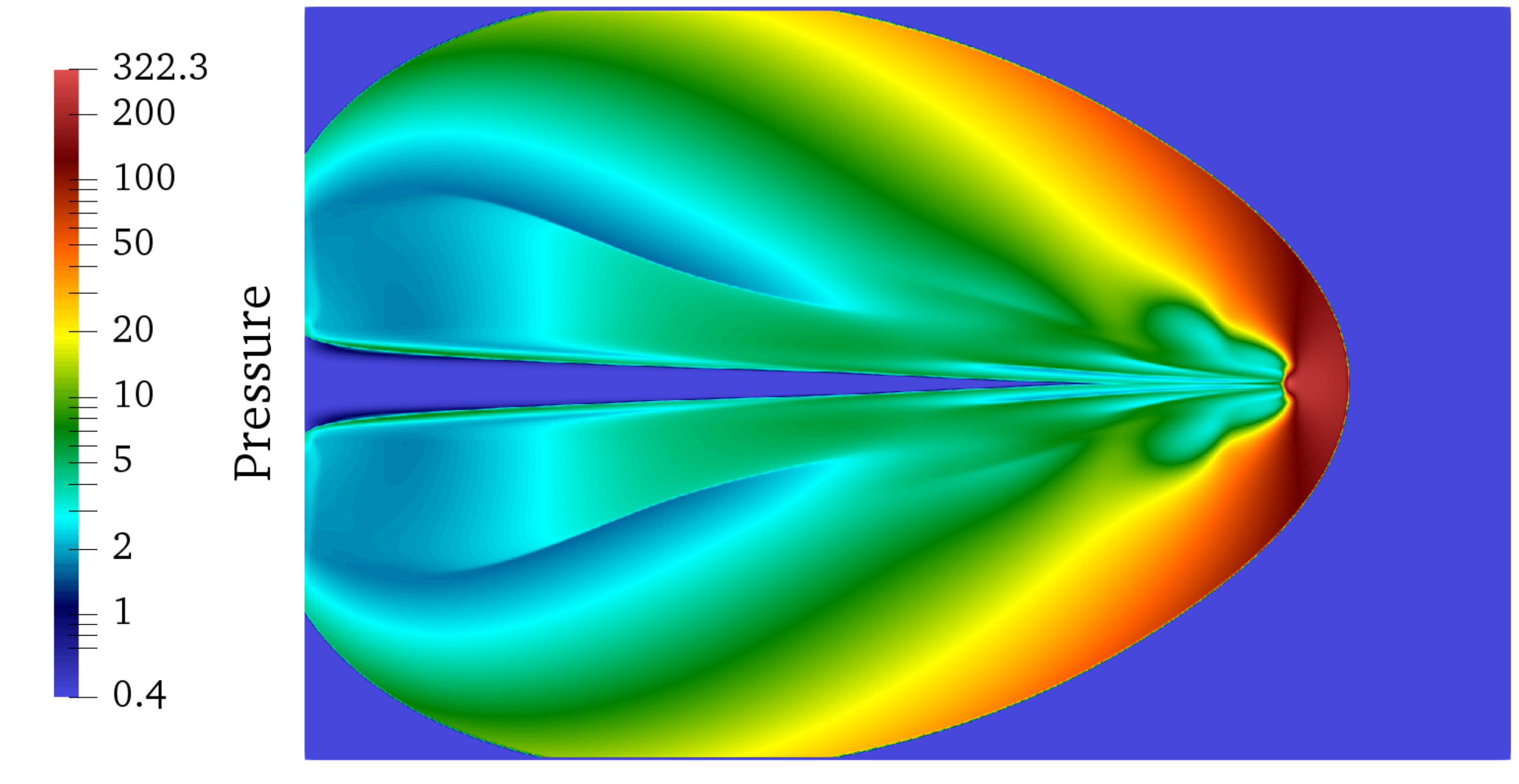}
      \includegraphics[width=0.48\textwidth]{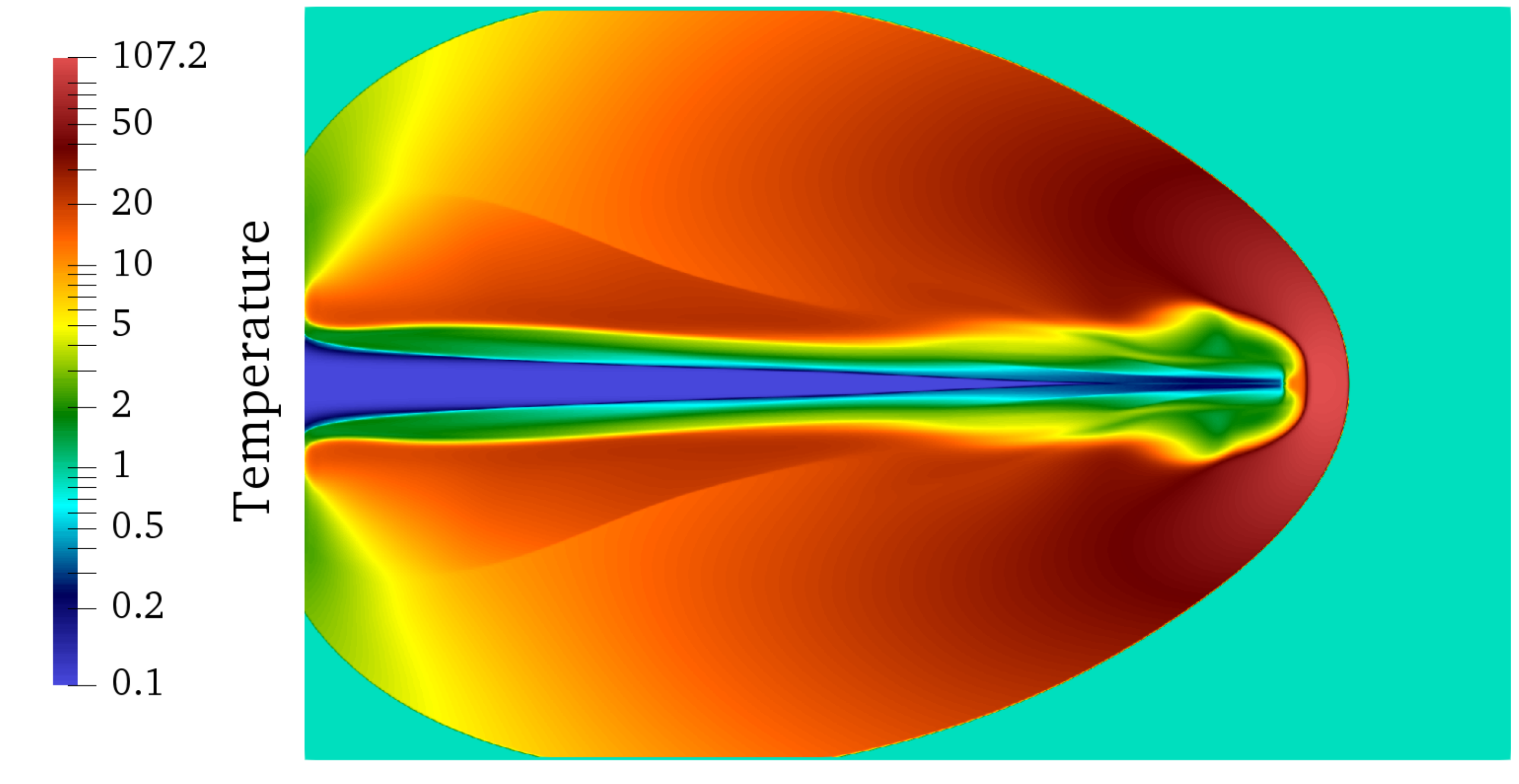}
       \includegraphics[width=0.48\textwidth]{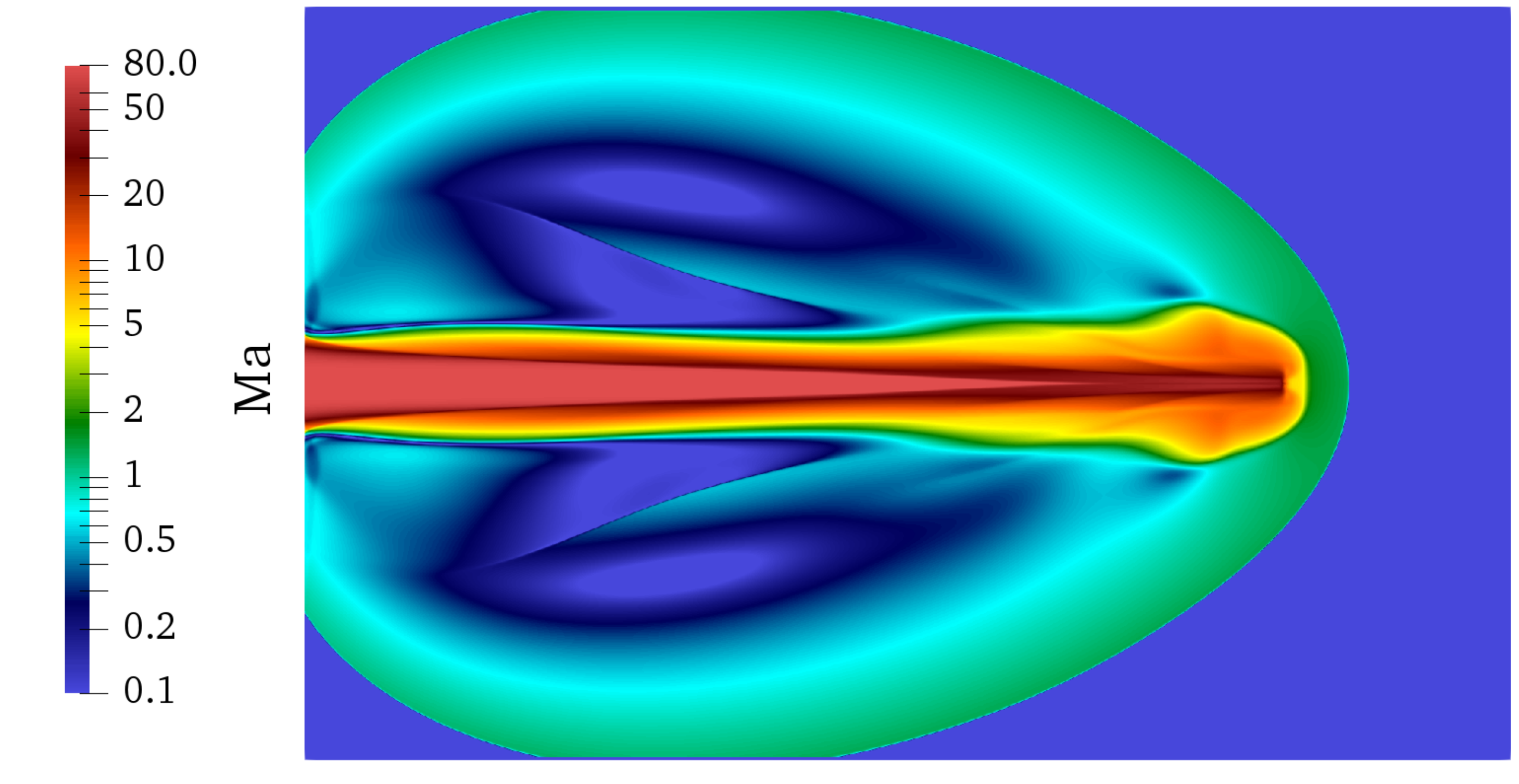}      
     \caption{Astrophysical jet problem. The following fields are plotted: density (top left), pressure (top right), temperature (bottom left)  and Mach number (bottom right).}
     \label{fig:Jet}
 \end{figure*}

As a final test case, we consider an astrophysical jet of Mach 80, without radiative cooling \cite{ZhangShu2010}. This case is an example of actual gas flows revealed from images of the Hubble Space Telescope and therefore is of high scientific interest. Following \cite{1DJetPaper}, we first present a 1D "jet" Riemann problem in a domain $[0,2.0]$, with the following initial conditions,
\begin{equation}
    (\rho,u,p)=\begin{cases}
    (5, 30, 0.4127),& 0 \leq x<0.1, \\
    (0.5, 0, 0.4127),& 0.1 \leq x<2. \\
    \end{cases}
\end{equation}
The results of the simulation with a resolution of 1500 points and $\gamma=5/3$ are shown in Fig. \ref{fig:1D Jet}. We continue with the 2D case, according to the configuration in \cite{ZhangShu2010}. The computational domain $[0,2]\times [-0.5,0.5]$ is initialized with the following conditions,
 \begin{equation}
 \begin{split}
    & (\rho,u_x,u_y,p) \\& =\begin{cases}
     (5, 30, 0, 0.4127),& \text{if } x=0, \ -0.05\leq y \leq 0.05, \\
     (0.5, 0, 0, 0.4127),& \text{ otherwise}. \\
     \end{cases}
      \end{split}
 \end{equation}
Outflow BCs are used around the domain, except the left boundary, where the prescribed fixed conditions are imposed. The simulation was performed with resolution $[1000,500]$. The density, pressure, temperature and Mach number are shown in Fig. \ref{fig:Jet}, where the bow shock propagating into the surrounding medium is captured.


\section{Conclusion}
\label{Conclusion}

In this work, we have presented a PonD model with a revised reference frame transformation using Grad's projection to enhance stability and accuracy. The resulting scheme was discretized using both a semi-Lagrangian approach as well as a finite volume realization. For validation, we have selected a number of challenging 1D and 2D test cases to probe accuracy and robustness for flows including very strong pressure and temperature discontinuities, large Mach numbers and near-vacuum regions. The results show that the proposed kinetic scheme, which is tightly connected to the classical LBM, can indeed capture the highly complex and nonlinear dynamics of high-speed compressible flows with strong discontinuities.

With these encouraging results, a number of possible directions for future work arise. For instance, using an adaptively refined velocity space, in the spirit of \cite{Multiscale2021}, will not only increase
efficiency but also extend the scheme to non-equilibrium flows.
Furthermore, performance in three dimensions and for flows involving complex geometries shall be assessed in future work.

\begin{acknowledgments}
This work was supported by European Research Council (ERC) Advanced Grant  834763-PonD. 
Computational resources at the Swiss National  Super  Computing  Center  CSCS  were  provided  under the grant  s1066.
\end{acknowledgments}

\bibliography{bibliogr}

\providecommand{\noopsort}[1]{}\providecommand{\singleletter}[1]{#1}%
\begin{thebibliography}{76}%
\makeatletter
\providecommand \@ifxundefined [1]{%
 \@ifx{#1\undefined}
}%
\providecommand \@ifnum [1]{%
 \ifnum #1\expandafter \@firstoftwo
 \else \expandafter \@secondoftwo
 \fi
}%
\providecommand \@ifx [1]{%
 \ifx #1\expandafter \@firstoftwo
 \else \expandafter \@secondoftwo
 \fi
}%
\providecommand \natexlab [1]{#1}%
\providecommand \enquote  [1]{``#1''}%
\providecommand \bibnamefont  [1]{#1}%
\providecommand \bibfnamefont [1]{#1}%
\providecommand \citenamefont [1]{#1}%
\providecommand \href@noop [0]{\@secondoftwo}%
\providecommand \href [0]{\begingroup \@sanitize@url \@href}%
\providecommand \@href[1]{\@@startlink{#1}\@@href}%
\providecommand \@@href[1]{\endgroup#1\@@endlink}%
\providecommand \@sanitize@url [0]{\catcode `\\12\catcode `\$12\catcode
  `\&12\catcode `\#12\catcode `\^12\catcode `\_12\catcode `\%12\relax}%
\providecommand \@@startlink[1]{}%
\providecommand \@@endlink[0]{}%
\providecommand \url  [0]{\begingroup\@sanitize@url \@url }%
\providecommand \@url [1]{\endgroup\@href {#1}{\urlprefix }}%
\providecommand \urlprefix  [0]{URL }%
\providecommand \Eprint [0]{\href }%
\providecommand \doibase [0]{https://doi.org/}%
\providecommand \selectlanguage [0]{\@gobble}%
\providecommand \bibinfo  [0]{\@secondoftwo}%
\providecommand \bibfield  [0]{\@secondoftwo}%
\providecommand \translation [1]{[#1]}%
\providecommand \BibitemOpen [0]{}%
\providecommand \bibitemStop [0]{}%
\providecommand \bibitemNoStop [0]{.\EOS\space}%
\providecommand \EOS [0]{\spacefactor3000\relax}%
\providecommand \BibitemShut  [1]{\csname bibitem#1\endcsname}%
\let\auto@bib@innerbib\@empty
\bibitem [{\citenamefont {Higuera}\ \emph {et~al.}(1989)\citenamefont
  {Higuera}, \citenamefont {Succi},\ and\ \citenamefont
  {Benzi}}]{Higuera_1989_SucciBenzi}%
  \BibitemOpen
  \bibfield  {author} {\bibinfo {author} {\bibfnamefont {F.~J.}\ \bibnamefont
  {Higuera}}, \bibinfo {author} {\bibfnamefont {S.}~\bibnamefont {Succi}},\
  and\ \bibinfo {author} {\bibfnamefont {R.}~\bibnamefont {Benzi}},\ }\bibfield
   {title} {\bibinfo {title} {Lattice gas dynamics with enhanced collisions},\
  }\href {https://doi.org/10.1209/0295-5075/9/4/008} {\bibfield  {journal}
  {\bibinfo  {journal} {Europhysics Letters}\ }\textbf {\bibinfo {volume}
  {9}},\ \bibinfo {pages} {345} (\bibinfo {year} {1989})}\BibitemShut {NoStop}%
\bibitem [{\citenamefont {Higuera}\ and\ \citenamefont
  {Jim{\'{e}}nez}(1989)}]{Higuera_1989}%
  \BibitemOpen
  \bibfield  {author} {\bibinfo {author} {\bibfnamefont {F.~J.}\ \bibnamefont
  {Higuera}}\ and\ \bibinfo {author} {\bibfnamefont {J.}~\bibnamefont
  {Jim{\'{e}}nez}},\ }\bibfield  {title} {\bibinfo {title} {Boltzmann approach
  to lattice gas simulations},\ }\href
  {https://doi.org/10.1209/0295-5075/9/7/009} {\bibfield  {journal} {\bibinfo
  {journal} {Europhysics Letters}\ }\textbf {\bibinfo {volume} {9}},\ \bibinfo
  {pages} {663} (\bibinfo {year} {1989})}\BibitemShut {NoStop}%
\bibitem [{\citenamefont {Dorschner}\ \emph {et~al.}(2017)\citenamefont
  {Dorschner}, \citenamefont {Chikatamarla},\ and\ \citenamefont
  {Karlin}}]{Dorschner2017JFM}%
  \BibitemOpen
  \bibfield  {author} {\bibinfo {author} {\bibfnamefont {B.}~\bibnamefont
  {Dorschner}}, \bibinfo {author} {\bibfnamefont {S.~S.}\ \bibnamefont
  {Chikatamarla}},\ and\ \bibinfo {author} {\bibfnamefont {I.~V.}\ \bibnamefont
  {Karlin}},\ }\bibfield  {title} {\bibinfo {title} {Transitional flows with
  the entropic lattice {B}oltzmann method},\ }\href@noop {} {\bibfield
  {journal} {\bibinfo  {journal} {J. Fluid Mech.}\ }\textbf {\bibinfo {volume}
  {824}},\ \bibinfo {pages} {388} (\bibinfo {year} {2017})}\BibitemShut
  {NoStop}%
\bibitem [{\citenamefont {Dorschner}\ \emph {et~al.}(2016)\citenamefont
  {Dorschner}, \citenamefont {B{\"o}sch}, \citenamefont {Chikatamarla},
  \citenamefont {Boulouchos},\ and\ \citenamefont
  {Karlin}}]{dorschner2016entropic}%
  \BibitemOpen
  \bibfield  {author} {\bibinfo {author} {\bibfnamefont {B.}~\bibnamefont
  {Dorschner}}, \bibinfo {author} {\bibfnamefont {F.}~\bibnamefont
  {B{\"o}sch}}, \bibinfo {author} {\bibfnamefont {S.~S.}\ \bibnamefont
  {Chikatamarla}}, \bibinfo {author} {\bibfnamefont {K.}~\bibnamefont
  {Boulouchos}},\ and\ \bibinfo {author} {\bibfnamefont {I.~V.}\ \bibnamefont
  {Karlin}},\ }\bibfield  {title} {\bibinfo {title} {Entropic multi-relaxation
  time lattice {B}oltzmann model for complex flows},\ }\href@noop {} {\bibfield
   {journal} {\bibinfo  {journal} {J. Fluid Mech.}\ }\textbf {\bibinfo {volume}
  {801}},\ \bibinfo {pages} {623} (\bibinfo {year} {2016})}\BibitemShut
  {NoStop}%
\bibitem [{\citenamefont {He}\ \emph {et~al.}(1998{\natexlab{a}})\citenamefont
  {He}, \citenamefont {Chen},\ and\ \citenamefont {Doolen}}]{he1998}%
  \BibitemOpen
  \bibfield  {author} {\bibinfo {author} {\bibfnamefont {X.}~\bibnamefont
  {He}}, \bibinfo {author} {\bibfnamefont {S.}~\bibnamefont {Chen}},\ and\
  \bibinfo {author} {\bibfnamefont {G.~D.}\ \bibnamefont {Doolen}},\ }\bibfield
   {title} {\bibinfo {title} {A novel thermal model for the lattice {B}oltzmann
  method in incompressible limit},\ }\href
  {https://doi.org/https://doi.org/10.1006/jcph.1998.6057} {\bibfield
  {journal} {\bibinfo  {journal} {J. Comp. Phys.}\ }\textbf {\bibinfo {volume}
  {146}},\ \bibinfo {pages} {282} (\bibinfo {year}
  {1998}{\natexlab{a}})}\BibitemShut {NoStop}%
\bibitem [{\citenamefont {Guo}\ \emph {et~al.}(2007)\citenamefont {Guo},
  \citenamefont {Zheng}, \citenamefont {Shi},\ and\ \citenamefont
  {Zhao}}]{guo2007twopop}%
  \BibitemOpen
  \bibfield  {author} {\bibinfo {author} {\bibfnamefont {Z.}~\bibnamefont
  {Guo}}, \bibinfo {author} {\bibfnamefont {C.}~\bibnamefont {Zheng}}, \bibinfo
  {author} {\bibfnamefont {B.}~\bibnamefont {Shi}},\ and\ \bibinfo {author}
  {\bibfnamefont {T.~S.}\ \bibnamefont {Zhao}},\ }\bibfield  {title} {\bibinfo
  {title} {Thermal lattice {B}oltzmann equation for low {M}ach number flows:
  decoupling model},\ }\href@noop {} {\bibfield  {journal} {\bibinfo  {journal}
  {Physical Review E}\ }\textbf {\bibinfo {volume} {75}},\ \bibinfo {pages}
  {036704} (\bibinfo {year} {2007})}\BibitemShut {NoStop}%
\bibitem [{\citenamefont {Karlin}\ \emph {et~al.}(2013)\citenamefont {Karlin},
  \citenamefont {Sichau},\ and\ \citenamefont
  {Chikatamarla}}]{karlinConsistent}%
  \BibitemOpen
  \bibfield  {author} {\bibinfo {author} {\bibfnamefont {I.~V.}\ \bibnamefont
  {Karlin}}, \bibinfo {author} {\bibfnamefont {D.}~\bibnamefont {Sichau}},\
  and\ \bibinfo {author} {\bibfnamefont {S.~S.}\ \bibnamefont {Chikatamarla}},\
  }\bibfield  {title} {\bibinfo {title} {Consistent two-population lattice
  {B}oltzmann model for thermal flows},\ }\href
  {https://doi.org/10.1103/PhysRevE.88.063310} {\bibfield  {journal} {\bibinfo
  {journal} {Phys. Rev. E}\ }\textbf {\bibinfo {volume} {88}},\ \bibinfo
  {pages} {063310} (\bibinfo {year} {2013})}\BibitemShut {NoStop}%
\bibitem [{\citenamefont {Mazloomi}\ \emph {et~al.}(2015)\citenamefont
  {Mazloomi}, \citenamefont {Chikatamarla},\ and\ \citenamefont
  {Karlin}}]{Mazloomi2015prl}%
  \BibitemOpen
  \bibfield  {author} {\bibinfo {author} {\bibfnamefont {A.~M.}\ \bibnamefont
  {Mazloomi}}, \bibinfo {author} {\bibfnamefont {S.~S.}\ \bibnamefont
  {Chikatamarla}},\ and\ \bibinfo {author} {\bibfnamefont {I.~V.}\ \bibnamefont
  {Karlin}},\ }\bibfield  {title} {\bibinfo {title} {Entropic lattice
  {B}oltzmann method for multiphase flows},\ }\href@noop {} {\bibfield
  {journal} {\bibinfo  {journal} {Phys. Rev. Lett.}\ }\textbf {\bibinfo
  {volume} {114}},\ \bibinfo {pages} {174502} (\bibinfo {year}
  {2015})}\BibitemShut {NoStop}%
\bibitem [{\citenamefont {Mazloomi}\ \emph {et~al.}(2017)\citenamefont
  {Mazloomi}, \citenamefont {Chikatamarla},\ and\ \citenamefont
  {Karlin}}]{Mazloomi2017JFM}%
  \BibitemOpen
  \bibfield  {author} {\bibinfo {author} {\bibfnamefont {A.~M.}\ \bibnamefont
  {Mazloomi}}, \bibinfo {author} {\bibfnamefont {S.~S.}\ \bibnamefont
  {Chikatamarla}},\ and\ \bibinfo {author} {\bibfnamefont {I.~V.}\ \bibnamefont
  {Karlin}},\ }\bibfield  {title} {\bibinfo {title} {Drops bouncing off
  macro-textured superhydrophobic surfaces},\ }\href@noop {} {\bibfield
  {journal} {\bibinfo  {journal} {J. Fluid Mech.}\ }\textbf {\bibinfo {volume}
  {824}},\ \bibinfo {pages} {866} (\bibinfo {year} {2017})}\BibitemShut
  {NoStop}%
\bibitem [{\citenamefont {W\"ohrwag}\ \emph {et~al.}(2018)\citenamefont
  {W\"ohrwag}, \citenamefont {Semprebon}, \citenamefont {Mazloomi~Moqaddam},
  \citenamefont {Karlin},\ and\ \citenamefont
  {Kusumaatmaja}}]{MultiPhase_Wohrwag}%
  \BibitemOpen
  \bibfield  {author} {\bibinfo {author} {\bibfnamefont {M.}~\bibnamefont
  {W\"ohrwag}}, \bibinfo {author} {\bibfnamefont {C.}~\bibnamefont
  {Semprebon}}, \bibinfo {author} {\bibfnamefont {A.}~\bibnamefont
  {Mazloomi~Moqaddam}}, \bibinfo {author} {\bibfnamefont {I.}~\bibnamefont
  {Karlin}},\ and\ \bibinfo {author} {\bibfnamefont {H.}~\bibnamefont
  {Kusumaatmaja}},\ }\bibfield  {title} {\bibinfo {title} {Ternary free-energy
  entropic lattice {B}oltzmann model with a high density ratio},\ }\href
  {https://doi.org/10.1103/PhysRevLett.120.234501} {\bibfield  {journal}
  {\bibinfo  {journal} {Phys. Rev. Lett.}\ }\textbf {\bibinfo {volume} {120}},\
  \bibinfo {pages} {234501} (\bibinfo {year} {2018})}\BibitemShut {NoStop}%
\bibitem [{\citenamefont {Sawant}\ \emph
  {et~al.}(2021{\natexlab{a}})\citenamefont {Sawant}, \citenamefont
  {Dorschner},\ and\ \citenamefont {Karlin}}]{NileshJFM1}%
  \BibitemOpen
  \bibfield  {author} {\bibinfo {author} {\bibfnamefont {N.}~\bibnamefont
  {Sawant}}, \bibinfo {author} {\bibfnamefont {B.}~\bibnamefont {Dorschner}},\
  and\ \bibinfo {author} {\bibfnamefont {I.~V.}\ \bibnamefont {Karlin}},\
  }\bibfield  {title} {\bibinfo {title} {Consistent lattice {B}oltzmann model
  for multicomponent mixtures},\ }\href {https://doi.org/10.1017/jfm.2020.853}
  {\bibfield  {journal} {\bibinfo  {journal} {Journal of Fluid Mechanics}\
  }\textbf {\bibinfo {volume} {909}},\ \bibinfo {pages} {A1} (\bibinfo {year}
  {2021}{\natexlab{a}})}\BibitemShut {NoStop}%
\bibitem [{\citenamefont {Sawant}\ \emph
  {et~al.}(2021{\natexlab{b}})\citenamefont {Sawant}, \citenamefont
  {Dorschner},\ and\ \citenamefont {Karlin}}]{NileshReactive}%
  \BibitemOpen
  \bibfield  {author} {\bibinfo {author} {\bibfnamefont {N.}~\bibnamefont
  {Sawant}}, \bibinfo {author} {\bibfnamefont {B.}~\bibnamefont {Dorschner}},\
  and\ \bibinfo {author} {\bibfnamefont {I.~V.}\ \bibnamefont {Karlin}},\
  }\bibfield  {title} {\bibinfo {title} {A lattice {B}oltzmann model for
  reactive mixtures},\ }\href {https://doi.org/10.1098/rsta.2020.0402}
  {\bibfield  {journal} {\bibinfo  {journal} {Philosophical Transactions of the
  Royal Society A: Mathematical, Physical and Engineering Sciences}\ }\textbf
  {\bibinfo {volume} {379}},\ \bibinfo {pages} {20200402} (\bibinfo {year}
  {2021}{\natexlab{b}})}\BibitemShut {NoStop}%
\bibitem [{\citenamefont {Shan}\ \emph {et~al.}(2006)\citenamefont {Shan},
  \citenamefont {Yuan},\ and\ \citenamefont {Chen}}]{shan2006kinetic}%
  \BibitemOpen
  \bibfield  {author} {\bibinfo {author} {\bibfnamefont {X.}~\bibnamefont
  {Shan}}, \bibinfo {author} {\bibfnamefont {X.-F.}\ \bibnamefont {Yuan}},\
  and\ \bibinfo {author} {\bibfnamefont {H.}~\bibnamefont {Chen}},\ }\bibfield
  {title} {\bibinfo {title} {Kinetic theory representation of hydrodynamics: a
  way beyond the {N}avier--{S}tokes equation},\ }\href@noop {} {\bibfield
  {journal} {\bibinfo  {journal} {J. Fluid Mech.}\ }\textbf {\bibinfo {volume}
  {550}},\ \bibinfo {pages} {413} (\bibinfo {year} {2006})}\BibitemShut
  {NoStop}%
\bibitem [{\citenamefont {Sharma}\ \emph {et~al.}(2020)\citenamefont {Sharma},
  \citenamefont {Straka},\ and\ \citenamefont {Tavares}}]{SHARMAReview}%
  \BibitemOpen
  \bibfield  {author} {\bibinfo {author} {\bibfnamefont {K.~V.}\ \bibnamefont
  {Sharma}}, \bibinfo {author} {\bibfnamefont {R.}~\bibnamefont {Straka}},\
  and\ \bibinfo {author} {\bibfnamefont {F.~W.}\ \bibnamefont {Tavares}},\
  }\bibfield  {title} {\bibinfo {title} {Current status of lattice {B}oltzmann
  methods applied to aerodynamic, aeroacoustic, and thermal flows},\ }\href
  {https://doi.org/https://doi.org/10.1016/j.paerosci.2020.100616} {\bibfield
  {journal} {\bibinfo  {journal} {Progress in Aerospace Sciences}\ }\textbf
  {\bibinfo {volume} {115}},\ \bibinfo {pages} {100616} (\bibinfo {year}
  {2020})}\BibitemShut {NoStop}%
\bibitem [{\citenamefont {Krueger}\ \emph {et~al.}(2016)\citenamefont
  {Krueger}, \citenamefont {Kusumaatmaja}, \citenamefont {Kuzmin},
  \citenamefont {Shardt}, \citenamefont {Silva},\ and\ \citenamefont
  {Viggen}}]{KrugerBook}%
  \BibitemOpen
  \bibfield  {author} {\bibinfo {author} {\bibfnamefont {T.}~\bibnamefont
  {Krueger}}, \bibinfo {author} {\bibfnamefont {H.}~\bibnamefont
  {Kusumaatmaja}}, \bibinfo {author} {\bibfnamefont {A.}~\bibnamefont
  {Kuzmin}}, \bibinfo {author} {\bibfnamefont {O.}~\bibnamefont {Shardt}},
  \bibinfo {author} {\bibfnamefont {G.}~\bibnamefont {Silva}},\ and\ \bibinfo
  {author} {\bibfnamefont {E.}~\bibnamefont {Viggen}},\ }\href@noop {} {\emph
  {\bibinfo {title} {The Lattice {B}oltzmann Method: Principles and
  Practice}}},\ Graduate Texts in Physics\ (\bibinfo  {publisher} {Springer},\
  \bibinfo {year} {2016})\BibitemShut {NoStop}%
\bibitem [{\citenamefont {Succi}(2018)}]{SucciBook}%
  \BibitemOpen
  \bibfield  {author} {\bibinfo {author} {\bibfnamefont {S.}~\bibnamefont
  {Succi}},\ }\href@noop {} {\emph {\bibinfo {title} {The Lattice {B}oltzmann
  Equation: For Complex States of Flowing Matter}}}\ (\bibinfo  {publisher}
  {Oxford University Press},\ \bibinfo {year} {2018})\BibitemShut {NoStop}%
\bibitem [{\citenamefont {Chikatamarla}\ and\ \citenamefont
  {Karlin}(2006)}]{Chikatamarla2006}%
  \BibitemOpen
  \bibfield  {author} {\bibinfo {author} {\bibfnamefont {S.~S.}\ \bibnamefont
  {Chikatamarla}}\ and\ \bibinfo {author} {\bibfnamefont {I.~V.}\ \bibnamefont
  {Karlin}},\ }\bibfield  {title} {\bibinfo {title} {Entropy and {G}alilean
  invariance of lattice {B}oltzmann theories},\ }\href
  {https://doi.org/10.1103/PhysRevLett.97.190601} {\bibfield  {journal}
  {\bibinfo  {journal} {Phys. Rev. Lett.}\ }\textbf {\bibinfo {volume} {97}},\
  \bibinfo {pages} {190601} (\bibinfo {year} {2006})}\BibitemShut {NoStop}%
\bibitem [{\citenamefont {Chikatamarla}\ and\ \citenamefont
  {Karlin}(2009)}]{Chikatamarla2009}%
  \BibitemOpen
  \bibfield  {author} {\bibinfo {author} {\bibfnamefont {S.~S.}\ \bibnamefont
  {Chikatamarla}}\ and\ \bibinfo {author} {\bibfnamefont {I.~V.}\ \bibnamefont
  {Karlin}},\ }\bibfield  {title} {\bibinfo {title} {Lattices for the lattice
  {B}oltzmann method},\ }\href {https://doi.org/10.1103/PhysRevE.79.046701}
  {\bibfield  {journal} {\bibinfo  {journal} {Phys. Rev. E}\ }\textbf {\bibinfo
  {volume} {79}},\ \bibinfo {pages} {046701} (\bibinfo {year}
  {2009})}\BibitemShut {NoStop}%
\bibitem [{\citenamefont {Alexander}\ \emph {et~al.}(1993)\citenamefont
  {Alexander}, \citenamefont {Chen},\ and\ \citenamefont
  {Sterling}}]{Alexander1993}%
  \BibitemOpen
  \bibfield  {author} {\bibinfo {author} {\bibfnamefont {F.~J.}\ \bibnamefont
  {Alexander}}, \bibinfo {author} {\bibfnamefont {S.}~\bibnamefont {Chen}},\
  and\ \bibinfo {author} {\bibfnamefont {J.~D.}\ \bibnamefont {Sterling}},\
  }\bibfield  {title} {\bibinfo {title} {{Lattice {B}oltzmann
  thermohydrodynamics}},\ }\href {https://doi.org/10.1103/PhysRevE.47.R2249}
  {\bibfield  {journal} {\bibinfo  {journal} {Phys. Rev. E}\ }\textbf {\bibinfo
  {volume} {47}},\ \bibinfo {pages} {R2249} (\bibinfo {year}
  {1993})}\BibitemShut {NoStop}%
\bibitem [{\citenamefont {Frapolli}\ \emph {et~al.}(2015)\citenamefont
  {Frapolli}, \citenamefont {Chikatamarla},\ and\ \citenamefont
  {Karlin}}]{Frappoli_2015}%
  \BibitemOpen
  \bibfield  {author} {\bibinfo {author} {\bibfnamefont {N.}~\bibnamefont
  {Frapolli}}, \bibinfo {author} {\bibfnamefont {S.~S.}\ \bibnamefont
  {Chikatamarla}},\ and\ \bibinfo {author} {\bibfnamefont {I.~V.}\ \bibnamefont
  {Karlin}},\ }\bibfield  {title} {\bibinfo {title} {Entropic lattice
  {B}oltzmann model for compressible flows},\ }\href
  {https://doi.org/10.1103/PhysRevE.92.061301} {\bibfield  {journal} {\bibinfo
  {journal} {Phys. Rev. E}\ }\textbf {\bibinfo {volume} {92}},\ \bibinfo
  {pages} {061301} (\bibinfo {year} {2015})}\BibitemShut {NoStop}%
\bibitem [{\citenamefont {Frapolli}\ \emph
  {et~al.}(2016{\natexlab{a}})\citenamefont {Frapolli}, \citenamefont
  {Chikatamarla},\ and\ \citenamefont {Karlin}}]{Frappoli_2016}%
  \BibitemOpen
  \bibfield  {author} {\bibinfo {author} {\bibfnamefont {N.}~\bibnamefont
  {Frapolli}}, \bibinfo {author} {\bibfnamefont {S.~S.}\ \bibnamefont
  {Chikatamarla}},\ and\ \bibinfo {author} {\bibfnamefont {I.~V.}\ \bibnamefont
  {Karlin}},\ }\bibfield  {title} {\bibinfo {title} {Entropic lattice
  {B}oltzmann model for gas dynamics: Theory, boundary conditions, and
  implementation},\ }\href {https://doi.org/10.1103/PhysRevE.93.063302}
  {\bibfield  {journal} {\bibinfo  {journal} {Phys. Rev. E}\ }\textbf {\bibinfo
  {volume} {93}},\ \bibinfo {pages} {063302} (\bibinfo {year}
  {2016}{\natexlab{a}})}\BibitemShut {NoStop}%
\bibitem [{\citenamefont {Prasianakis}\ and\ \citenamefont
  {Karlin}(2007)}]{Prasianakis2007}%
  \BibitemOpen
  \bibfield  {author} {\bibinfo {author} {\bibfnamefont {N.~I.}\ \bibnamefont
  {Prasianakis}}\ and\ \bibinfo {author} {\bibfnamefont {I.~V.}\ \bibnamefont
  {Karlin}},\ }\bibfield  {title} {\bibinfo {title} {Lattice {B}oltzmann method
  for thermal flow simulation on standard lattices},\ }\href
  {https://doi.org/10.1103/PhysRevE.76.016702} {\bibfield  {journal} {\bibinfo
  {journal} {Phys. Rev. E}\ }\textbf {\bibinfo {volume} {76}},\ \bibinfo
  {pages} {016702} (\bibinfo {year} {2007})}\BibitemShut {NoStop}%
\bibitem [{\citenamefont {Saadat}\ \emph {et~al.}(2019)\citenamefont {Saadat},
  \citenamefont {B\"osch},\ and\ \citenamefont {Karlin}}]{Saadat2019}%
  \BibitemOpen
  \bibfield  {author} {\bibinfo {author} {\bibfnamefont {M.~H.}\ \bibnamefont
  {Saadat}}, \bibinfo {author} {\bibfnamefont {F.}~\bibnamefont {B\"osch}},\
  and\ \bibinfo {author} {\bibfnamefont {I.~V.}\ \bibnamefont {Karlin}},\
  }\bibfield  {title} {\bibinfo {title} {Lattice {B}oltzmann model for
  compressible flows on standard lattices: Variable {P}randtl number and
  adiabatic exponent},\ }\href {https://doi.org/10.1103/PhysRevE.99.013306}
  {\bibfield  {journal} {\bibinfo  {journal} {Phys. Rev. E}\ }\textbf {\bibinfo
  {volume} {99}},\ \bibinfo {pages} {013306} (\bibinfo {year}
  {2019})}\BibitemShut {NoStop}%
\bibitem [{\citenamefont {Saadat}\ \emph
  {et~al.}(2021{\natexlab{a}})\citenamefont {Saadat}, \citenamefont
  {Dorschner},\ and\ \citenamefont {Karlin}}]{HosseinExtended1}%
  \BibitemOpen
  \bibfield  {author} {\bibinfo {author} {\bibfnamefont {M.~H.}\ \bibnamefont
  {Saadat}}, \bibinfo {author} {\bibfnamefont {B.}~\bibnamefont {Dorschner}},\
  and\ \bibinfo {author} {\bibfnamefont {I.}~\bibnamefont {Karlin}},\
  }\bibfield  {title} {\bibinfo {title} {Extended lattice {B}oltzmann model},\
  }\bibfield  {journal} {\bibinfo  {journal} {Entropy}\ }\textbf {\bibinfo
  {volume} {23}},\ \href {https://doi.org/10.3390/e23040475}
  {10.3390/e23040475} (\bibinfo {year} {2021}{\natexlab{a}})\BibitemShut
  {NoStop}%
\bibitem [{\citenamefont {Saadat}\ \emph
  {et~al.}(2021{\natexlab{b}})\citenamefont {Saadat}, \citenamefont {Hosseini},
  \citenamefont {Dorschner},\ and\ \citenamefont {Karlin}}]{HosseinExtended2}%
  \BibitemOpen
  \bibfield  {author} {\bibinfo {author} {\bibfnamefont {M.~H.}\ \bibnamefont
  {Saadat}}, \bibinfo {author} {\bibfnamefont {S.~A.}\ \bibnamefont
  {Hosseini}}, \bibinfo {author} {\bibfnamefont {B.}~\bibnamefont
  {Dorschner}},\ and\ \bibinfo {author} {\bibfnamefont {I.~V.}\ \bibnamefont
  {Karlin}},\ }\bibfield  {title} {\bibinfo {title} {Extended lattice
  {B}oltzmann model for gas dynamics},\ }\href
  {https://doi.org/10.1063/5.0048029} {\bibfield  {journal} {\bibinfo
  {journal} {Physics of Fluids}\ }\textbf {\bibinfo {volume} {33}},\ \bibinfo
  {pages} {046104} (\bibinfo {year} {2021}{\natexlab{b}})},\ \Eprint
  {https://arxiv.org/abs/https://doi.org/10.1063/5.0048029}
  {https://doi.org/10.1063/5.0048029} \BibitemShut {NoStop}%
\bibitem [{\citenamefont {Bardow}\ \emph {et~al.}(2006)\citenamefont {Bardow},
  \citenamefont {Karlin},\ and\ \citenamefont {Gusev}}]{Bardow_2006}%
  \BibitemOpen
  \bibfield  {author} {\bibinfo {author} {\bibfnamefont {A.}~\bibnamefont
  {Bardow}}, \bibinfo {author} {\bibfnamefont {I.~V.}\ \bibnamefont {Karlin}},\
  and\ \bibinfo {author} {\bibfnamefont {A.~A.}\ \bibnamefont {Gusev}},\
  }\bibfield  {title} {\bibinfo {title} {General characteristic-based algorithm
  for off-lattice {B}oltzmann simulations},\ }\href
  {https://doi.org/10.1209/epl/i2006-10138-1} {\bibfield  {journal} {\bibinfo
  {journal} {Europhysics Letters ({EPL})}\ }\textbf {\bibinfo {volume} {75}},\
  \bibinfo {pages} {434} (\bibinfo {year} {2006})}\BibitemShut {NoStop}%
\bibitem [{\citenamefont {Shu}\ \emph {et~al.}(2002)\citenamefont {Shu},
  \citenamefont {Niu},\ and\ \citenamefont {Chew}}]{InterBasedLB_Shu}%
  \BibitemOpen
  \bibfield  {author} {\bibinfo {author} {\bibfnamefont {C.}~\bibnamefont
  {Shu}}, \bibinfo {author} {\bibfnamefont {X.~D.}\ \bibnamefont {Niu}},\ and\
  \bibinfo {author} {\bibfnamefont {Y.~T.}\ \bibnamefont {Chew}},\ }\bibfield
  {title} {\bibinfo {title} {Taylor-series expansion and least-squares-based
  lattice {B}oltzmann method: Two-dimensional formulation and its
  applications},\ }\href {https://doi.org/10.1103/PhysRevE.65.036708}
  {\bibfield  {journal} {\bibinfo  {journal} {Phys. Rev. E}\ }\textbf {\bibinfo
  {volume} {65}},\ \bibinfo {pages} {036708} (\bibinfo {year}
  {2002})}\BibitemShut {NoStop}%
\bibitem [{\citenamefont {Cheng}\ and\ \citenamefont
  {Hung}(2004)}]{InterpoBasedLB_cHENG_HUNG}%
  \BibitemOpen
  \bibfield  {author} {\bibinfo {author} {\bibfnamefont {M.}~\bibnamefont
  {Cheng}}\ and\ \bibinfo {author} {\bibfnamefont {K.~C.}\ \bibnamefont
  {Hung}},\ }\bibfield  {title} {\bibinfo {title} {Lattice {B}oltzmann method
  on nonuniform mesh},\ }\href {https://doi.org/10.1142/S1465876304002381}
  {\bibfield  {journal} {\bibinfo  {journal} {International Journal of
  Computational Engineering Science}\ }\textbf {\bibinfo {volume} {05}},\
  \bibinfo {pages} {291} (\bibinfo {year} {2004})},\ \Eprint
  {https://arxiv.org/abs/https://doi.org/10.1142/S1465876304002381}
  {https://doi.org/10.1142/S1465876304002381} \BibitemShut {NoStop}%
\bibitem [{\citenamefont {Kr\"amer}\ \emph {et~al.}(2017)\citenamefont
  {Kr\"amer}, \citenamefont {K\"ullmer}, \citenamefont {Reith}, \citenamefont
  {Joppich},\ and\ \citenamefont {Foysi}}]{SemiLag2017}%
  \BibitemOpen
  \bibfield  {author} {\bibinfo {author} {\bibfnamefont {A.}~\bibnamefont
  {Kr\"amer}}, \bibinfo {author} {\bibfnamefont {K.}~\bibnamefont {K\"ullmer}},
  \bibinfo {author} {\bibfnamefont {D.}~\bibnamefont {Reith}}, \bibinfo
  {author} {\bibfnamefont {W.}~\bibnamefont {Joppich}},\ and\ \bibinfo {author}
  {\bibfnamefont {H.}~\bibnamefont {Foysi}},\ }\bibfield  {title} {\bibinfo
  {title} {Semi-{L}agrangian off-lattice {B}oltzmann method for weakly
  compressible flows},\ }\href {https://doi.org/10.1103/PhysRevE.95.023305}
  {\bibfield  {journal} {\bibinfo  {journal} {Phys. Rev. E}\ }\textbf {\bibinfo
  {volume} {95}},\ \bibinfo {pages} {023305} (\bibinfo {year}
  {2017})}\BibitemShut {NoStop}%
\bibitem [{\citenamefont {Wilde}\ \emph {et~al.}(2020)\citenamefont {Wilde},
  \citenamefont {Kr\"amer}, \citenamefont {Reith},\ and\ \citenamefont
  {Foysi}}]{Kramer_SemiLagrangian2020}%
  \BibitemOpen
  \bibfield  {author} {\bibinfo {author} {\bibfnamefont {D.}~\bibnamefont
  {Wilde}}, \bibinfo {author} {\bibfnamefont {A.}~\bibnamefont {Kr\"amer}},
  \bibinfo {author} {\bibfnamefont {D.}~\bibnamefont {Reith}},\ and\ \bibinfo
  {author} {\bibfnamefont {H.}~\bibnamefont {Foysi}},\ }\bibfield  {title}
  {\bibinfo {title} {Semi-{L}agrangian lattice {B}oltzmann method for
  compressible flows},\ }\href {https://doi.org/10.1103/PhysRevE.101.053306}
  {\bibfield  {journal} {\bibinfo  {journal} {Phys. Rev. E}\ }\textbf {\bibinfo
  {volume} {101}},\ \bibinfo {pages} {053306} (\bibinfo {year}
  {2020})}\BibitemShut {NoStop}%
\bibitem [{\citenamefont {Wilde}\ \emph {et~al.}(2021)\citenamefont {Wilde},
  \citenamefont {Kr\"amer}, \citenamefont {Reith},\ and\ \citenamefont
  {Foysi}}]{Wilde2020_Turbulence}%
  \BibitemOpen
  \bibfield  {author} {\bibinfo {author} {\bibfnamefont {D.}~\bibnamefont
  {Wilde}}, \bibinfo {author} {\bibfnamefont {A.}~\bibnamefont {Kr\"amer}},
  \bibinfo {author} {\bibfnamefont {D.}~\bibnamefont {Reith}},\ and\ \bibinfo
  {author} {\bibfnamefont {H.}~\bibnamefont {Foysi}},\ }\bibfield  {title}
  {\bibinfo {title} {High-order semi-{L}agrangian kinetic scheme for
  compressible turbulence},\ }\href
  {https://doi.org/10.1103/PhysRevE.104.025301} {\bibfield  {journal} {\bibinfo
   {journal} {Phys. Rev. E}\ }\textbf {\bibinfo {volume} {104}},\ \bibinfo
  {pages} {025301} (\bibinfo {year} {2021})}\BibitemShut {NoStop}%
\bibitem [{\citenamefont {Guo}\ \emph {et~al.}(2013)\citenamefont {Guo},
  \citenamefont {Xu},\ and\ \citenamefont {Wang}}]{Guo_DUGKS_IsoT}%
  \BibitemOpen
  \bibfield  {author} {\bibinfo {author} {\bibfnamefont {Z.}~\bibnamefont
  {Guo}}, \bibinfo {author} {\bibfnamefont {K.}~\bibnamefont {Xu}},\ and\
  \bibinfo {author} {\bibfnamefont {R.}~\bibnamefont {Wang}},\ }\bibfield
  {title} {\bibinfo {title} {Discrete unified gas kinetic scheme for all
  {K}nudsen number flows: Low-speed isothermal case},\ }\href
  {https://doi.org/10.1103/PhysRevE.88.033305} {\bibfield  {journal} {\bibinfo
  {journal} {Phys. Rev. E}\ }\textbf {\bibinfo {volume} {88}},\ \bibinfo
  {pages} {033305} (\bibinfo {year} {2013})}\BibitemShut {NoStop}%
\bibitem [{\citenamefont {Guo}\ \emph {et~al.}(2015)\citenamefont {Guo},
  \citenamefont {Wang},\ and\ \citenamefont {Xu}}]{Guo_DUGKS_Compres}%
  \BibitemOpen
  \bibfield  {author} {\bibinfo {author} {\bibfnamefont {Z.}~\bibnamefont
  {Guo}}, \bibinfo {author} {\bibfnamefont {R.}~\bibnamefont {Wang}},\ and\
  \bibinfo {author} {\bibfnamefont {K.}~\bibnamefont {Xu}},\ }\bibfield
  {title} {\bibinfo {title} {Discrete unified gas kinetic scheme for all
  {K}nudsen number flows. ii. thermal compressible case},\ }\href
  {https://doi.org/10.1103/PhysRevE.91.033313} {\bibfield  {journal} {\bibinfo
  {journal} {Phys. Rev. E}\ }\textbf {\bibinfo {volume} {91}},\ \bibinfo
  {pages} {033313} (\bibinfo {year} {2015})}\BibitemShut {NoStop}%
\bibitem [{\citenamefont {Gan}\ \emph {et~al.}(2018)\citenamefont {Gan},
  \citenamefont {Xu}, \citenamefont {Zhang}, \citenamefont {Zhang},\ and\
  \citenamefont {Succi}}]{DBM_2018_XU}%
  \BibitemOpen
  \bibfield  {author} {\bibinfo {author} {\bibfnamefont {Y.}~\bibnamefont
  {Gan}}, \bibinfo {author} {\bibfnamefont {A.}~\bibnamefont {Xu}}, \bibinfo
  {author} {\bibfnamefont {G.}~\bibnamefont {Zhang}}, \bibinfo {author}
  {\bibfnamefont {Y.}~\bibnamefont {Zhang}},\ and\ \bibinfo {author}
  {\bibfnamefont {S.}~\bibnamefont {Succi}},\ }\bibfield  {title} {\bibinfo
  {title} {Discrete {B}oltzmann trans-scale modeling of high-speed compressible
  flows},\ }\href {https://doi.org/10.1103/PhysRevE.97.053312} {\bibfield
  {journal} {\bibinfo  {journal} {Phys. Rev. E}\ }\textbf {\bibinfo {volume}
  {97}},\ \bibinfo {pages} {053312} (\bibinfo {year} {2018})}\BibitemShut
  {NoStop}%
\bibitem [{\citenamefont {Ji}\ \emph {et~al.}(2021)\citenamefont {Ji},
  \citenamefont {Lin},\ and\ \citenamefont {Luo}}]{DBM_2021_LUO}%
  \BibitemOpen
  \bibfield  {author} {\bibinfo {author} {\bibfnamefont {Y.~å.}\ \bibnamefont
  {Ji}}, \bibinfo {author} {\bibfnamefont {C.~æ.}\ \bibnamefont {Lin}},\ and\
  \bibinfo {author} {\bibfnamefont {K.~H.~ç.}\ \bibnamefont {Luo}},\ }\bibfield
   {title} {\bibinfo {title} {Three-dimensional multiple-relaxation-time
  discrete {B}oltzmann model of compressible reactive flows with nonequilibrium
  effects},\ }\href {https://doi.org/10.1063/5.0047480} {\bibfield  {journal}
  {\bibinfo  {journal} {AIP Advances}\ }\textbf {\bibinfo {volume} {11}},\
  \bibinfo {pages} {045217} (\bibinfo {year} {2021})},\ \Eprint
  {https://arxiv.org/abs/https://doi.org/10.1063/5.0047480}
  {https://doi.org/10.1063/5.0047480} \BibitemShut {NoStop}%
\bibitem [{\citenamefont {Gan}\ \emph {et~al.}(2015)\citenamefont {Gan},
  \citenamefont {Xu}, \citenamefont {Zhang},\ and\ \citenamefont
  {Succi}}]{DBM_SUCCI_XU2015}%
  \BibitemOpen
  \bibfield  {author} {\bibinfo {author} {\bibfnamefont {Y.}~\bibnamefont
  {Gan}}, \bibinfo {author} {\bibfnamefont {A.}~\bibnamefont {Xu}}, \bibinfo
  {author} {\bibfnamefont {G.}~\bibnamefont {Zhang}},\ and\ \bibinfo {author}
  {\bibfnamefont {S.}~\bibnamefont {Succi}},\ }\bibfield  {title} {\bibinfo
  {title} {Discrete {B}oltzmann modeling of multiphase flows: hydrodynamic and
  thermodynamic non-equilibrium effects},\ }\href
  {https://doi.org/10.1039/C5SM01125F} {\bibfield  {journal} {\bibinfo
  {journal} {Soft Matter}\ }\textbf {\bibinfo {volume} {11}},\ \bibinfo {pages}
  {5336} (\bibinfo {year} {2015})}\BibitemShut {NoStop}%
\bibitem [{\citenamefont {Lai}\ \emph {et~al.}(2016)\citenamefont {Lai},
  \citenamefont {Xu}, \citenamefont {Zhang}, \citenamefont {Gan}, \citenamefont
  {Ying},\ and\ \citenamefont {Succi}}]{DBM_SUCCI_XU2016}%
  \BibitemOpen
  \bibfield  {author} {\bibinfo {author} {\bibfnamefont {H.}~\bibnamefont
  {Lai}}, \bibinfo {author} {\bibfnamefont {A.}~\bibnamefont {Xu}}, \bibinfo
  {author} {\bibfnamefont {G.}~\bibnamefont {Zhang}}, \bibinfo {author}
  {\bibfnamefont {Y.}~\bibnamefont {Gan}}, \bibinfo {author} {\bibfnamefont
  {Y.}~\bibnamefont {Ying}},\ and\ \bibinfo {author} {\bibfnamefont
  {S.}~\bibnamefont {Succi}},\ }\bibfield  {title} {\bibinfo {title}
  {Nonequilibrium thermohydrodynamic effects on the {R}ayleigh-{T}aylor
  instability in compressible flows},\ }\href
  {https://doi.org/10.1103/PhysRevE.94.023106} {\bibfield  {journal} {\bibinfo
  {journal} {Phys. Rev. E}\ }\textbf {\bibinfo {volume} {94}},\ \bibinfo
  {pages} {023106} (\bibinfo {year} {2016})}\BibitemShut {NoStop}%
\bibitem [{\citenamefont {{La Rocca}}\ \emph {et~al.}(2015)\citenamefont {{La
  Rocca}}, \citenamefont {Montessori}, \citenamefont {Prestininzi},\ and\
  \citenamefont {Succi}}]{DBM_SUCCI2015}%
  \BibitemOpen
  \bibfield  {author} {\bibinfo {author} {\bibfnamefont {M.}~\bibnamefont {{La
  Rocca}}}, \bibinfo {author} {\bibfnamefont {A.}~\bibnamefont {Montessori}},
  \bibinfo {author} {\bibfnamefont {P.}~\bibnamefont {Prestininzi}},\ and\
  \bibinfo {author} {\bibfnamefont {S.}~\bibnamefont {Succi}},\ }\bibfield
  {title} {\bibinfo {title} {A multispeed discrete {B}oltzmann model for
  transcritical 2d shallow water flows},\ }\href
  {https://doi.org/https://doi.org/10.1016/j.jcp.2014.12.029} {\bibfield
  {journal} {\bibinfo  {journal} {Journal of Computational Physics}\ }\textbf
  {\bibinfo {volume} {284}},\ \bibinfo {pages} {117} (\bibinfo {year}
  {2015})}\BibitemShut {NoStop}%
\bibitem [{\citenamefont {Frapolli}\ \emph
  {et~al.}(2016{\natexlab{b}})\citenamefont {Frapolli}, \citenamefont
  {Chikatamarla},\ and\ \citenamefont {Karlin}}]{Frapolli_ShiftedLattices}%
  \BibitemOpen
  \bibfield  {author} {\bibinfo {author} {\bibfnamefont {N.}~\bibnamefont
  {Frapolli}}, \bibinfo {author} {\bibfnamefont {S.~S.}\ \bibnamefont
  {Chikatamarla}},\ and\ \bibinfo {author} {\bibfnamefont {I.~V.}\ \bibnamefont
  {Karlin}},\ }\bibfield  {title} {\bibinfo {title} {Lattice kinetic theory in
  a comoving {G}alilean reference frame},\ }\href
  {https://doi.org/10.1103/PhysRevLett.117.010604} {\bibfield  {journal}
  {\bibinfo  {journal} {Phys. Rev. Lett.}\ }\textbf {\bibinfo {volume} {117}},\
  \bibinfo {pages} {010604} (\bibinfo {year} {2016}{\natexlab{b}})}\BibitemShut
  {NoStop}%
\bibitem [{\citenamefont {Dorschner}\ \emph {et~al.}(2018)\citenamefont
  {Dorschner}, \citenamefont {B\"osch},\ and\ \citenamefont {Karlin}}]{Pond}%
  \BibitemOpen
  \bibfield  {author} {\bibinfo {author} {\bibfnamefont {B.}~\bibnamefont
  {Dorschner}}, \bibinfo {author} {\bibfnamefont {F.}~\bibnamefont {B\"osch}},\
  and\ \bibinfo {author} {\bibfnamefont {I.~V.}\ \bibnamefont {Karlin}},\
  }\bibfield  {title} {\bibinfo {title} {Particles on demand for kinetic
  theory},\ }\href {https://doi.org/10.1103/PhysRevLett.121.130602} {\bibfield
  {journal} {\bibinfo  {journal} {Phys. Rev. Lett.}\ }\textbf {\bibinfo
  {volume} {121}},\ \bibinfo {pages} {130602} (\bibinfo {year}
  {2018})}\BibitemShut {NoStop}%
\bibitem [{\citenamefont {Zipunova}\ \emph
  {et~al.}(2021{\natexlab{a}})\citenamefont {Zipunova}, \citenamefont
  {Perepelkina}, \citenamefont {Zakirov},\ and\ \citenamefont
  {Khilkov}}]{RegPond}%
  \BibitemOpen
  \bibfield  {author} {\bibinfo {author} {\bibfnamefont {E.}~\bibnamefont
  {Zipunova}}, \bibinfo {author} {\bibfnamefont {A.}~\bibnamefont
  {Perepelkina}}, \bibinfo {author} {\bibfnamefont {A.}~\bibnamefont
  {Zakirov}},\ and\ \bibinfo {author} {\bibfnamefont {S.}~\bibnamefont
  {Khilkov}},\ }\bibfield  {title} {\bibinfo {title} {Regularization and the
  particles-on-demand method for the solution of the discrete {B}oltzmann
  equation},\ }\href
  {https://doi.org/https://doi.org/10.1016/j.jocs.2021.101376} {\bibfield
  {journal} {\bibinfo  {journal} {Journal of Computational Science}\ }\textbf
  {\bibinfo {volume} {53}},\ \bibinfo {pages} {101376} (\bibinfo {year}
  {2021}{\natexlab{a}})}\BibitemShut {NoStop}%
\bibitem [{\citenamefont {Fu}(2019)}]{LinFuAllSpeed}%
  \BibitemOpen
  \bibfield  {author} {\bibinfo {author} {\bibfnamefont {L.}~\bibnamefont
  {Fu}},\ }\bibfield  {title} {\bibinfo {title} {A very-high-order {TENO}
  scheme for all-speed gas dynamics and turbulence},\ }\href
  {https://doi.org/https://doi.org/10.1016/j.cpc.2019.06.013} {\bibfield
  {journal} {\bibinfo  {journal} {Computer Physics Communications}\ }\textbf
  {\bibinfo {volume} {244}},\ \bibinfo {pages} {117} (\bibinfo {year}
  {2019})}\BibitemShut {NoStop}%
\bibitem [{\citenamefont {Zhang}\ and\ \citenamefont
  {Shu}(2010)}]{ZhangShu2010}%
  \BibitemOpen
  \bibfield  {author} {\bibinfo {author} {\bibfnamefont {X.}~\bibnamefont
  {Zhang}}\ and\ \bibinfo {author} {\bibfnamefont {C.-W.}\ \bibnamefont
  {Shu}},\ }\bibfield  {title} {\bibinfo {title} {On positivity-preserving high
  order discontinuous {G}alerkin schemes for compressible {E}uler equations on
  rectangular meshes},\ }\href
  {https://doi.org/https://doi.org/10.1016/j.jcp.2010.08.016} {\bibfield
  {journal} {\bibinfo  {journal} {Journal of Computational Physics}\ }\textbf
  {\bibinfo {volume} {229}},\ \bibinfo {pages} {8918} (\bibinfo {year}
  {2010})}\BibitemShut {NoStop}%
\bibitem [{\citenamefont {Zhang}\ and\ \citenamefont
  {Shu}(2012)}]{ZhangShu2012}%
  \BibitemOpen
  \bibfield  {author} {\bibinfo {author} {\bibfnamefont {X.}~\bibnamefont
  {Zhang}}\ and\ \bibinfo {author} {\bibfnamefont {C.-W.}\ \bibnamefont
  {Shu}},\ }\bibfield  {title} {\bibinfo {title} {Positivity-preserving high
  order finite difference {WENO} schemes for compressible {E}uler equations},\
  }\href {https://doi.org/https://doi.org/10.1016/j.jcp.2011.11.020} {\bibfield
   {journal} {\bibinfo  {journal} {Journal of Computational Physics}\ }\textbf
  {\bibinfo {volume} {231}},\ \bibinfo {pages} {2245} (\bibinfo {year}
  {2012})}\BibitemShut {NoStop}%
\bibitem [{\citenamefont {Kallikounis}\ \emph {et~al.}(2021)\citenamefont
  {Kallikounis}, \citenamefont {Dorschner},\ and\ \citenamefont
  {Karlin}}]{Multiscale2021}%
  \BibitemOpen
  \bibfield  {author} {\bibinfo {author} {\bibfnamefont {N.~G.}\ \bibnamefont
  {Kallikounis}}, \bibinfo {author} {\bibfnamefont {B.}~\bibnamefont
  {Dorschner}},\ and\ \bibinfo {author} {\bibfnamefont {I.~V.}\ \bibnamefont
  {Karlin}},\ }\bibfield  {title} {\bibinfo {title} {Multiscale
  semi-{L}agrangian lattice {B}oltzmann method},\ }\href
  {https://doi.org/10.1103/PhysRevE.103.063305} {\bibfield  {journal} {\bibinfo
   {journal} {Phys. Rev. E}\ }\textbf {\bibinfo {volume} {103}},\ \bibinfo
  {pages} {063305} (\bibinfo {year} {2021})}\BibitemShut {NoStop}%
\bibitem [{\citenamefont {Zipunova}\ \emph
  {et~al.}(2021{\natexlab{b}})\citenamefont {Zipunova}, \citenamefont
  {Perepelkina},\ and\ \citenamefont {Zakirov}}]{PondReg2}%
  \BibitemOpen
  \bibfield  {author} {\bibinfo {author} {\bibfnamefont {E.}~\bibnamefont
  {Zipunova}}, \bibinfo {author} {\bibfnamefont {A.}~\bibnamefont
  {Perepelkina}},\ and\ \bibinfo {author} {\bibfnamefont {A.}~\bibnamefont
  {Zakirov}},\ }\bibfield  {title} {\bibinfo {title} {Applicability of
  regularized particles-on-demand method to solve {R}iemann problem},\ }\href
  {https://doi.org/10.1088/1742-6596/1740/1/012024} {\bibfield  {journal}
  {\bibinfo  {journal} {Journal of Physics: Conference Series}\ }\textbf
  {\bibinfo {volume} {1740}},\ \bibinfo {pages} {012024} (\bibinfo {year}
  {2021}{\natexlab{b}})}\BibitemShut {NoStop}%
\bibitem [{\citenamefont {Gorban}\ and\ \citenamefont
  {Karlin}(2005)}]{GorbanKarlin}%
  \BibitemOpen
  \bibfield  {author} {\bibinfo {author} {\bibfnamefont {A.~N.}\ \bibnamefont
  {Gorban}}\ and\ \bibinfo {author} {\bibfnamefont {I.~V.}\ \bibnamefont
  {Karlin}},\ }\href@noop {} {\emph {\bibinfo {title} {Invariant Manifolds for
  Physical and Chemical Kinetics}}}\ (\bibinfo  {publisher} {Springer-Verlag
  Berlin Heidelberg},\ \bibinfo {year} {2005})\BibitemShut {NoStop}%
\bibitem [{\citenamefont {Reyhanian}\ \emph {et~al.}(2020)\citenamefont
  {Reyhanian}, \citenamefont {Dorschner},\ and\ \citenamefont
  {Karlin}}]{Ehsan2020}%
  \BibitemOpen
  \bibfield  {author} {\bibinfo {author} {\bibfnamefont {E.}~\bibnamefont
  {Reyhanian}}, \bibinfo {author} {\bibfnamefont {B.}~\bibnamefont
  {Dorschner}},\ and\ \bibinfo {author} {\bibfnamefont {I.~V.}\ \bibnamefont
  {Karlin}},\ }\bibfield  {title} {\bibinfo {title} {Thermokinetic lattice
  {B}oltzmann model of nonideal fluids},\ }\href
  {https://doi.org/10.1103/PhysRevE.102.020103} {\bibfield  {journal} {\bibinfo
   {journal} {Phys. Rev. E}\ }\textbf {\bibinfo {volume} {102}},\ \bibinfo
  {pages} {020103} (\bibinfo {year} {2020})}\BibitemShut {NoStop}%
\bibitem [{\citenamefont {Reyhanian}\ \emph {et~al.}(2021)\citenamefont
  {Reyhanian}, \citenamefont {Dorschner},\ and\ \citenamefont
  {Karlin}}]{Ehsan2021}%
  \BibitemOpen
  \bibfield  {author} {\bibinfo {author} {\bibfnamefont {E.}~\bibnamefont
  {Reyhanian}}, \bibinfo {author} {\bibfnamefont {B.}~\bibnamefont
  {Dorschner}},\ and\ \bibinfo {author} {\bibfnamefont {I.}~\bibnamefont
  {Karlin}},\ }\bibfield  {title} {\bibinfo {title} {Kinetic simulations of
  compressible non-ideal fluids: From supercritical flows to phase-change and
  exotic behavior},\ }\bibfield  {journal} {\bibinfo  {journal} {Computation}\
  }\textbf {\bibinfo {volume} {9}},\ \href
  {https://doi.org/10.3390/computation9020013} {10.3390/computation9020013}
  (\bibinfo {year} {2021})\BibitemShut {NoStop}%
\bibitem [{\citenamefont {He}\ \emph {et~al.}(1998{\natexlab{b}})\citenamefont
  {He}, \citenamefont {Shan},\ and\ \citenamefont
  {Doolen}}]{LBM_VarTransform1}%
  \BibitemOpen
  \bibfield  {author} {\bibinfo {author} {\bibfnamefont {X.}~\bibnamefont
  {He}}, \bibinfo {author} {\bibfnamefont {X.}~\bibnamefont {Shan}},\ and\
  \bibinfo {author} {\bibfnamefont {G.~D.}\ \bibnamefont {Doolen}},\ }\bibfield
   {title} {\bibinfo {title} {Discrete {B}oltzmann equation model for nonideal
  gases},\ }\href {https://doi.org/10.1103/PhysRevE.57.R13} {\bibfield
  {journal} {\bibinfo  {journal} {Phys. Rev. E}\ }\textbf {\bibinfo {volume}
  {57}},\ \bibinfo {pages} {R13} (\bibinfo {year}
  {1998}{\natexlab{b}})}\BibitemShut {NoStop}%
\bibitem [{\citenamefont {He}\ \emph {et~al.}(1998{\natexlab{c}})\citenamefont
  {He}, \citenamefont {Chen},\ and\ \citenamefont
  {Doolen}}]{LBM_VarTransform2}%
  \BibitemOpen
  \bibfield  {author} {\bibinfo {author} {\bibfnamefont {X.}~\bibnamefont
  {He}}, \bibinfo {author} {\bibfnamefont {S.}~\bibnamefont {Chen}},\ and\
  \bibinfo {author} {\bibfnamefont {G.~D.}\ \bibnamefont {Doolen}},\ }\bibfield
   {title} {\bibinfo {title} {A novel thermal model for the lattice {B}oltzmann
  method in incompressible limit},\ }\href
  {https://doi.org/https://doi.org/10.1006/jcph.1998.6057} {\bibfield
  {journal} {\bibinfo  {journal} {Journal of Computational Physics}\ }\textbf
  {\bibinfo {volume} {146}},\ \bibinfo {pages} {282} (\bibinfo {year}
  {1998}{\natexlab{c}})}\BibitemShut {NoStop}%
\bibitem [{\citenamefont {Reyhanian}(2021)}]{EhsanThesis}%
  \BibitemOpen
  \bibfield  {author} {\bibinfo {author} {\bibfnamefont {E.}~\bibnamefont
  {Reyhanian}},\ }\emph {\bibinfo {title} {Thermokinetic Model for Compressible
  Generic Fluids}},\ \href@noop {} {Ph.D. thesis},\ \bibinfo  {school} {ETH
  Z\"{u}rich} (\bibinfo {year} {2021})\BibitemShut {NoStop}%
\bibitem [{\citenamefont {Lentine}\ \emph {et~al.}(2011)\citenamefont
  {Lentine}, \citenamefont {Grétarsson},\ and\ \citenamefont
  {Fedkiw}}]{ConservSemiLag}%
  \BibitemOpen
  \bibfield  {author} {\bibinfo {author} {\bibfnamefont {M.}~\bibnamefont
  {Lentine}}, \bibinfo {author} {\bibfnamefont {J.~T.}\ \bibnamefont
  {Grétarsson}},\ and\ \bibinfo {author} {\bibfnamefont {R.}~\bibnamefont
  {Fedkiw}},\ }\bibfield  {title} {\bibinfo {title} {An unconditionally stable
  fully conservative semi-{L}agrangian method},\ }\href
  {https://doi.org/https://doi.org/10.1016/j.jcp.2010.12.036} {\bibfield
  {journal} {\bibinfo  {journal} {Journal of Computational Physics}\ }\textbf
  {\bibinfo {volume} {230}},\ \bibinfo {pages} {2857} (\bibinfo {year}
  {2011})}\BibitemShut {NoStop}%
\bibitem [{\citenamefont {Xiao}\ and\ \citenamefont
  {Yabe}(2001)}]{ConserSemiLag2}%
  \BibitemOpen
  \bibfield  {author} {\bibinfo {author} {\bibfnamefont {F.}~\bibnamefont
  {Xiao}}\ and\ \bibinfo {author} {\bibfnamefont {T.}~\bibnamefont {Yabe}},\
  }\bibfield  {title} {\bibinfo {title} {Completely conservative and
  oscillationless semi-{L}agrangian schemes for advection transportation},\
  }\href {https://doi.org/https://doi.org/10.1006/jcph.2001.6746} {\bibfield
  {journal} {\bibinfo  {journal} {Journal of Computational Physics}\ }\textbf
  {\bibinfo {volume} {170}},\ \bibinfo {pages} {498} (\bibinfo {year}
  {2001})}\BibitemShut {NoStop}%
\bibitem [{\citenamefont {Guo}\ and\ \citenamefont {Xu}(2021)}]{Guo_DUGKS_Rev}%
  \BibitemOpen
  \bibfield  {author} {\bibinfo {author} {\bibfnamefont {Z.}~\bibnamefont
  {Guo}}\ and\ \bibinfo {author} {\bibfnamefont {K.}~\bibnamefont {Xu}},\
  }\bibfield  {title} {\bibinfo {title} {Progress of discrete unified
  gas-kinetic scheme for multiscale flows},\ }\href
  {https://doi.org/10.1186/s42774-020-00058-3} {\bibfield  {journal} {\bibinfo
  {journal} {Advances in Aerodynamics}\ }\textbf {\bibinfo {volume} {3}},\
  \bibinfo {pages} {6} (\bibinfo {year} {2021})}\BibitemShut {NoStop}%
\bibitem [{\citenamefont {{Van Leer}}(1977)}]{VanLeerLimiter}%
  \BibitemOpen
  \bibfield  {author} {\bibinfo {author} {\bibfnamefont {B.}~\bibnamefont {{Van
  Leer}}},\ }\bibfield  {title} {\bibinfo {title} {Towards the ultimate
  conservative difference scheme. iv. a new approach to numerical convection},\
  }\href {https://doi.org/https://doi.org/10.1016/0021-9991(77)90095-X}
  {\bibfield  {journal} {\bibinfo  {journal} {Journal of Computational
  Physics}\ }\textbf {\bibinfo {volume} {23}},\ \bibinfo {pages} {276}
  (\bibinfo {year} {1977})}\BibitemShut {NoStop}%
\bibitem [{\citenamefont {Roe}(1986)}]{RoeLimiter}%
  \BibitemOpen
  \bibfield  {author} {\bibinfo {author} {\bibfnamefont {P.~L.}\ \bibnamefont
  {Roe}},\ }\bibfield  {title} {\bibinfo {title} {Characteristic-based schemes
  for the {E}uler equations},\ }\href
  {https://doi.org/10.1146/annurev.fl.18.010186.002005} {\bibfield  {journal}
  {\bibinfo  {journal} {Annual Review of Fluid Mechanics}\ }\textbf {\bibinfo
  {volume} {18}},\ \bibinfo {pages} {337} (\bibinfo {year} {1986})}\BibitemShut
  {NoStop}%
\bibitem [{\citenamefont {{van Leer}}(1979)}]{VanLeerMuscl}%
  \BibitemOpen
  \bibfield  {author} {\bibinfo {author} {\bibfnamefont {B.}~\bibnamefont {{van
  Leer}}},\ }\bibfield  {title} {\bibinfo {title} {Towards the ultimate
  conservative difference scheme. {V}. {A} second-order sequel to {G}odunov's
  method},\ }\href
  {https://doi.org/https://doi.org/10.1016/0021-9991(79)90145-1} {\bibfield
  {journal} {\bibinfo  {journal} {Journal of Computational Physics}\ }\textbf
  {\bibinfo {volume} {32}},\ \bibinfo {pages} {101} (\bibinfo {year}
  {1979})}\BibitemShut {NoStop}%
\bibitem [{\citenamefont {Sod}(1978)}]{SodTube}%
  \BibitemOpen
  \bibfield  {author} {\bibinfo {author} {\bibfnamefont {G.~A.}\ \bibnamefont
  {Sod}},\ }\bibfield  {title} {\bibinfo {title} {A survey of several finite
  difference methods for systems of nonlinear hyperbolic conservation laws},\
  }\href {https://doi.org/https://doi.org/10.1016/0021-9991(78)90023-2}
  {\bibfield  {journal} {\bibinfo  {journal} {Journal of Computational
  Physics}\ }\textbf {\bibinfo {volume} {27}},\ \bibinfo {pages} {1} (\bibinfo
  {year} {1978})}\BibitemShut {NoStop}%
\bibitem [{\citenamefont {Lax}(1954)}]{LaxTube}%
  \BibitemOpen
  \bibfield  {author} {\bibinfo {author} {\bibfnamefont {P.~D.}\ \bibnamefont
  {Lax}},\ }\bibfield  {title} {\bibinfo {title} {Weak solutions of nonlinear
  hyperbolic equations and their numerical computation},\ }\href
  {https://doi.org/https://doi.org/10.1002/cpa.3160070112} {\bibfield
  {journal} {\bibinfo  {journal} {Communications on Pure and Applied
  Mathematics}\ }\textbf {\bibinfo {volume} {7}},\ \bibinfo {pages} {159}
  (\bibinfo {year} {1954})}\BibitemShut {NoStop}%
\bibitem [{\citenamefont {Shu}\ and\ \citenamefont
  {Osher}(1989)}]{ShuOsherProblem}%
  \BibitemOpen
  \bibfield  {author} {\bibinfo {author} {\bibfnamefont {C.-W.}\ \bibnamefont
  {Shu}}\ and\ \bibinfo {author} {\bibfnamefont {S.}~\bibnamefont {Osher}},\
  }\bibfield  {title} {\bibinfo {title} {Efficient implementation of
  essentially non-oscillatory shock-capturing schemes, ii},\ }\href
  {https://doi.org/https://doi.org/10.1016/0021-9991(89)90222-2} {\bibfield
  {journal} {\bibinfo  {journal} {Journal of Computational Physics}\ }\textbf
  {\bibinfo {volume} {83}},\ \bibinfo {pages} {32} (\bibinfo {year}
  {1989})}\BibitemShut {NoStop}%
\bibitem [{\citenamefont {Jiang}\ and\ \citenamefont
  {Shu}(1996)}]{ShuOsherRefSol}%
  \BibitemOpen
  \bibfield  {author} {\bibinfo {author} {\bibfnamefont {G.-S.}\ \bibnamefont
  {Jiang}}\ and\ \bibinfo {author} {\bibfnamefont {C.-W.}\ \bibnamefont
  {Shu}},\ }\bibfield  {title} {\bibinfo {title} {Efficient implementation of
  weighted {ENO} schemes},\ }\href
  {https://doi.org/https://doi.org/10.1006/jcph.1996.0130} {\bibfield
  {journal} {\bibinfo  {journal} {Journal of Computational Physics}\ }\textbf
  {\bibinfo {volume} {126}},\ \bibinfo {pages} {202} (\bibinfo {year}
  {1996})}\BibitemShut {NoStop}%
\bibitem [{\citenamefont {Toro}\ and\ \citenamefont
  {Vázquez-Cendón}(2012)}]{StrongShock}%
  \BibitemOpen
  \bibfield  {author} {\bibinfo {author} {\bibfnamefont {E.}~\bibnamefont
  {Toro}}\ and\ \bibinfo {author} {\bibfnamefont {M.}~\bibnamefont
  {Vázquez-Cendón}},\ }\bibfield  {title} {\bibinfo {title} {Flux splitting
  schemes for the {E}uler equations},\ }\href
  {https://doi.org/https://doi.org/10.1016/j.compfluid.2012.08.023} {\bibfield
  {journal} {\bibinfo  {journal} {Computers \& Fluids}\ }\textbf {\bibinfo
  {volume} {70}},\ \bibinfo {pages} {1} (\bibinfo {year} {2012})}\BibitemShut
  {NoStop}%
\bibitem [{\citenamefont {Woodward}\ and\ \citenamefont
  {Colella}(1984)}]{WoodwardCollela_DoubleMachReflection}%
  \BibitemOpen
  \bibfield  {author} {\bibinfo {author} {\bibfnamefont {P.}~\bibnamefont
  {Woodward}}\ and\ \bibinfo {author} {\bibfnamefont {P.}~\bibnamefont
  {Colella}},\ }\bibfield  {title} {\bibinfo {title} {The numerical simulation
  of two-dimensional fluid flow with strong shocks},\ }\href
  {https://doi.org/https://doi.org/10.1016/0021-9991(84)90142-6} {\bibfield
  {journal} {\bibinfo  {journal} {Journal of Computational Physics}\ }\textbf
  {\bibinfo {volume} {54}},\ \bibinfo {pages} {115} (\bibinfo {year}
  {1984})}\BibitemShut {NoStop}%
\bibitem [{\citenamefont {Hu}\ \emph {et~al.}(2013)\citenamefont {Hu},
  \citenamefont {Adams},\ and\ \citenamefont {Shu}}]{DoubleRaref}%
  \BibitemOpen
  \bibfield  {author} {\bibinfo {author} {\bibfnamefont {X.~Y.}\ \bibnamefont
  {Hu}}, \bibinfo {author} {\bibfnamefont {N.~A.}\ \bibnamefont {Adams}},\ and\
  \bibinfo {author} {\bibfnamefont {C.-W.}\ \bibnamefont {Shu}},\ }\bibfield
  {title} {\bibinfo {title} {Positivity-preserving method for high-order
  conservative schemes solving compressible {E}uler equations},\ }\href
  {https://doi.org/https://doi.org/10.1016/j.jcp.2013.01.024} {\bibfield
  {journal} {\bibinfo  {journal} {Journal of Computational Physics}\ }\textbf
  {\bibinfo {volume} {242}},\ \bibinfo {pages} {169} (\bibinfo {year}
  {2013})}\BibitemShut {NoStop}%
\bibitem [{\citenamefont {Loubère}\ and\ \citenamefont
  {Shashkov}(2005)}]{LeBlanc}%
  \BibitemOpen
  \bibfield  {author} {\bibinfo {author} {\bibfnamefont {R.}~\bibnamefont
  {Loubère}}\ and\ \bibinfo {author} {\bibfnamefont {M.~J.}\ \bibnamefont
  {Shashkov}},\ }\bibfield  {title} {\bibinfo {title} {A subcell remapping
  method on staggered polygonal grids for arbitrary-{L}agrangian–{E}ulerian
  methods},\ }\href {https://doi.org/https://doi.org/10.1016/j.jcp.2005.03.019}
  {\bibfield  {journal} {\bibinfo  {journal} {Journal of Computational
  Physics}\ }\textbf {\bibinfo {volume} {209}},\ \bibinfo {pages} {105}
  (\bibinfo {year} {2005})}\BibitemShut {NoStop}%
\bibitem [{\citenamefont {Sedov}(1993)}]{Sedov}%
  \BibitemOpen
  \bibfield  {author} {\bibinfo {author} {\bibfnamefont {L.~I.}\ \bibnamefont
  {Sedov}},\ }\href@noop {} {\emph {\bibinfo {title} {Similarity and
  Dimensional Methods in Mechanics}}}\ (\bibinfo  {publisher} {CRC Press},\
  \bibinfo {year} {1993})\BibitemShut {NoStop}%
\bibitem [{\citenamefont {Lax}\ and\ \citenamefont
  {Liu}(1998)}]{Lax_RiemannReference}%
  \BibitemOpen
  \bibfield  {author} {\bibinfo {author} {\bibfnamefont {P.~D.}\ \bibnamefont
  {Lax}}\ and\ \bibinfo {author} {\bibfnamefont {X.-D.}\ \bibnamefont {Liu}},\
  }\bibfield  {title} {\bibinfo {title} {Solution of two-dimensional {R}iemann
  problems of gas dynamics by positive schemes},\ }\href
  {https://doi.org/10.1137/S1064827595291819} {\bibfield  {journal} {\bibinfo
  {journal} {SIAM Journal on Scientific Computing}\ }\textbf {\bibinfo {volume}
  {19}},\ \bibinfo {pages} {319} (\bibinfo {year} {1998})},\ \Eprint
  {https://arxiv.org/abs/https://doi.org/10.1137/S1064827595291819}
  {https://doi.org/10.1137/S1064827595291819} \BibitemShut {NoStop}%
\bibitem [{\citenamefont {Kurganov}\ and\ \citenamefont
  {Tadmor}(2002)}]{2DRiem_ref}%
  \BibitemOpen
  \bibfield  {author} {\bibinfo {author} {\bibfnamefont {A.}~\bibnamefont
  {Kurganov}}\ and\ \bibinfo {author} {\bibfnamefont {E.}~\bibnamefont
  {Tadmor}},\ }\bibfield  {title} {\bibinfo {title} {Solution of
  two-dimensional {R}iemann problems for gas dynamics without {R}iemann problem
  solvers},\ }\href {https://doi.org/https://doi.org/10.1002/num.10025}
  {\bibfield  {journal} {\bibinfo  {journal} {Numerical Methods for Partial
  Differential Equations}\ }\textbf {\bibinfo {volume} {18}},\ \bibinfo {pages}
  {584} (\bibinfo {year} {2002})}\BibitemShut {NoStop}%
\bibitem [{Exp()}]{ExplosionBoxRef}%
  \BibitemOpen
  \href@noop {} {\bibinfo {title} {Amroc-blockstructured adaptive mesh
  refinement in object-oriented c\texttt{++}}},\ \bibinfo {howpublished}
  {\url{http://amroc.sourceforge.net/examples/euler/2d/html/box_n.htm}}\BibitemShut
  {NoStop}%
\bibitem [{\citenamefont {Brouillette}(2002)}]{RMI}%
  \BibitemOpen
  \bibfield  {author} {\bibinfo {author} {\bibfnamefont {M.}~\bibnamefont
  {Brouillette}},\ }\bibfield  {title} {\bibinfo {title} {The
  {R}ichtmyer-{M}eshkov instability},\ }\href
  {https://doi.org/10.1146/annurev.fluid.34.090101.162238} {\bibfield
  {journal} {\bibinfo  {journal} {Annual Review of Fluid Mechanics}\ }\textbf
  {\bibinfo {volume} {34}},\ \bibinfo {pages} {445} (\bibinfo {year} {2002})},\
  \Eprint
  {https://arxiv.org/abs/https://doi.org/10.1146/annurev.fluid.34.090101.162238}
  {https://doi.org/10.1146/annurev.fluid.34.090101.162238} \BibitemShut
  {NoStop}%
\bibitem [{\citenamefont {Chen}\ \emph {et~al.}(2011)\citenamefont {Chen},
  \citenamefont {Xu}, \citenamefont {Zhang},\ and\ \citenamefont
  {Li}}]{RMI_Ref}%
  \BibitemOpen
  \bibfield  {author} {\bibinfo {author} {\bibfnamefont {F.}~\bibnamefont
  {Chen}}, \bibinfo {author} {\bibfnamefont {A.-G.}\ \bibnamefont {Xu}},
  \bibinfo {author} {\bibfnamefont {G.-C.}\ \bibnamefont {Zhang}},\ and\
  \bibinfo {author} {\bibfnamefont {Y.-J.}\ \bibnamefont {Li}},\ }\bibfield
  {title} {\bibinfo {title} {Multiple-relaxation-time lattice {B}oltzmann
  approach to {R}ichtmyer{\textemdash}{M}eshkov instability},\ }\href
  {https://doi.org/10.1088/0253-6102/55/2/23} {\bibfield  {journal} {\bibinfo
  {journal} {Communications in Theoretical Physics}\ }\textbf {\bibinfo
  {volume} {55}},\ \bibinfo {pages} {325} (\bibinfo {year} {2011})}\BibitemShut
  {NoStop}%
\bibitem [{\citenamefont {Ben-Dor}\ and\ \citenamefont
  {Glass}(1979)}]{DoubleMachRef_Exp}%
  \BibitemOpen
  \bibfield  {author} {\bibinfo {author} {\bibfnamefont {G.}~\bibnamefont
  {Ben-Dor}}\ and\ \bibinfo {author} {\bibfnamefont {I.~I.}\ \bibnamefont
  {Glass}},\ }\bibfield  {title} {\bibinfo {title} {Domains and boundaries of
  non-stationary oblique shock-wave reflexions. 1. diatomic gas},\ }\href
  {https://doi.org/10.1017/S0022112079000732} {\bibfield  {journal} {\bibinfo
  {journal} {Journal of Fluid Mechanics}\ }\textbf {\bibinfo {volume} {92}},\
  \bibinfo {pages} {459–496} (\bibinfo {year} {1979})}\BibitemShut {NoStop}%
\bibitem [{\citenamefont {Ben-Dor}(2007)}]{DoubleMachRef_Theory}%
  \BibitemOpen
  \bibfield  {author} {\bibinfo {author} {\bibfnamefont {G.}~\bibnamefont
  {Ben-Dor}},\ }\href {https://doi.org/10.1007/978-3-540-71382-1} {\emph
  {\bibinfo {title} {Shock Wave Reflection Phenomena}}}\ (\bibinfo  {publisher}
  {Springer-Verlag Berlin Heidelberg},\ \bibinfo {year} {2007})\BibitemShut
  {NoStop}%
\bibitem [{\citenamefont {Vevek}\ \emph {et~al.}(2019)\citenamefont {Vevek},
  \citenamefont {Zang},\ and\ \citenamefont {New}}]{DoubleMachRef_Conf}%
  \BibitemOpen
  \bibfield  {author} {\bibinfo {author} {\bibfnamefont {U.~S.}\ \bibnamefont
  {Vevek}}, \bibinfo {author} {\bibfnamefont {B.}~\bibnamefont {Zang}},\ and\
  \bibinfo {author} {\bibfnamefont {T.~H.}\ \bibnamefont {New}},\ }\bibfield
  {title} {\bibinfo {title} {On alternative setups of the double {M}ach
  reflection problem},\ }\href {https://doi.org/10.1007/s10915-018-0803-x}
  {\bibfield  {journal} {\bibinfo  {journal} {Journal of Scientific Computing}\
  }\textbf {\bibinfo {volume} {78}},\ \bibinfo {pages} {1291–1303} (\bibinfo
  {year} {2019})}\BibitemShut {NoStop}%
\bibitem [{\citenamefont {Ha}\ and\ \citenamefont
  {Gardner}(2010)}]{1DJetPaper}%
  \BibitemOpen
  \bibfield  {author} {\bibinfo {author} {\bibfnamefont {Y.}~\bibnamefont
  {Ha}}\ and\ \bibinfo {author} {\bibfnamefont {C.~L.}\ \bibnamefont
  {Gardner}},\ }\bibfield  {title} {\bibinfo {title} {Positive scheme numerical
  simulation of high {M}ach number astrophysical jets},\ }\href
  {https://doi.org/10.1007/s10915-007-9165-5} {\bibfield  {journal} {\bibinfo
  {journal} {Journal of Scientific Computing}\ }\textbf {\bibinfo {volume}
  {34}},\ \bibinfo {pages} {247} (\bibinfo {year} {2010})}\BibitemShut
  {NoStop}%
\end{thebibliography}%

\end{document}